\documentclass[aps,prc,reprint,superscriptaddress,showpacs,amsmath,amssymb,lengthcheck]{revtex4-2}
\pdfoutput=1
\usepackage{graphicx}
\usepackage{newtxtext,newtxmath}

\begin{document}
\preprint{YITP}
\title{Optimization of basis functions for the multi-configuration mixing using the replica exchange Monte-Carlo method and its application to $^{12}$C}

\author{Takatoshi Ichikawa}%
\affiliation{SIGMAXYZ Inc., 4-1-28 Toranomon, Minato-ku, Tokyo 105-0001, Japan}
\affiliation{Yukawa Institute for Theoretical Physics, Kyoto University,
Kyoto 606-8502, Japan}
\author{Naoyuki Itagaki}
\affiliation{Yukawa Institute for Theoretical Physics, Kyoto University,
Kyoto 606-8502, Japan}

\begin{abstract}
\begin{description}
\item[Background] To calculate excited states in quantum many-body systems, multi-configuration mixing has often been employed. However, it has been still unclear how to choose important Slater determinants from a huge model space.
\item[Purpose] We propose a novel efficient method as the replica exchange Monte-Carlo (RXMC) method to sample important Slater determinants and optimize and analyze the obtained results. As an application, we apply it to the ground and excited states of $^{12}$C based on the Bloch-Brink $\alpha$ cluster model and show the detailed structure of the obtained states.
\item[Methods] The RXMC method enables us to efficiently sample Slater determinants following the Boltzmann distribution on the multi-dimensional potential energy surface (PES) under a given model space. To analyze the obtained excited states, we embed sampled basis functions onto the PES calculated with the $\beta$-$\gamma$ constraint method and discuss the main component in each state.
\item[Results] The RXMC method can efficiently perform the samplings with a temperature parameter of $T_L=2.5$ MeV in $^{12}$C. We obtain the gas-like state with a wide density distribution in the tail part in the second $0^+$ state. We also obtain the linear-chain-like states with the bending and stretching vibrational modes in the third and fourth $0^+$ states, respectively. In the fifth $0^+$ state, the main component of the basis functions contains expanded equilateral triangle configurations.
\item[Conclusions] The second $0^+$ gas-like state emerges at the local minimum in the PES, which is the beginning of the valley structure connected to the linear-chain breakup channel. The third and fourth linear-chain-like states emerge in this valley structure. We conclude that the RXMC method is a powerful method to calculate the excited states of nuclei, which would be applied to much complicated nuclear fission dynamics in heavier nuclei. 
\end{description}
\end{abstract}


\maketitle

\section{Introduction}
A wide variety of excited states emerges with complicated dynamics in quantum many-body systems. An essential problem in nuclear physics is how to describe these dynamics beyond the description of the single Slater determinant based on the mean-field picture. A successful method to describe excited states has been the multi-configuration mixing, that is, the superposition of many Slater determinants. It is also often called the generator coordinate method (GCM)~\cite{PhysRev.89.1102,SV98}. However, it has been still unclear how to choose important Slater determinants from a huge model space. Unless we find a good prescription to choose those, it will be difficult to describe entirely complicated nuclear fission dynamics in heavy-mass nuclei. 

In order to resolve this, many prescriptions have been proposed. In the few-body model, Suzuki and Varga have proposed stochastic sampling ~\cite{10.1143_PTPS.146.413}. In the nuclear shell model, the Monte-Carlo shell model has been successful in describing excited states, especially in medium-heavy nuclei \cite{MCSM01}. In this method, single Slater determinants are chosen by adding fluctuations to the energy minimum obtained by the $\beta$-$\gamma$ constraint \cite{MCSM12}. In the nuclear cluster model, Suhara and Kanada-En'yo have proposed the $\beta$-$\gamma$ constraint to choose many Slater determinants, while parameters contained in each Slater determinant are variationally optimized under the constraints~\cite{Suh10}.
An advantage of the cluster model can clearly define important collective coordinates such as the inter-cluster distance, that is, easily coupled with the GCM method by giving the well-defined collective path.
In the mean-field model, the imaginary-time evolution method has been proposed by Fukuoka {\it et al.} \cite{Fuk13}. 

We have also developed the Bloch-Brink $\alpha$ cluster model with the stochastic sampling method~\cite{PhysRevC.83.061301,PhysRevC.86.031303}. In this method, many Slater determinants are sampled by randomly generating center-of-mass coordinates of $\alpha$ clusters based on a gaussian distribution. However, this method has a defect that the more significant part of Slater determinants less contributes to lowering the total energy due to random sampling. 
This defect is crucial when dealing with the unbound states; the inclusion of many basis states creates many continuum states, which shade the physical resonance states. 
We have suffered from contaminations of unphysical spurious states in the calculations. In addition, many Slater determinants identical under the rotational symmetry are contaminated, causing unphysical small splits of a state due to the numerical error of the angular momentum projection. Therefore, it is necessary to perform screening for randomly generated Slater determinants with an appropriate prescription.

In this paper, we propose a novel sampling method to generate Slater determinants based on the replica exchange Monte-Carlo (RXMC) method \cite{PhysRevLett.57.2607,Geyer,doi:10.1143/JPSJ.65.1604,doi:10.1063/1.477812}.
The RXMC method has been widely used in the sampling of the Monte-Carlo integration to calculate a posterior probability in the Bayesian inference \cite{Nag08,Nag12}.
In material physics, the RXMC method has been also used in the optimization of ion configurations coupled with large-scale density functional calculations \cite{Kasamatsu_Sugino_2019}.
The RXMC method enables us to obtain the distribution of sampling points following a Boltzmann distribution on an arbitrary multi-dimensional PES. 
The main purpose of this study is to show the efficiency of this method in the calculations of excited states in quantum many-body systems and perform an application for this method in the $^{12}$C nucleus. We also show how to analyze the properties of the obtained excited states using this method.

As an application, the $^{12}$C nucleus is the most suitable system.
The second $0^+$ state, called the Hoyle state, has the character of spatially extended distribution of three $\alpha$ clusters. This prominent character has been attracting interest for decades because this state has been considered to play a crucial role in forming Carbon in stars.
According to the threshold rule, many of the cluster states, including the second $0^+$ state of $^{12}$C, emerge around the threshold energies for corresponding cluster decay channels.
It means that they emerge in energies close to the continuum states. 

This fact causes two severe problems in the calculations;
one is that the description of such weakly bound states requires the superposition of many different configurations, which is necessary for huge model space in numerical calculations. 
Suppose that the first problem is overcome; another problem is that the real state is embedded in the continuum states.
Thus, it is necessary to develop an additional filtering technique to obtain physical states. We will show later that our method performs the screening of important components efficiently, resulting in drastic noise reductions in the calculations.

The rest of the paper is organized as follows. In Sec. II, we describe the theoretical framework and how to generate many Slater determinants. In Sec. III, We present the results of the calculations using the RXMC method and the analysis of each excited state. We summarize our studies in Sec. IV.

\section{Framework}
\subsection{Bloch-Brink $\alpha$ cluster wave function and Hamiltonian}
We here describe the $^{12}$C nucleus using the Bloch-Brink $\alpha$ cluster model~\cite{Brink} to assess the performance of the Markov-chain Monte-Carlo method for optimizing base functions. Although we attempt to adapt our method for a simple three $\alpha$'s structure in this paper, where the contribution of the spin-orbit interaction cannot be taken into account, our method can be easily extended to other models such as the antisymmetrized molecular dynamics (AMD) model and the stochastic-variation correlated Gaussian model.

In the conventional Bloch-Brink $\alpha$ cluster model, the spacial part of the wave function for an $\alpha$ cluster is described by a simple Gaussian packet, given by $\phi_\alpha({\vec
R})\propto\prod_{i=1}^4\exp[-\nu(\vec r_i-\vec R)^2]$, where $\vec R$
is the center position vector of the Gaussian function, $\vec r_i$
is the spatial coordinate of each nucleon, and $\nu$ is the
size parameter. We take $\nu=1/2b^2$ with $b=1.46$ fm in the
calculations. To describe the $\alpha$ cluster, we take the same $\vec
R$ for all four nucleons.
Thus, the wave function of $^{12}$C can be described by the anti-symmetrized product of three $\alpha$ clusters, given by $\Phi=[\mathcal{A} \phi_\alpha(\vec{R_1})\phi_\alpha(\vec{R_2})\phi_\alpha(\vec{R_3})]$, where $\mathcal{A}$ is the anti-symmetrization operator.
Here, we always impose the center-of-mass position, $\vec{R}_G$, of all clusters at the origin, {\it i.e.}, $\vec{R}_G=(4\vec{R}_1+4\vec{R}_2+4\vec{R}_3)/A=0$, where $A$ is the mass number.

The wave function, $\Phi$, of $^{12}$C is thus given by
\begin{equation}
|\Psi\rangle =\hat{P}^{\pi}\hat{P}^{J}_{MK}|\Phi\rangle,
\end{equation}
where $\hat{P}^{\pi}$ is the parity projection operator and $\hat{P}^{J}_{MK}$ is the angular momentum projection operator.
In the calculations, the angular momentum projection onto $J=0$, $M=0$, and $K=0$ is performed by the numerical integration with $32\times 32\times 32$ grid points for the $\alpha$, $\beta$, and $\gamma$ directions of the Euler angle. 

For the Hamiltonian operator, $\hat{H}$, we take the following form:
\begin{equation}
\hat{H}=\sum_{i}^{A}\hat{t}_i-\hat{T}_{\rm c.m.}+\sum_{i>j}^{A}\hat{v}_{ij}.
\end{equation} 
where $\hat{t}_i$ is the kinetic energy of the $i$th nucleon, $\hat{T}_{\rm c.m.}$ is the center-of-mass kinetic energy, and $\hat{v}_{ij}$ is the two-body interaction consisting of the central and coulomb parts. In the calculations, the spurious center-of-mass motions are exactly removed. For $\hat{v}_{ij}$, we employ the Volkov No.~2 effective $N$-$N$ potential, given by 
\begin{equation}
v(r)=(W-MP^{\sigma}P^{\tau})\sum_{k=1,2}V_k\exp{(-r^2/c_k^2)},
\end{equation}
where $P^{\sigma}$ and $P^{\tau}$ are the spin and iso-spin exchange operators, respectively, and $W=1-M$ with the Majorana exchange parameter $M$.
In the Volkov No.~2, $V_k=\{-60.65,~61.14\}$ MeV and $c_k=\{0.308642,~0.980296\}$ fm are used. We choose $M=0.60$ so as to reproduce the $\alpha$-$\alpha$ scattering phase shift.
This Majorana parameter gives a reasonable value for the ground state energy of $^{12}$C. 

To calculate excited states of $^{12}$C, we superimpose many $\Psi$'s. The total wave function, $\bar{\Psi}$, consists of the linear superposition of $\Psi$'s given by $|\bar{\Psi}\rangle=\sum_i c_i|\Psi_i\rangle$, where $c_i$ is the real coefficient calculated by the diagonalizion of the Hamiltonian matrix elements. That is, we calculate $\sum_{ij}(h_{ij}-w_{ij})c_i c_j=0$, where $h_{ij}$ is the Hamiltonian matrix elements $h_{ij}=\langle \Psi_i|\hat{H}|\Psi_j\rangle$ and $w_{ij}$ is the overlap matrix elements $w_{ij}=\langle \Psi_i|\Psi_j\rangle$.

\subsection{Principal axes and $\beta$-$\gamma$ deformation parameter}
Before the angular momentum projection, we transform the coordinates of bases functions so that the $z$ axis coincides with the principal axis. This transformation is essential because even if two bases functions have the same configuration under the rotational isomorphism, their energies are slightly different due to the numerical accuracy of the angular momentum projection. This leads to unphysical splits of calculated excitation energies when we superimpose many bases.

To avoid this, we calculate the quadrupole moment tensor, $Q_{kl}$, of the basis function, given by
\begin{equation} 
\hat{Q}_{kl}=\sum_i 3\hat{r}_{i,k}\hat{r}_{i,l}-\delta_{kl} \hat{r}^2_i,
\end{equation}
where $k$ and $l$ indices the $x$, $y$, and $z$ axes. The operators $\hat{r}_{i,x}$, $\hat{r}_{i,y}$, and $\hat{r}_{i,z}$ indicate the position operators $\hat{x}_i$, $\hat{y}_i$, and $\hat{z}_i$ for the $i$th nucleon, respectively. The operator $\hat{r}^2_i$ indicates $\hat{x}^2_i+\hat{y}^2_i+\hat{z}^2_i$.
Since the matrix element $Q$, the expectation values $\langle\hat{Q}_{kl}\rangle$, is not invariant under rotations of the coordinate frame, there is a preferred coordinate system, {\it i.e.}, the system of principal axes. The coordinate system is defined by diagonalizing $P^{-1}QP=\bar{Q}$, where $\bar{Q}$ is the diagonal matrix of eigenvalues and $P$ is the matrix transforming into the principal-axis frame. Here, the trace condition of $\mathrm{tr}~\bar{Q}=0$ is always held. In transforming to the principal-axis frame, we choose the directions with the largest and smallest eigenvalues as the $z$ and $x$ axes, respectively. That is, the $y$ axis corresponds to the unstable rotational axis. The quadrupole matrix in the principal-axis frame thus has only three non-vanishing entries, $\langle\hat{Q}_{xx}\rangle$, $\langle\hat{Q}_{yy}\rangle$, and $\langle\hat{Q}_{zz}\rangle$. 

Next, we define the deformation parameter in this principal-axis frame.
All definitions shown here follow Ref.~\cite{MARUHN20142195}.
Using the quadrupole tensor in the principal-axis frame, we define the quadrupole moments, $Q_{20}$ and $Q_{22}$, given by
\begin{align}
Q_{20} &= \sqrt{\frac{5}{16\pi}}q_3=\sqrt{\frac{5}{16\pi}}(2\langle z^2\rangle-\langle x^2\rangle-\langle y^2\rangle),\\
Q_{22} &= \sqrt{\frac{5}{96\pi}}(q_2-q_1)=\sqrt{\frac{15}{32\pi}}(\langle y^2\rangle-\langle x^2\rangle),
\end{align}
where $q_1$, $q_2$, and $q_3$ are the eigenvalues of the multipole moment sorted by ascending order.
To remove a scale effect, they are often expressed as dimensionless quadrupole moments \cite{MARUHN20142195}:
\begin{equation}
a_m=\frac{4\pi}{5}\frac{Q_{2m}}{AR^2},
\label{defa}
\end{equation}
where $R = r_0A^{1/3}$ using a fixed radius derived from the total mass number A with $r_0=1.2$ fm.
In the principal-axis frame, we obtain a unique characterization of the shape of the nucleus with the conditions $a_{\pm 1}=0$ and $a_2=a_{-2}$. 
There remain only two shape parameters $a_0$ and $a_2$. Using these parameters, we can reexpress the total deformation $\beta$ and triaxiality $\gamma$, often called the Bohr-Mottelson deformation parameter \cite{Bo52,Bo54}, given by
\begin{align}
\beta&=\sqrt{a_0^2+2a_2^2},\\ 
\gamma&=\arctan \left(\frac{\sqrt{2}a_2}{a_0}\right).
\end{align}
The triaxiality $\gamma$ would be interpreted as an angle. It can, in principle, take all values between $0^\circ$ and $360^\circ$, but physically relevant parameters stay in the $0^\circ\dots60^\circ$ range. The other sectors correspond to equivalent configurations \cite{GR96}.

To eliminate spurious energy splits coming from the numerical error for reflection symmetry of each axis, we also calculate octupole moments, $Q_{30}^{(x)}$, $Q_{30}^{(y)}$, and $Q_{30}^{(z)}$ for the x, y, and z axes, given by
\begin{align}
Q_{30}^{(x)} = 2\langle x^3\rangle-3\langle xy^2\rangle-3\langle xz^2\rangle, \\
Q_{30}^{(y)} = 2\langle y^3\rangle-3\langle yz^2\rangle-3\langle yx^2\rangle, \\
Q_{30}^{(z)} = 2\langle z^3\rangle-3\langle zx^2\rangle-3\langle zy^2\rangle.
\end{align}
In the calculations, we always choose the sign of each axis so as to be $Q_{30}>0$.

\subsection{$\beta$-$\gamma$ constraint calculation}
To analyze the structure of cluster wave functions obtained by the RXMC method, we project basis functions onto the PES of the two-dimensional $\beta$-$\gamma$ plane as just an eye guide. Note that in reality, cluster dynamics arise in complicated multi-dimensional space where the degree of freedoms is $3A$. This often leads to incorrect mapping of the dynamics because the projection onto a restricted two-dimensional space is a specific cross-section of the multi-dimensional space. One of the authors (TI) has discussed such a problem of incorrect mapping in nuclear fission dynamics in Ref.~\cite{PhysRevC.79.064304}. 
Fortunately, three $\alpha$'s system has very simple triangular structures, and this misleading would be unlikely.

To obtain the PES of the $\beta$-$\gamma$ plane, we minimize the effective Hamiltonian, $H'$, by adding constraint terms to $H$ with the following form:
\begin{multline}
H'=H+\eta\left[ (a_0-\overline{a}_0)^2+(a_2-\overline{a}_2)^2\right] \\
+ \eta_0 |\vec{R}_G|^2+\eta_1\left[\langle\hat{Q}_{xy}\rangle^2+\langle\hat{Q}_{xz}\rangle^2+\langle\hat{Q}_{yz}\rangle^2\right],
\end{multline}
where $\overline{a}_0$ and $\overline{a}_2$ are desirable values obtained by minimization. 
The parameters $\eta_0$ and $\eta_1$ are constraint on the center-of-mass position and the principal-axis frame of the system, respectively.
In the calculations, we take $\eta_0\sim10$ and $\eta_1\sim1$. For the parameter $\eta$, we increase its value until desirable values are obtained. To minimize $H'$, we adapt the conjugate gradient method with the numerical derivative using the two-points formula.

\subsection{Replica exchange Monte-Carlo method}
The main purpose of this paper is how to efficiently sample basis functions from a huge model space with the coordinates $\vec{\bf R}=\{\vec{R}_{1}. \vec{R}_{2}\dots \vec{R}_{A/4}\}$. An ideas is sampling so that the distribution of basis functions, $D(\vec{\bf R})$, is in proportion to the Boltzmann distribution $D(\vec{\bf R}) \propto e^{-E(\vec{\bf R})/T}$ with the potential energy $E(\vec{\bf R})=\langle \Psi(\vec{\bf R})|\hat{H}|\Psi(\vec{\bf R})\rangle$ and a temperature parameter $T$. 
This is because we would like to intensively pick sample points up around minima in the multi-dimensional PES.
This can be achieved by the Markov-chain Monte Carlo (MCMC) method.
It is mathematically guaranteed that sampling points obtained by the MCMC method obey the Boltzmann distribution under the potential energy and a given temperature. 

However, the conventional MCMC method fails in samplings when a deep minimum exists in its potential energy. That is, a Markov-chain sampling point cannot escape from the deep minimum, resulting in distorted distributions of sampling points. In addition, other important components would be missing due to the lack of model space explored by Markov-chain samplings. In order to overcome this defect, the RXMC method has been proposed in the field of Bayesian inference. When the posterior probability is estimated, the RXMC method is often used in samplings of Monte-Carlo integration points necessary for evaluating the normalization factor. 

\begin{figure}[h]
\includegraphics[keepaspectratio,width=\linewidth]{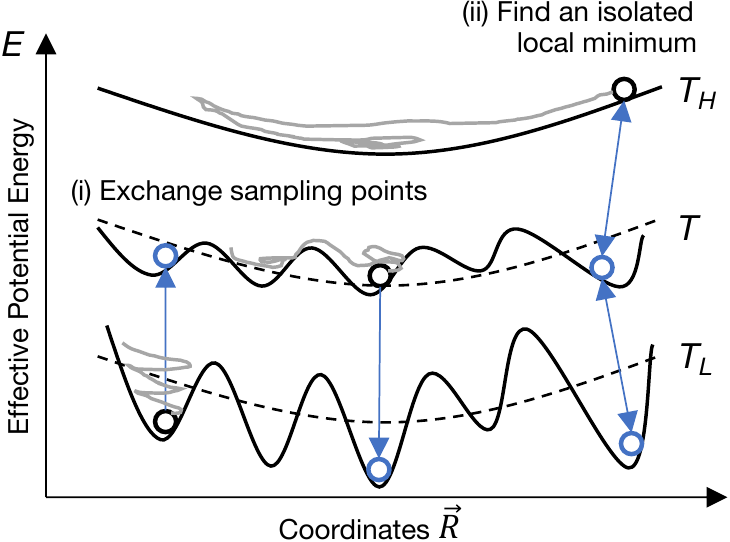}
\caption{Schismatic picture of the replica-exchange Malkov-chain Monte-Carlo method. The solid lines are depicted as effective potential energies with temperature $T_L$, $T$, and $T_H$. Here, we assume $T_L$<$T$<$T_H$. Malkov-chain steps are represented as open circles with gray lines.} 
\label{fig1} 
\end{figure}

Figure \ref{fig1} shows a schematic picture of how a Markov-chain sampling point trapped in a deep minimum escapes from there.
In the figure, the solid lines indicate the effective potential energies depending on the temperature, where the Markov-chain sampling point explores.
The effective potential energy becomes flat with increasing temperature because a Markov-chain sampling point similar to the Brownian particle can easily jump to the next position in higher temperature. 

In the RXMC method, independent Markov-chain samplings are simultaneously performed with several different temperatures.
Markov-chain sampling points are widely explored in model space with increasing temperature because the effective PES becomes flat with increasing temperature.
In contrast, sampling points in lower temperatures may be accidentally trapped in a deep minimum in the effective PES.
Then, sampling points in different temperatures should be exchanged between them with a certain probability. A Markov-chain step restarts from a position of one at a higher temperature, much widely exploring the effective PES [see (i) Fig.\ref{fig1}]. Another advantage is much easier to find an isolated local minimum due to a large barrier for reaching. Markov-chain steps at very high temperatures are freely explored in the effective PES. Thus, the exchange of sampling points allows us to effectively collect important sampling points by bypassing the barrier [see (ii) in Fig.\ref{fig1}]

We here show briefly the formulation of the RXMC method.
The details are shown in Refs. \cite{10.1109/tnn.2008.2000202,10.1016/j.neunet.2011.12.001}.
In the calculations, we independently perform MCMC samplings in different temperatures. The inverse temperature $\beta_l=1/T_l$ at the $l$th sampling is given by
\begin{equation}
\beta_l = \begin{cases}
0 & \text{($l=1$)}\\
\beta_L b^{l-L} & \text{(otherwise)},
\end{cases}
\end{equation}
where $\beta_L$ is our target temperature, $b$ is the common ratio parameter, and $L$ is the total number of samplings independently performed. In the calculations, we take $b=1.5$ and $L=24$, which is often used in the field of the Bayesian inference.

At each temperature, Markov-chain samplings are performed using the Metlopolis method. The next step of the coordinate vector $\vec{\bf R}_t^{(l)}$ is given by
\begin{equation}
\vec{\bf R}_{t+1}^{(l)}= \vec{\bf R}_{t}^{(l)} + \Delta_l\bar{\bf U}_t,
\end{equation}
where $\Delta_l$ is the diffusion parameter at each temperature and $\bar{\bf U}_t$ is the matrix of uniform random numbers generated from $-1$ to $1$.
After $\vec{\bf R}^{(l)}_{t+1}$ is generated, we determine whether it is acceptable or not with a probability $P_{\rm MC}$ given by $P_{\rm MC} = \min{(1,p_{\rm MC})}$ and $p_{\rm MC}$ has the following form: 
\begin{equation}
p_{\rm MC}=e^{-\beta_l\left(E(\vec{\bf R}^{(l)}_{t+1})-E(\vec{\bf R}^{(l)}_{t})\right)}.
\end{equation}
If $\vec{\bf R}^{(l)}_{t+1}$ is rejected, we accept $\vec{\bf R}^{(l)}_t$ as the next positions instead.
We also reject $\vec{\bf R}^{(l)}_{t+1}$, if $|\vec{R_i}^{(l)}_{t+1}| > r_{\max}$ for any $i$, where $r_{\rm max}$ is the maximum radius taken as $r_{\rm max}=6.0$ fm.
Before the calculations, we fine-tune $\Delta_l$ so as to become the acceptance rate $r_{\rm acc}\sim0.5$.

After the next step of the Markov-chain samplings at each temperature is generated, we also exchange between $\vec{\bf R}^{(l)}_{t}$ and $\vec{\bf R}^{(l+1)}_{t}$ in different temperatures with a probability $P_{\rm ex}$. The probability $P_{\rm ex}$ is given by $P_{\rm ex}=\min(1, p_{\rm ex})$ and $p_{\rm ex}$ is chosen as 
\begin{equation}
p_{\rm ex}= e^{(\beta_{l+1}-\beta_l)\left(E(\vec{\bf R}^{(l+1)}_{t})-E(\vec{\bf R}^{(l)}_{t})\right)},
\end{equation}
where $l=\{1,3,5\dots\}$ and $\{2,4,6\dots\}$ are taken for odd and even $t$, respectively.

Before samplings, pre-samplings must be performed until the Markov-chain steps reach thermal equilibrium, which is called "burn-in".
In this paper, we perform burn-in at least more than 5,000 steps.
After the burn-in, we pick 5,000 samples up from the Markov-chain steps at $l=L$.

\subsection{Selection of optimum energy bases}
After 5,000 bases are generated using the RXMC method, we randomly extract 500 bases from them. 
(If the same base as the one already extracted is selected, it is rejected.)
We next calculate the Hamiltonian $h_{ij}$ and overlap $w_{ij}$ matrix elements of the extracted 500 bases. After that, we choose the optimum 100 bases to decrease the energies from the ground to the fifth states.
This additional optimization is very effective for analyzing excited states because the most important bases constructing a core structure of states are preferably collected. Later, we will show how to analyze the structure of excited states.

The procedure for selecting the optimum 100 bases is as follows: (i) take the basis function with the lowest energy from all bases and construct a sub-matrix based on it.
(ii) pick a basis function up from residual bases and add it to the sub matrix elements $h_{ij}$ and $w_{ij}$.
(iii) check eigen values of $w_{ij}$ of the sub matrix. If the lowest eigen value of $w_{ij}$ is smaller than $1.0 \times 10^{-12}$, it is rejected.
(iv) check eigenvalues of $h_{ij}$ of the submatrix. If it gives the lowest energy at the $i$th states compared to all other residual bases, we accept it.
(v) increase the dimension of the submatrix and continue the procedure until the total number of basis functions is satisfied, and all the desired states are optimized.

In the calculations, we choose 100 bases where the excited energies from the first to fifth states are optimized with 20 bases at each state.

\subsection{Kernel density estimation}
\label{KDE}
After sampling states, we estimate a population of the states on the $\beta$-$\gamma$ plane using the kernel density estimation often used in statistics. We calculate the population $f$ of basis functions as a function of the coordinates $a_0$ and $a_2$ defined in Eq.~\ref{defa}:
\begin{align}
    f(a_0,a_2)=\frac{1}{A}\sum_i^n w_i K_\sigma(a_0-\overline{a_0}^{(i)})K_\sigma(a_2-\overline{a_2}^{(i)}),
\end{align}
where $\overline{a_0}^{(i)}$ and $\overline{a_2}^{(i)}$ are the $i$-th sampled data, $n$ is the total number of sampled data, $w_i$ is the weight factor, and $A$ is the normalization factor $A=\sum_i^n w_i$. The symbol $K_{\sigma}$ is the kernel function with the width parameter $\sigma$, which is often given by the gaussian form $K_{\sigma}(x)=1/\sqrt{2\pi}\sigma\cdot e^{-x^2/2\sigma^2}$. The value of $\sigma$ is determined by Scott's Rule \cite{SCOTT92} given by $\sigma=n_{\rm eff}^{-1/(d+4)}$, where $n_{\rm eff}=n$ if $w_i=1$ for all $i$. Otherwise, $n_{\rm eff}=(\sum_i w_i)^2/\sum_i w_i^2$.
In the calculation, we use Python code \texttt{gaussian\_kde} implemented in \texttt{SciPy} \cite{2020SciPy-NMeth}.

\subsection{Density distribution in the laboratory frame}
We also estimate the radial density distribution in the laboratory frame to discuss its tail part of calculated states. The details of the calculation are given in Ref.~\cite{PhysRevC.104.034613}. 
In general, the matrix element of the density operator $ \sum_i \delta({\bf r}_i - {\bf R})$ between the states with the angular momentum  $\langle J'M|$ and $|JM\rangle$
is defined using multipole decomposition as
\begin{multline}
\langle J'M' | \sum_i \delta({\bf r}_i - {\bf R}) |JM\rangle \\
= \sqrt{4\pi} \sum_{\lambda \mu} (JM\lambda \mu|J'M') \rho^{J'J}_\lambda (R) Y^*_{\lambda \mu}(\Omega).
\end{multline}
Here, ${\bf r}_i$ is the physical coordinate of the $i$-th nucleon and $\lambda$ is the rank of the density,
and $\rho^{J'J}_\lambda (R) $ is called transition density. 
We now consider the matrix elements between the $0^+$ states, where
only $\lambda = 0$ contributes, which is the normal scalar density.
According to this definition, the integration of 
$\rho^{00}_0 (R) $ over the radius $R$ is normalized to the number of nucleons, $A$;
\begin{equation}
\int R^2 \rho^{00}_0 (R) dR = A.
\end{equation}

\section{Results and Discussion}
\subsection{Potential energy surface on the $\beta$-$\gamma$ plane for $^{12}$C}
\begin{figure}[htbp]
\begin{center}
\includegraphics[keepaspectratio,width=0.7\linewidth]{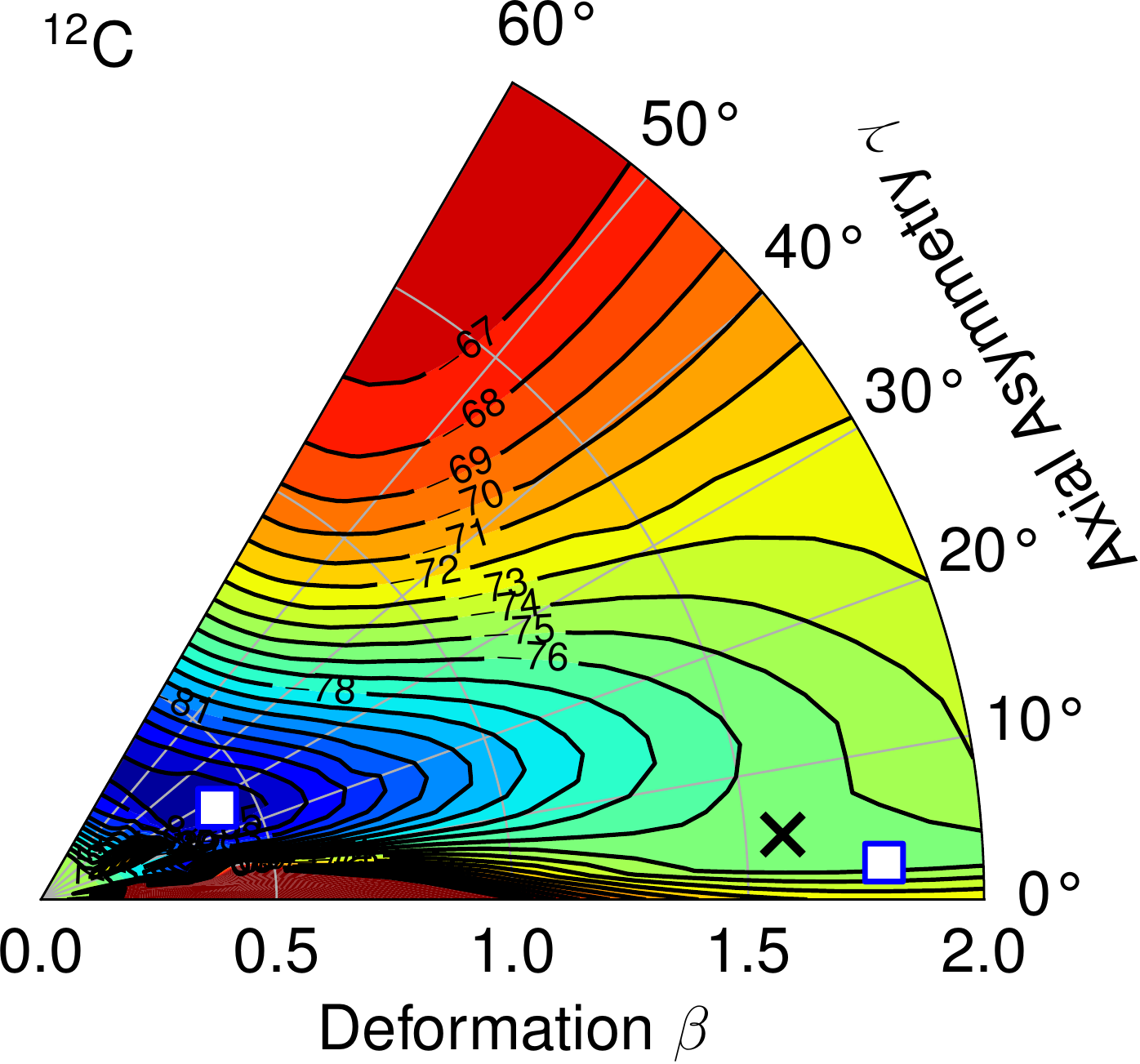}
\end{center}
\caption{Potential energy surface on the $\beta$-$\gamma$ plane for the $0^{+}$ state of $^{12}$C from $\beta=0.0$ to 2.0. The open squares indicate the local minima. The cross indicates the saddle point.}
\label{pot1}
\end{figure}

\begin{figure}[htbp]
\begin{center}
\includegraphics[keepaspectratio,width=0.7\linewidth]{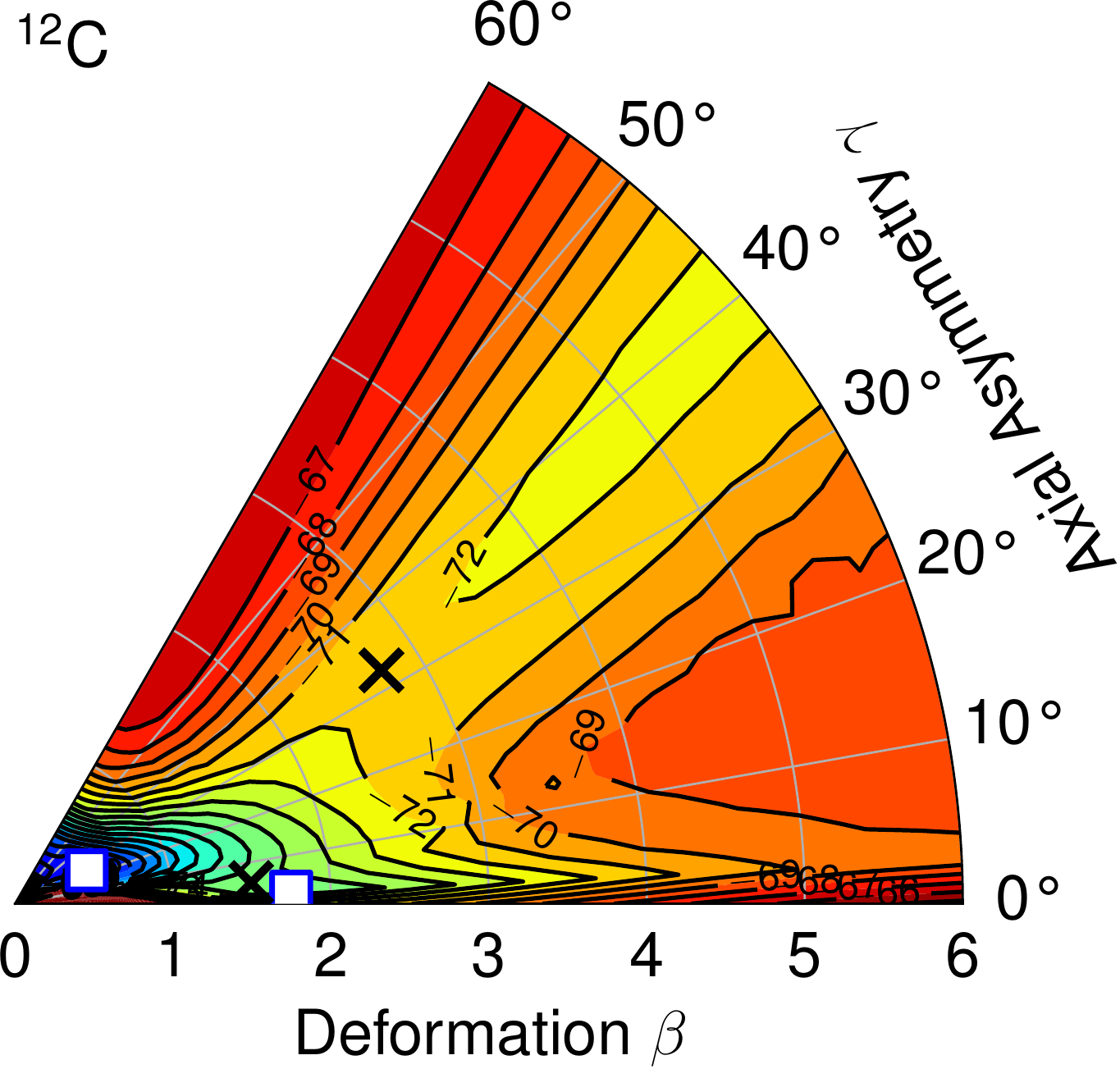}
\end{center}
\caption{Potential energy surface on the $\beta$-$\gamma$ plane for the $0^{+}$ state of $^{12}$C from $\beta=0.0$ to 6.0. The symbols are the same as in Fig.~\ref{pot1}.}
\label{pot2} 
\end{figure}

\begin{figure}[htbp]
\begin{tabular}{cc}
\begin{minipage}{0.5\hsize}
\begin{center}
\includegraphics[keepaspectratio,width=\linewidth]{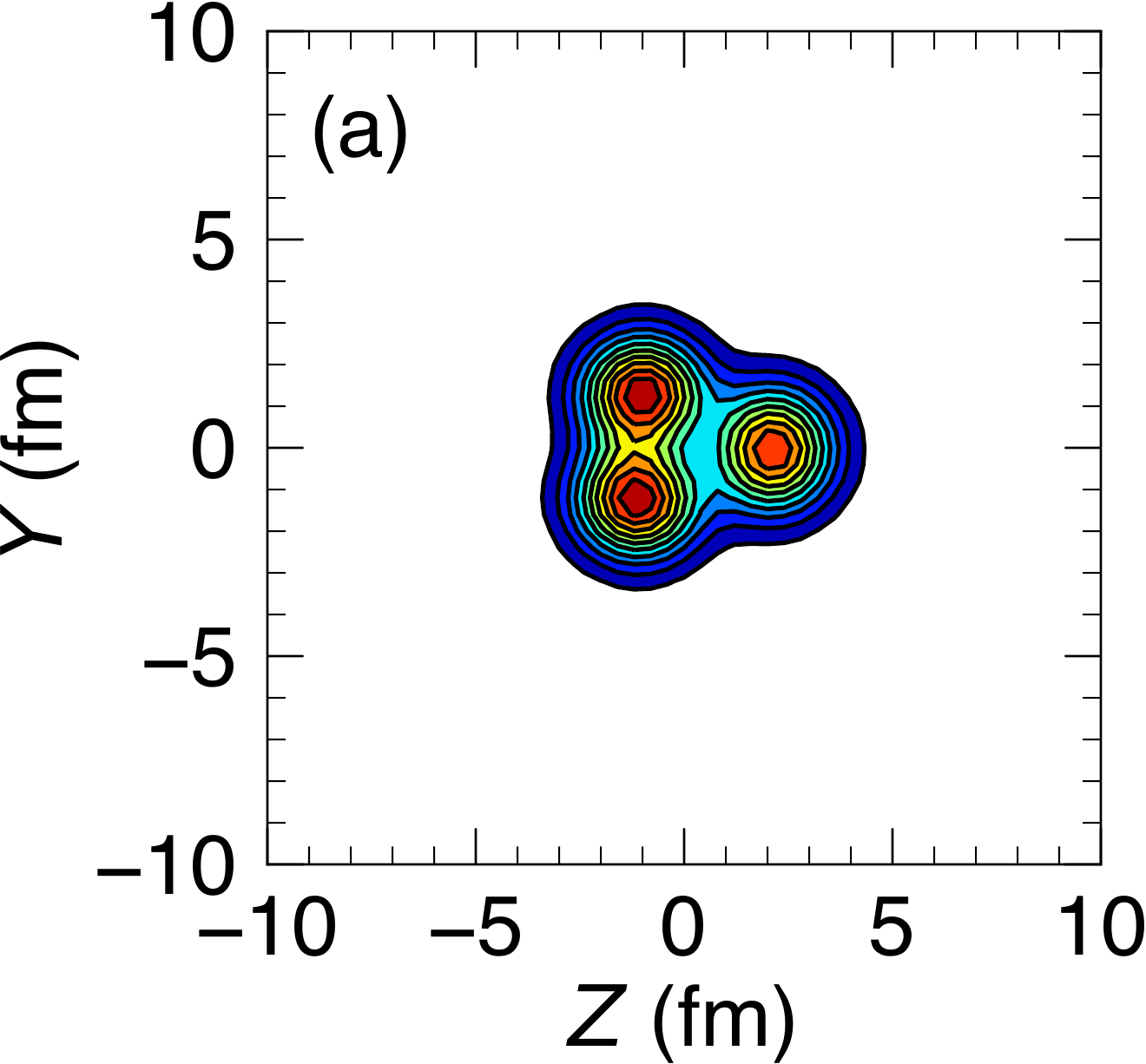}
\end{center}
\end{minipage}
\begin{minipage}{0.5\hsize}
\begin{center}
\includegraphics[keepaspectratio,width=\linewidth]{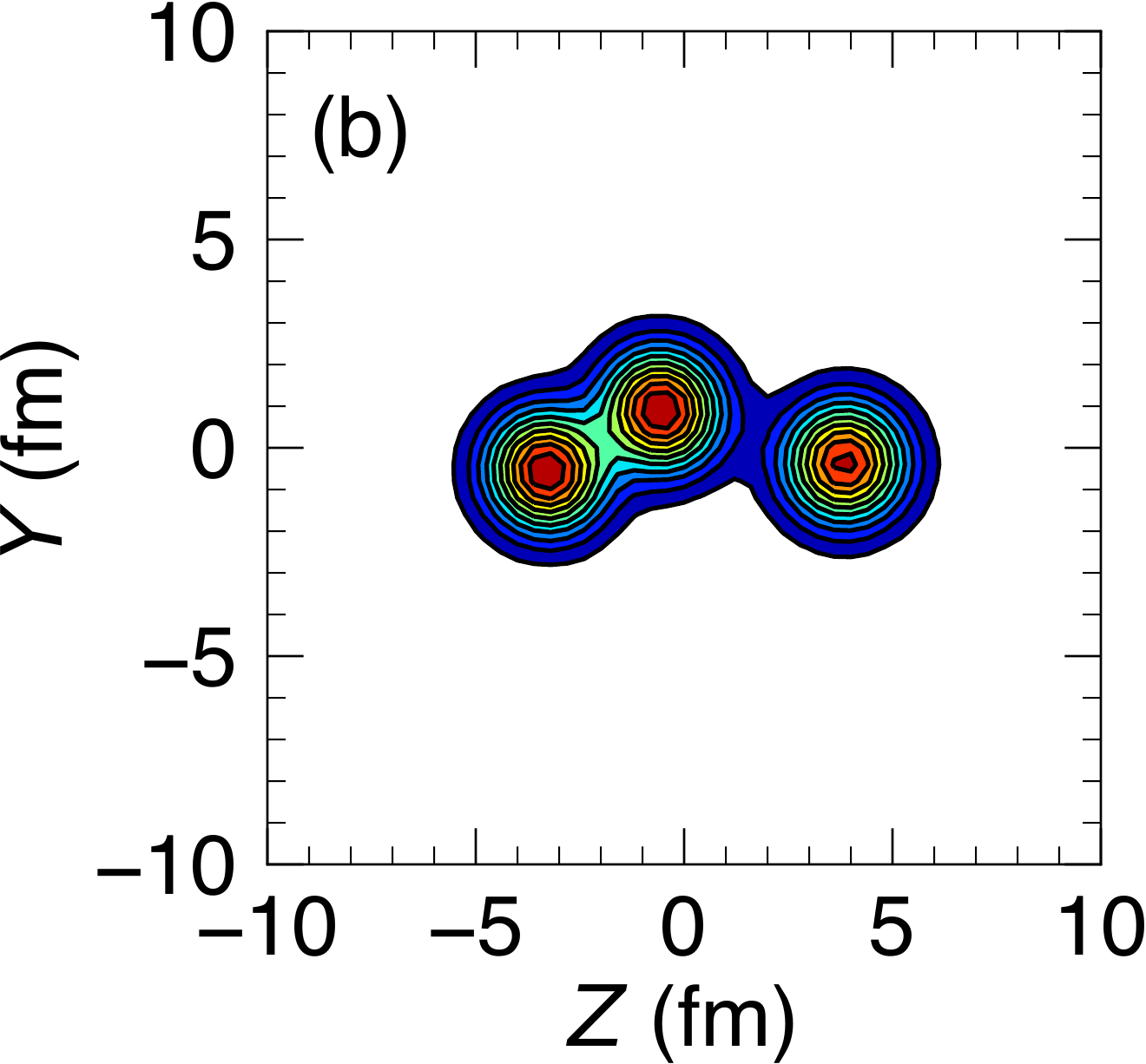}
\end{center}
\end{minipage} \\
\vspace{.1cm}\\
\begin{minipage}{0.5\hsize}
\begin{center}
\includegraphics[keepaspectratio,width=\linewidth]{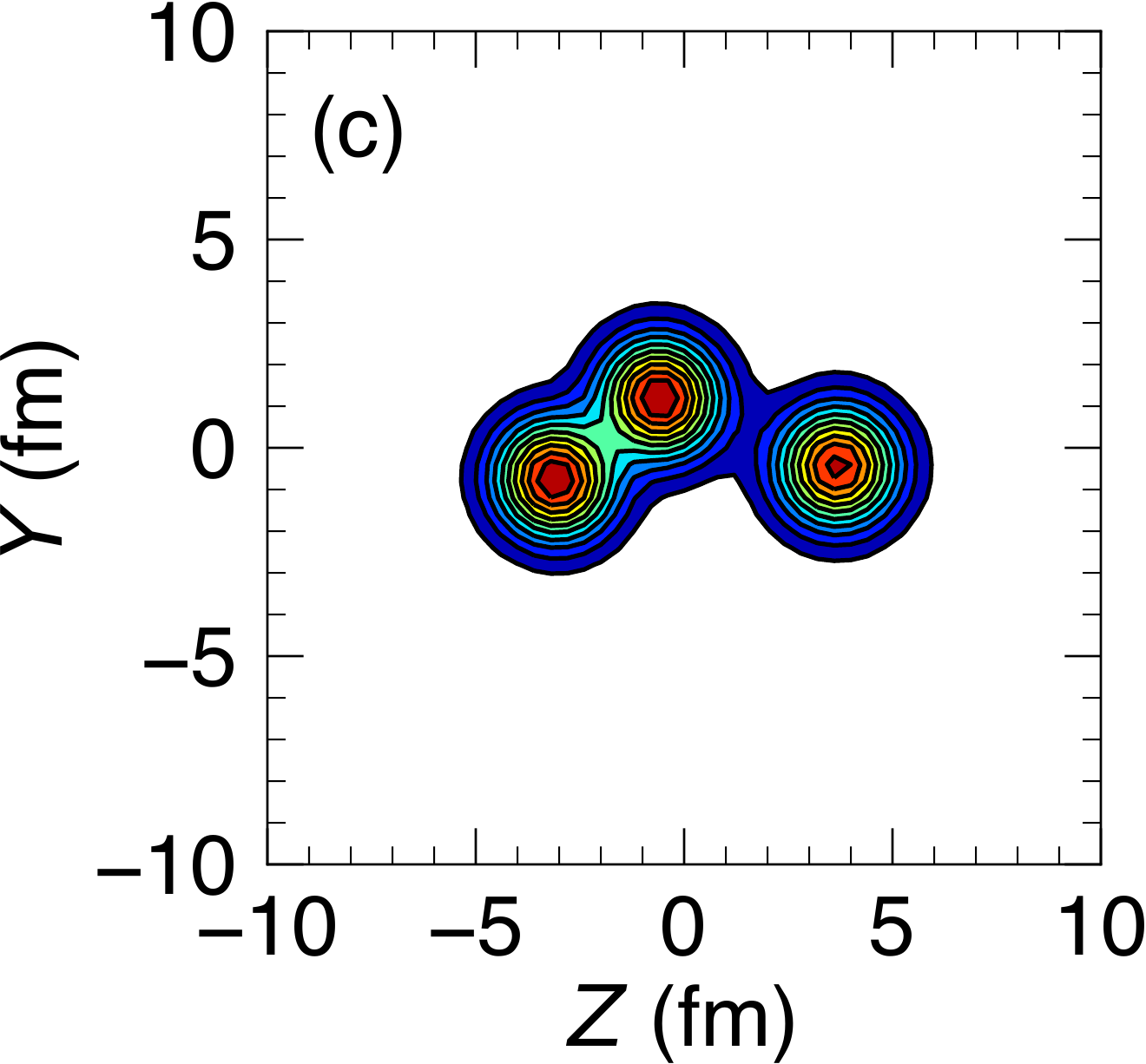}
\end{center}
\end{minipage}
\begin{minipage}{0.5\hsize}
\begin{center}
\includegraphics[keepaspectratio,width=\linewidth]{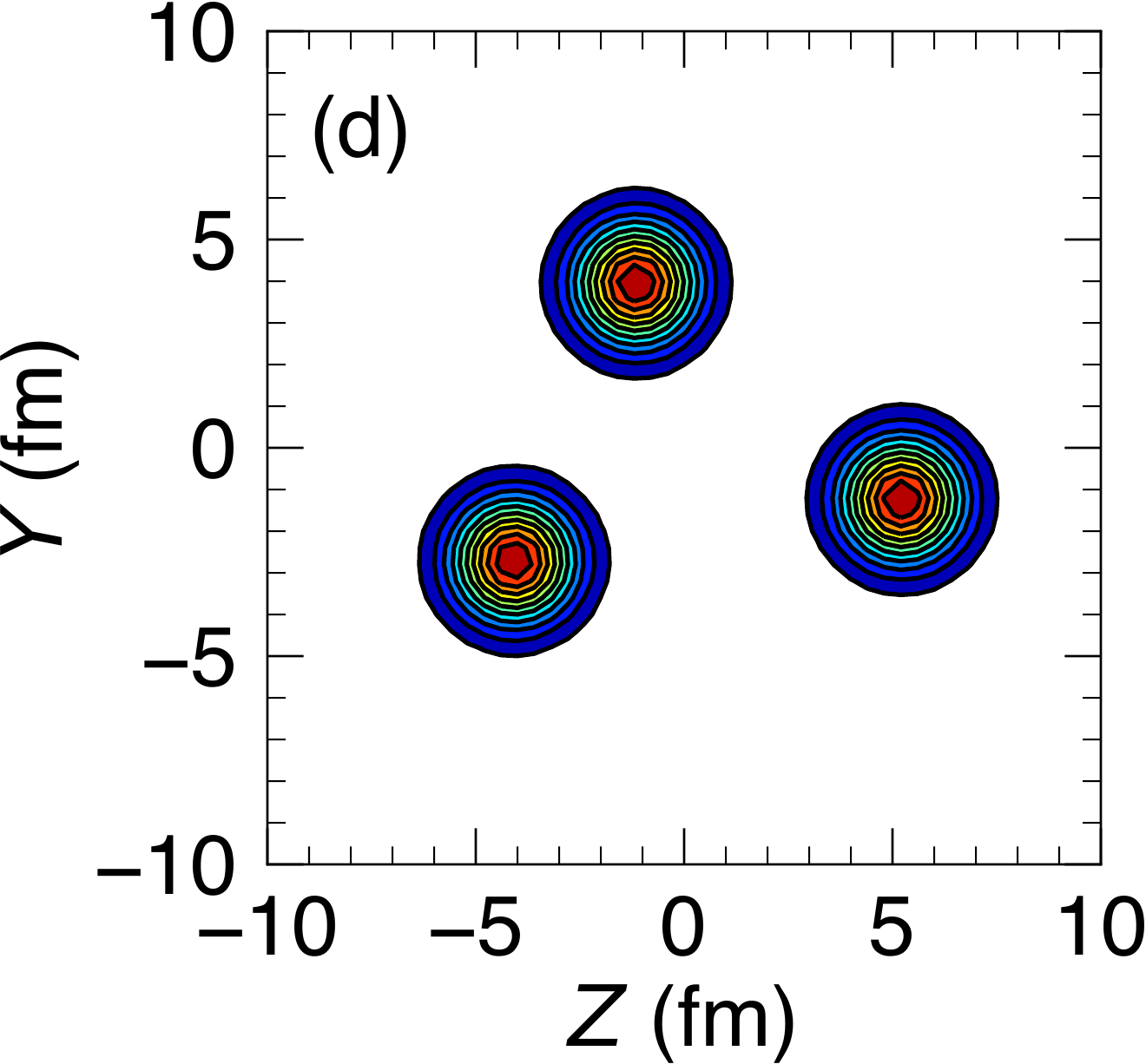}
\end{center}
\end{minipage} \\
\vspace{.1cm}\\
\begin{minipage}{0.5\hsize}
\begin{center}
\includegraphics[keepaspectratio,width=\linewidth]{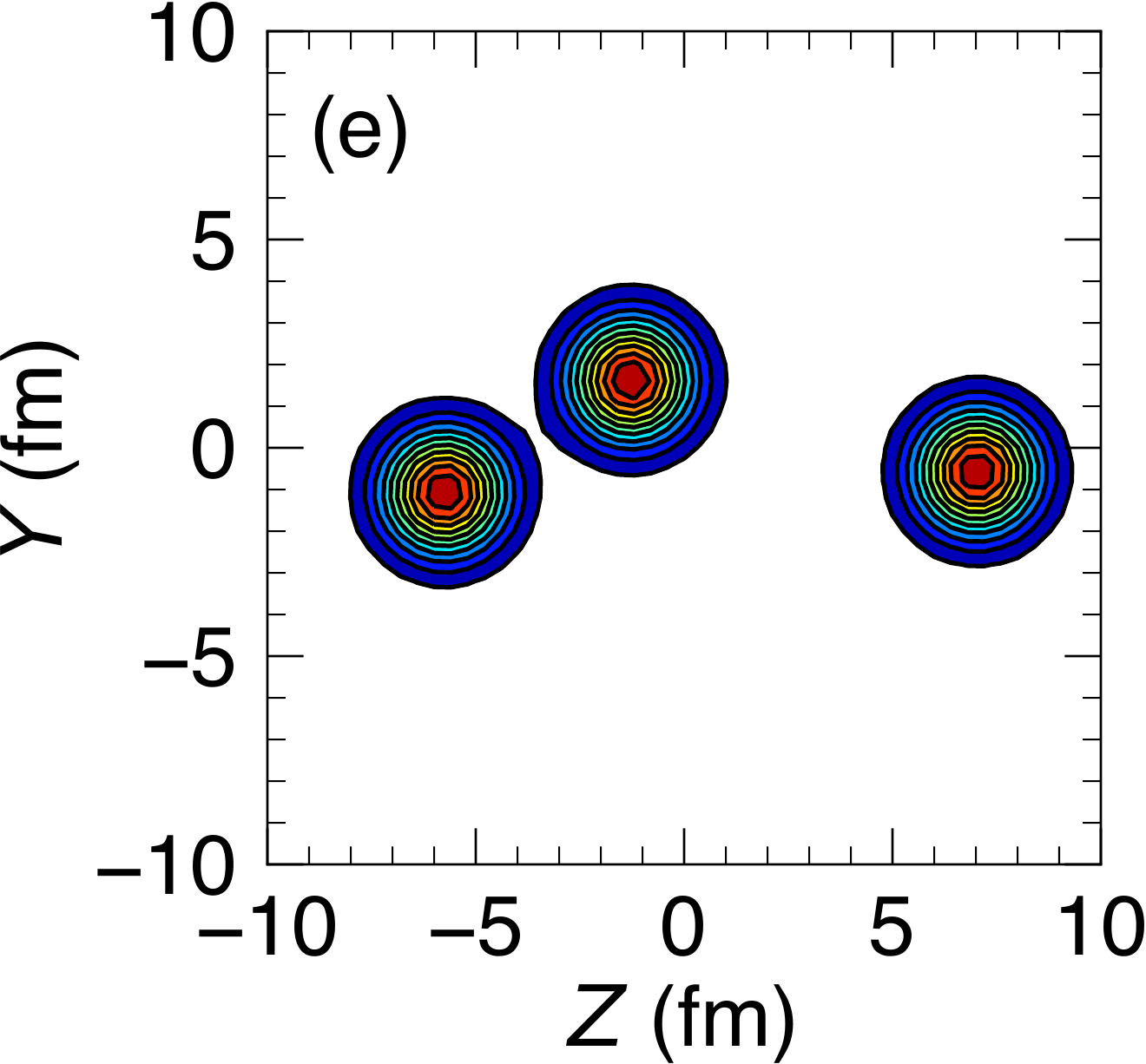}
\end{center}
\end{minipage} \\
\end{tabular}
\caption{Density distributions of (a) the lowest energy minimum at $\beta=0.421$ and $\gamma=27.5^{\circ}$, (b) the local energy minimum at $\beta=1.75$ and $\gamma=5.0^{\circ}$, (c) the inner saddle point at $\beta=1.58$ and $\gamma=5.0^{\circ}$, (d) the triangular saddle point at $\beta=2.75$ and $\gamma=32.5^{\circ}$, and (e) the linear-chain configuration at $\beta=5.75$ and $\gamma=2.5^\circ$.}
\label{den1}
\end{figure}

Before calculating excited states of $^{12}$C, we check the PES on the $\beta$-$\gamma$ plane for the $0^{+}$ state of $^{12}$C.
We perform the constraint calculations from $\beta=0.0$ to 6.0.
Note that we will never apply the wave functions obtained by the constraint calculations to the calculations of excited states shown later.

Figure \ref{pot1} shows the obtained PES from $\beta=0.0$ to 2.0. The open squares indicate the local minima thus obtained. 
The overall structure of the obtained PES is similar to that of Ref.~\cite{Suh10}.
The lowest energy minimum is obtained at $\beta=0.421$ and $\gamma=27.5^{\circ}$ with $E= -85.98$~MeV. The density distribution obtained at this minimum has the compact triangular configuration of three $\alpha$'s shown in Fig.~\ref{den1}(a).

There appears a shallow local minimum at $\beta=1.75$ and $\gamma=5.0^{\circ}$ with $E=-75.80$~MeV in the PES, whose density distribution is shown in Fig.~\ref{den1}(b). A characteristic shoulder structure can be seen around this local minimum. The depth of this minimum is 0.14 MeV and the inner saddle point relative to this minimum is located at $\beta=1.58$ and $\gamma=5.0^{\circ}$ with $E=-75.66$ MeV [the density is given in Fig.~\ref{den1}(c)].This shoulder structure around the shallow local minimum has been well discussed by Tohsaki {\it et al.} in connection to the gas-like state of $^{12}$C based on the Tohsaki-Horiuchi-Schuck-R\"{o}pke wave function~\cite{PhysRevLett.87.192501}.

Figure \ref{pot2} also shows the obtained PES for the $0^{+}$ state of $^{12}$C, but the range of $\beta$ is different. Here, it from $\beta=0.0$ to 6.0 is shown. We can see two prominent potential-energy valleys leading to the exit channels of the equilateral triangular ($\gamma\sim40^{\circ}$) and linear-chain ($\gamma\sim5^{\circ}$) configurations at larger $\beta$. They are well divided by separating ridges for the $\gamma$ direction. Later, we will discuss the height of these separating ridges in Sec.~III-F. The lowest-energy saddle (threshold) point is the equilateral triangular configuration [see Fig.~\ref{den1}(d)] at $\beta=2.75$ and $\gamma=32.5$ with $E=-71.71$~MeV. The height of this saddle point from the lowest energy minimum is $B_f^{\rm (tri)}=14.27$ MeV. It is interesting that a very flat area widely exists around this saddle point. 
In this calculation, we cannot find the saddle point for the linear-chain configuration due to the region of the $\beta$-$\gamma$ constraint calculations.
As a sample, the density distribution at $\beta=5.75$ and $\gamma=2.5^\circ$ is shown in Fig.~\ref{den1}(e).

\subsection{Samplings of basis functions}
We here perform samplings of basis functions for $^{12}$C using the RXMC method and analyze the distributions of the basis functions obtained with different temperatures on the $\beta$-$\gamma$ plane. 

In the calculations, we perform the RXMC samplings with four different lowest temperatures of $T_L=$ 1.00, 1.50, 2.00, and 2.50 MeV. If we assume the microcanonical ensemble and use the level density parameter obtained by the empirical data, $a = A/8$ MeV$^{-1}$, the excitation energies correspond to $E^*\sim$ 1.50, 3.38, 6.00, and 9.38 MeV, calculated with $E^*=aT^2$. We perform burn-in for 5,000 steps before the samplings. We estimate the distribution of the obtained sampling points on the $\gamma$-$\beta$ plane using the kernel density estimation method with the bandwidth taken from Scott's rule given in Sec.~\ref{KDE}. 

\begin{figure}[htbp]
\begin{tabular}{cc}
\begin{minipage}{0.5\hsize}
\begin{center}
\includegraphics[keepaspectratio,width=\linewidth]{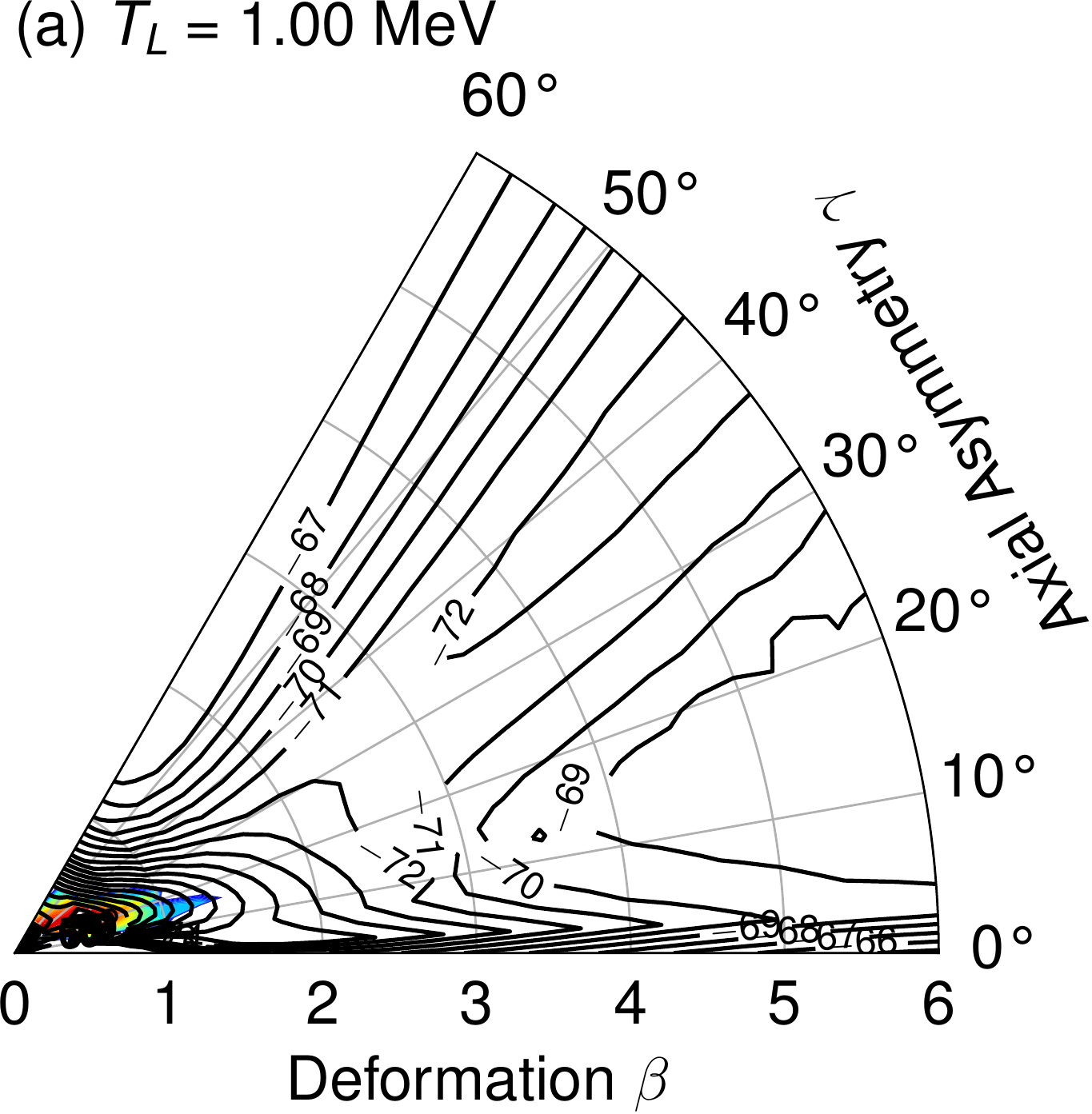}
\end{center}
\end{minipage}
\begin{minipage}{0.5\hsize}
\begin{center}
\includegraphics[keepaspectratio,width=\linewidth]{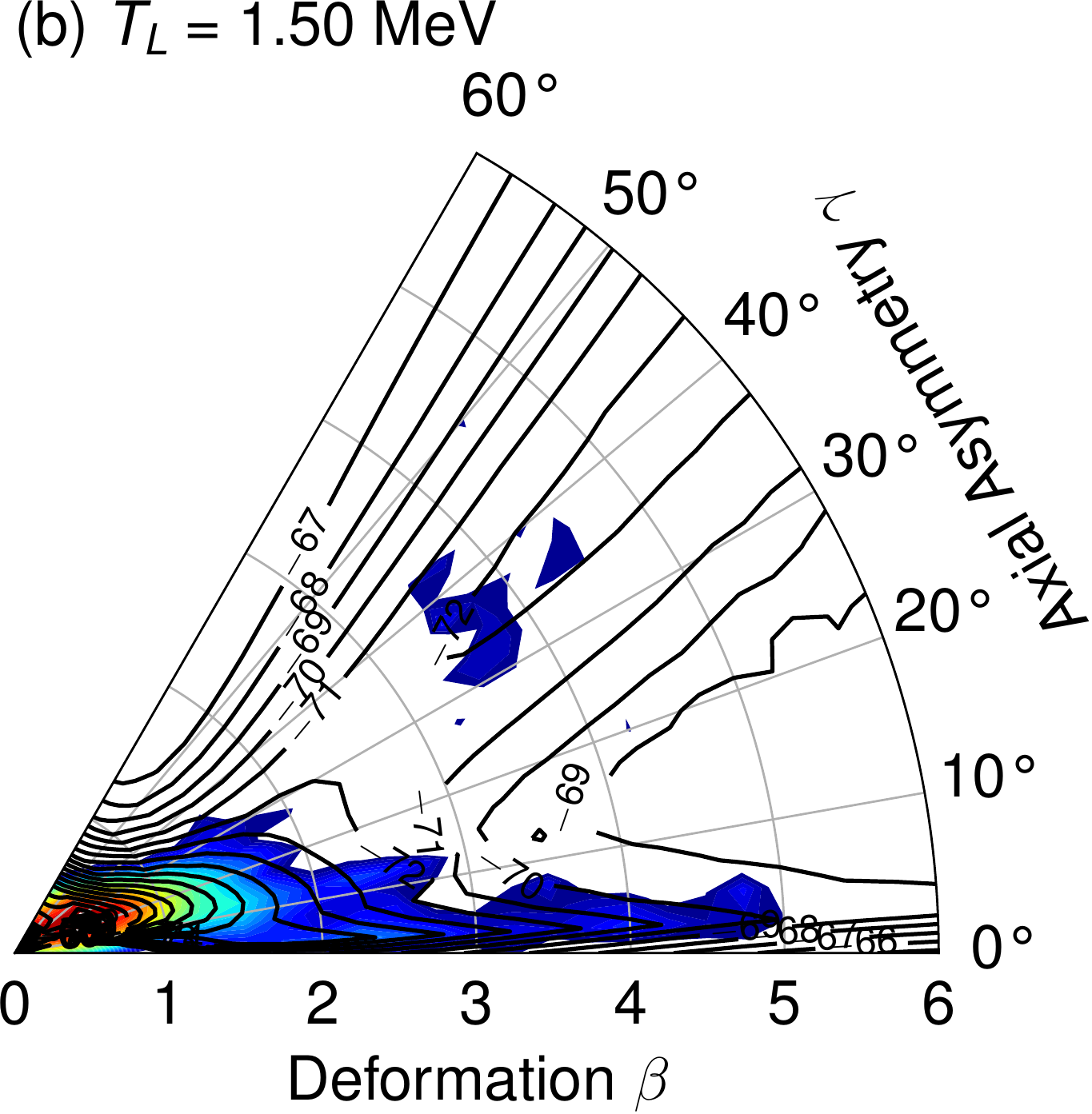}
\end{center}
\end{minipage} \\
\vspace{.1cm}\\
\begin{minipage}{0.5\hsize}
\begin{center}
\includegraphics[keepaspectratio,width=\linewidth]{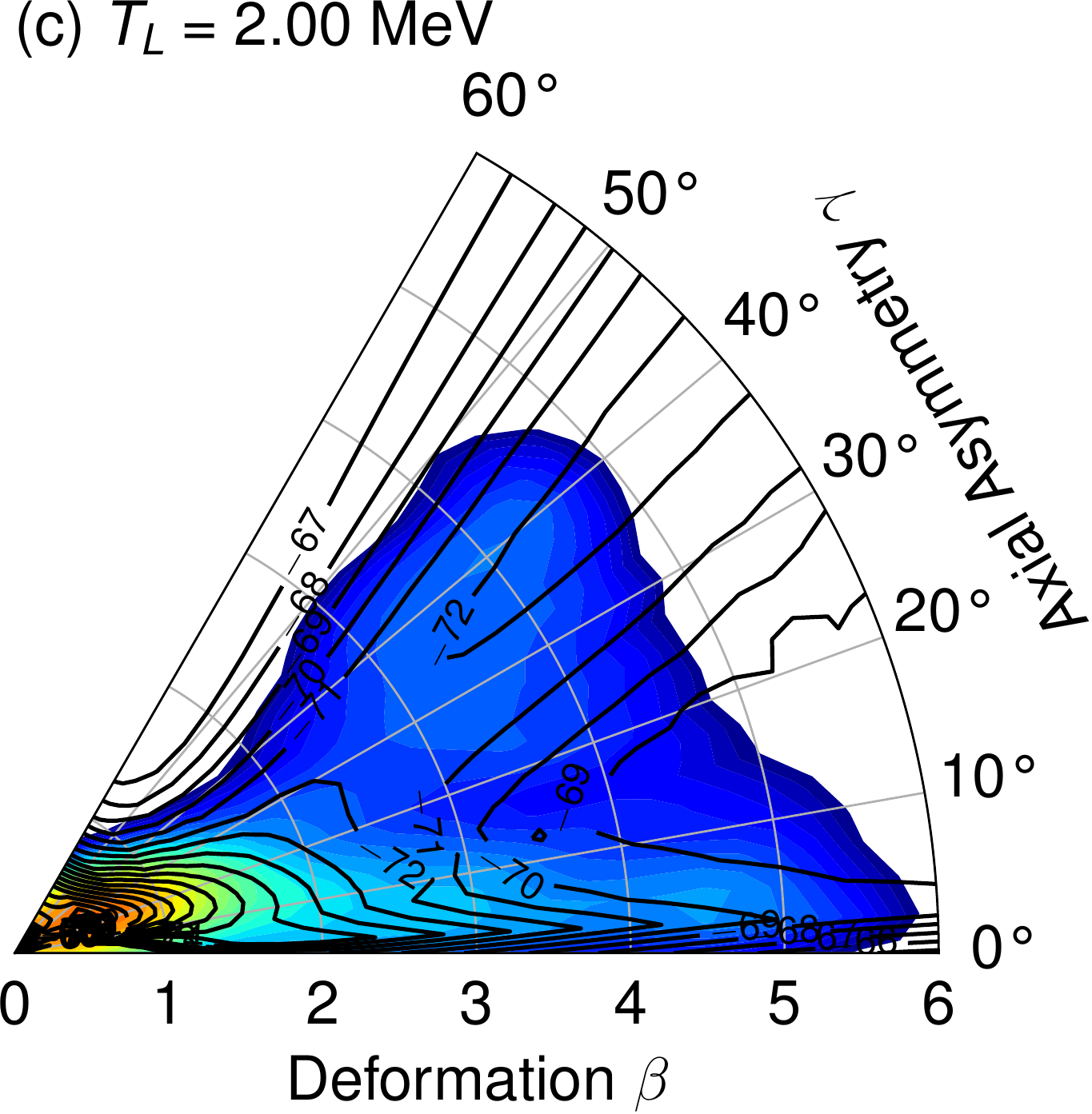}
\end{center}
\end{minipage}
\begin{minipage}{0.5\hsize}
\begin{center}
\includegraphics[keepaspectratio,width=\linewidth]{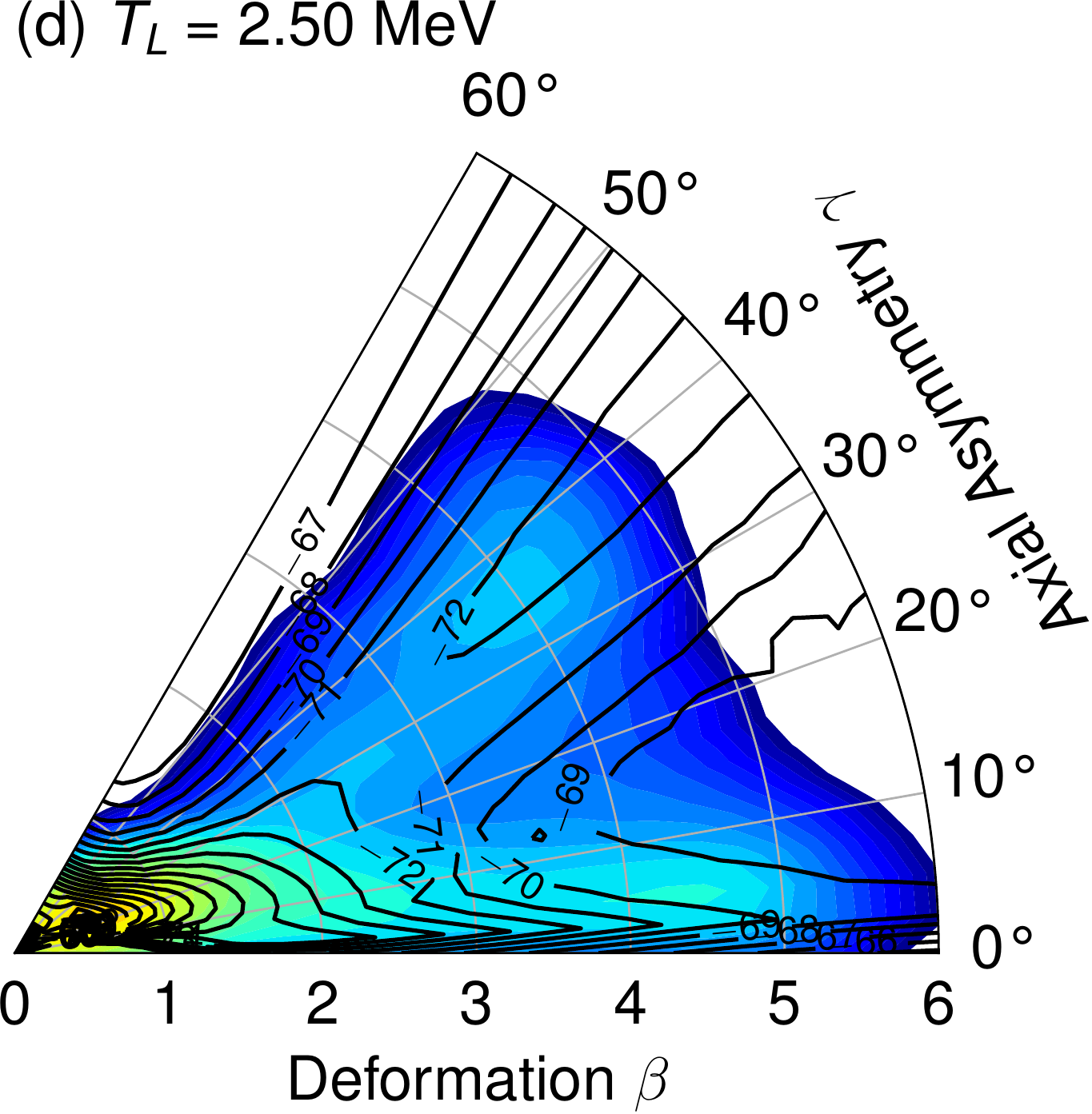}
\end{center}
\end{minipage} 
\end{tabular}
\caption{Distributions of 5,000 basis functions obtained by the RXMC method with temperatures $T_L=$ (a) 1.00, (b) 1.50, (c) 2.00, and (d) 2.50 MeV.
The contour corresponds to logarithm steps of 0.1 from $10^{-2}$ to 10.
}
\label{dis1}
\end{figure}

Figure \ref{dis1} shows the distributions thus obtained. We can see that the obtained distributions are spread along two valleys with increasing temperatures.
It is emphasized that we do not consider any results of the constraint calculations in the RXMC method. The Markov-chain steps automatically follow the energy-minimum paths with the Boltzmann distribution.

\begin{figure}[htbp]
\begin{tabular}{cc}
\begin{minipage}{0.5\hsize}
\begin{center}
\includegraphics[keepaspectratio,width=\linewidth]{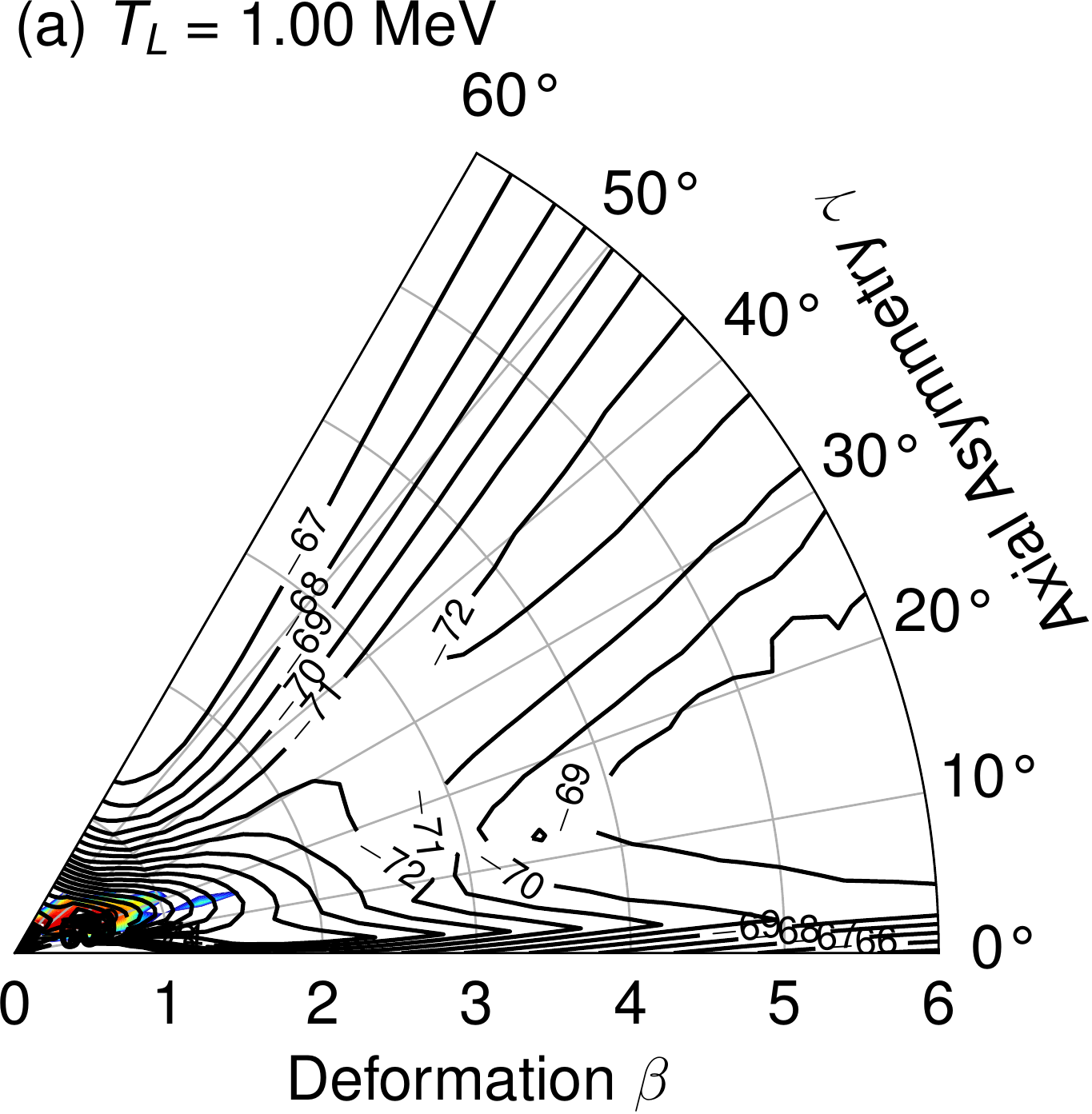}
\end{center}
\end{minipage}
\begin{minipage}{0.5\hsize}
\begin{center}
\includegraphics[keepaspectratio,width=\linewidth]{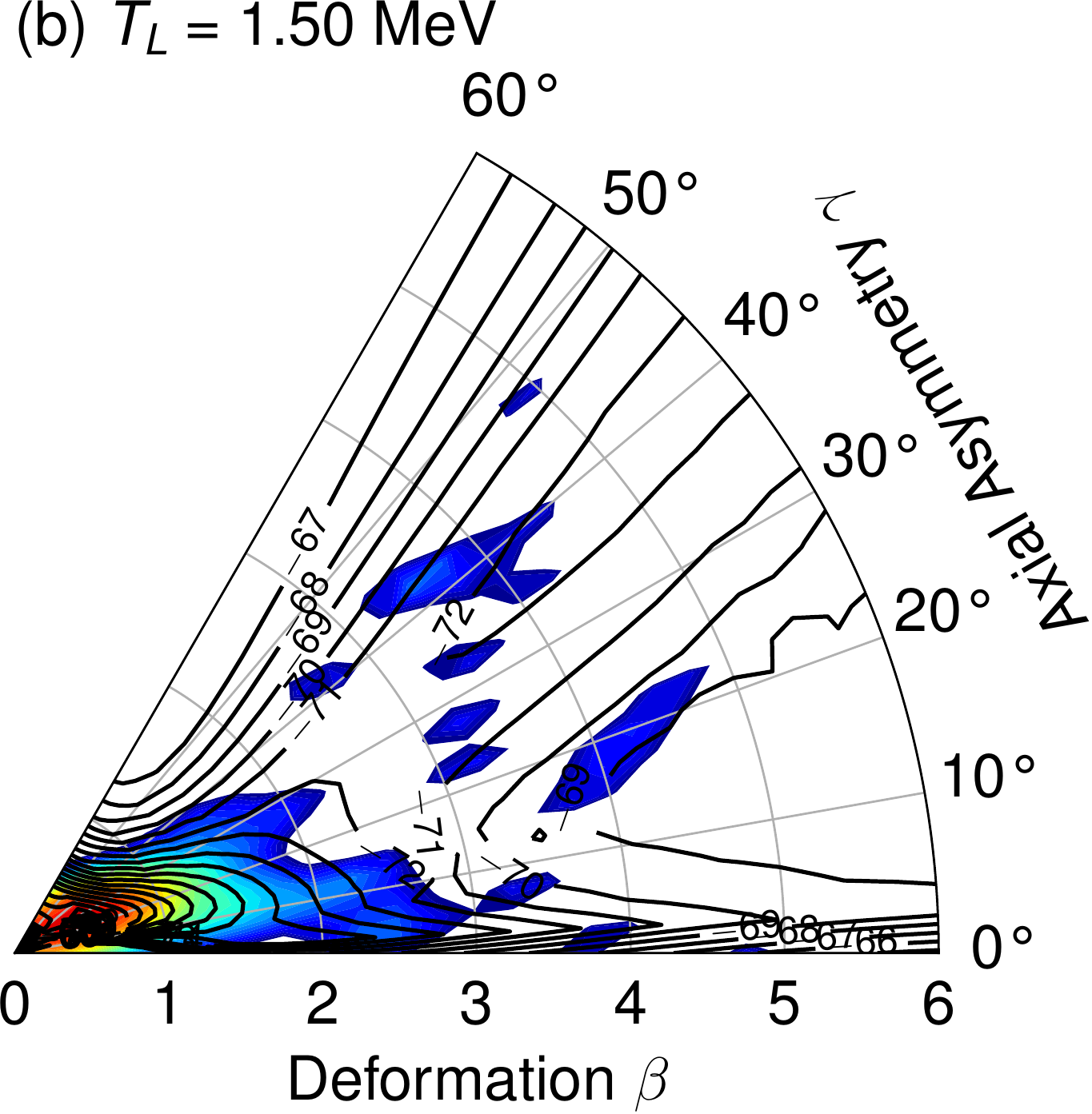}
\end{center}
\end{minipage} \\
\vspace{.1cm}\\
\begin{minipage}{0.5\hsize}
\begin{center}
\includegraphics[keepaspectratio,width=\linewidth]{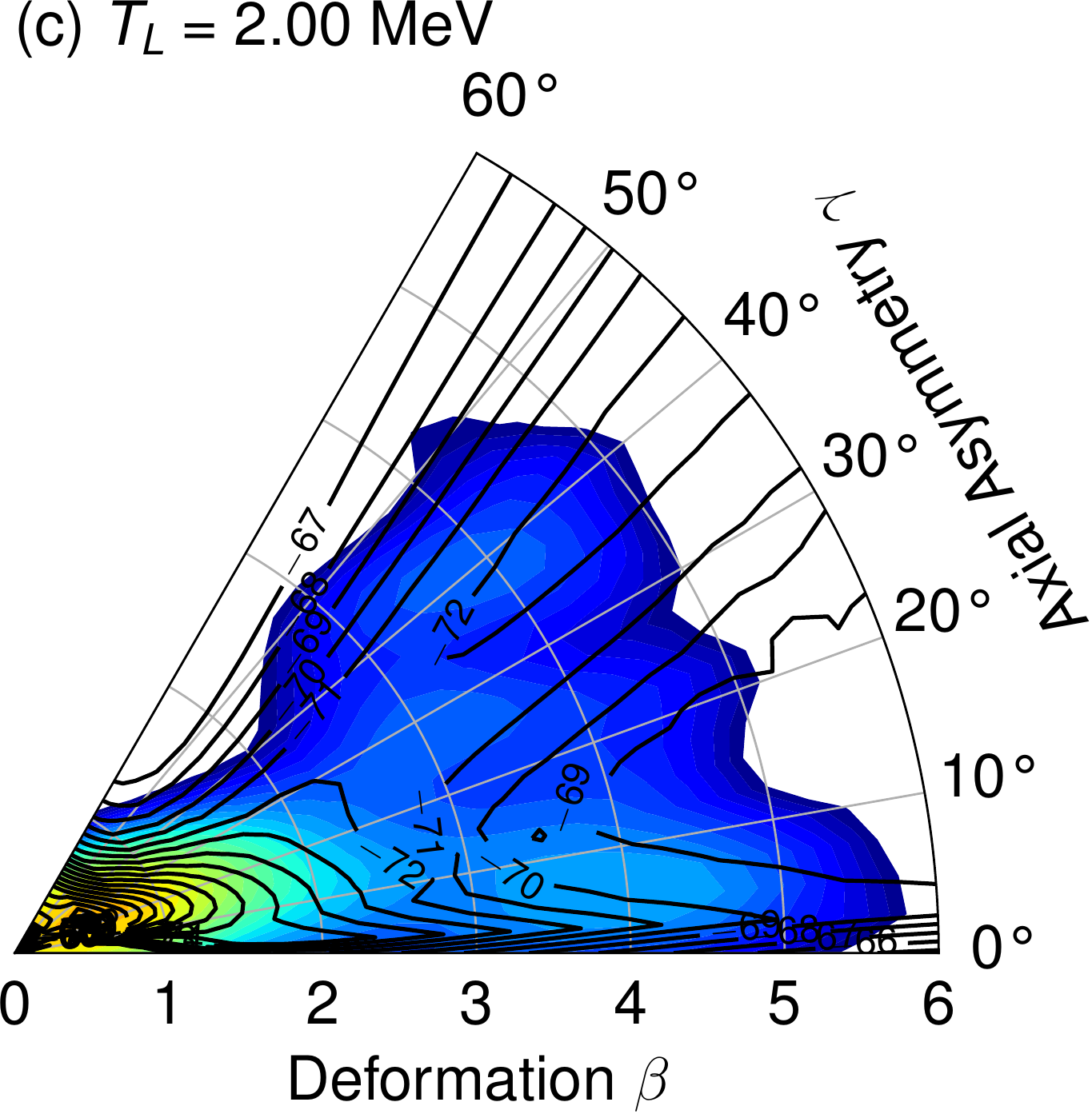}
\end{center}
\end{minipage}
\begin{minipage}{0.5\hsize}
\begin{center}
\includegraphics[keepaspectratio,width=\linewidth]{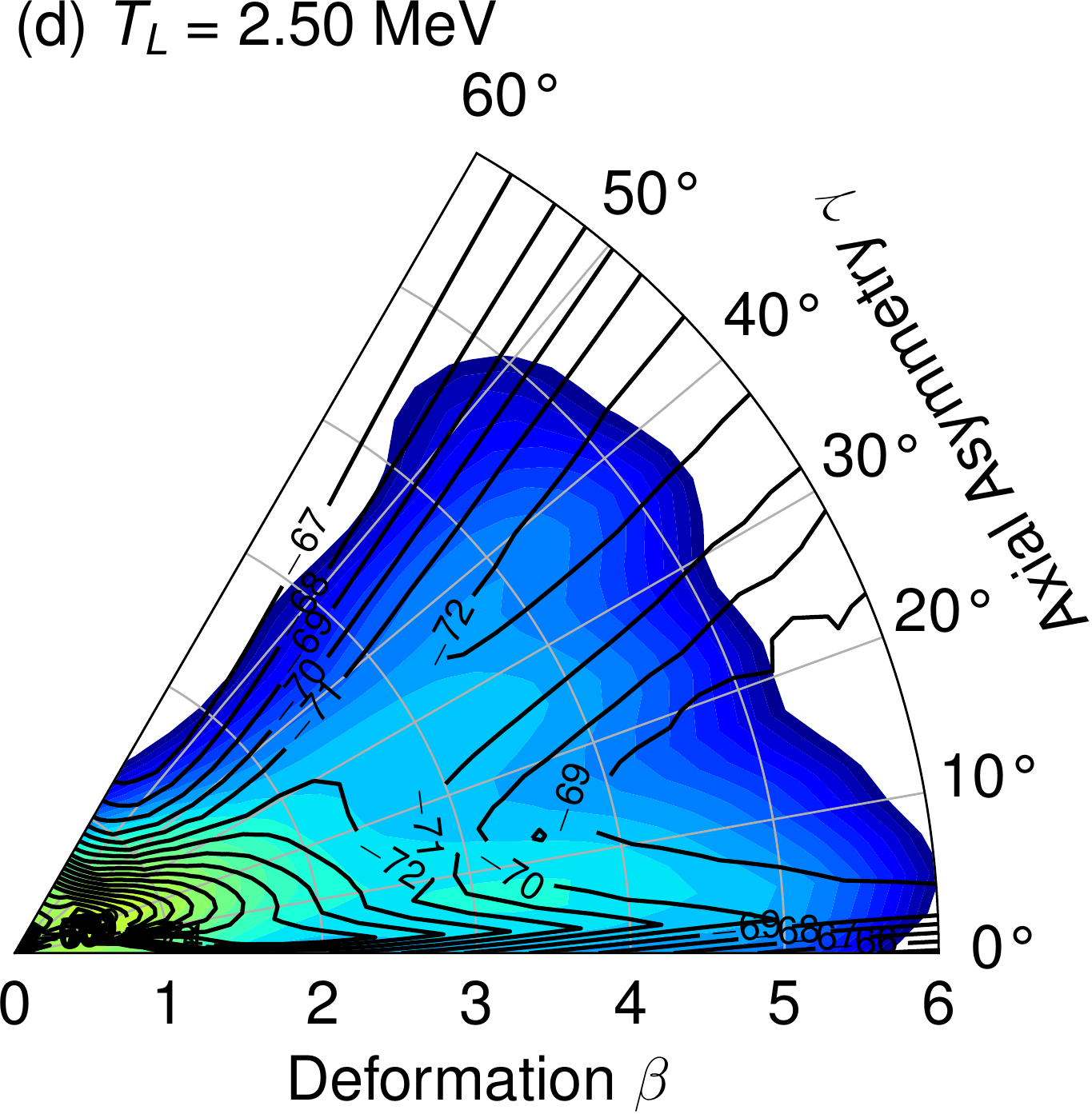}
\end{center}
\end{minipage} 
\end{tabular}
\caption{Distributions of 500 basis functions re-sampled from the obtained 5,000 states with temperatures of $T_L=$ (a) 1.00, (b) 1.50, (c) 2.00, and (d) 2.50 MeV. The contour corresponds to logarithm steps of 0.1 from $10^{-2}$ to 10.}
\label{dis2}
\end{figure}

Figure \ref{dis2} shows the distributions of 500 basis functions re-sampled from 5,000 ones obtained by the RXMC method with the different temperatures. 
In the figure, we can see that the distribution of the 500 re-sampled basis functions is well conserved from the 5,000 basis functions obtained by the RXMC method.

\subsection{Optimization of energies for $0^+$ excited states in $^{12}$C}

\begin{figure}[htbp]
\begin{center}
\includegraphics[keepaspectratio,width=0.9\linewidth]{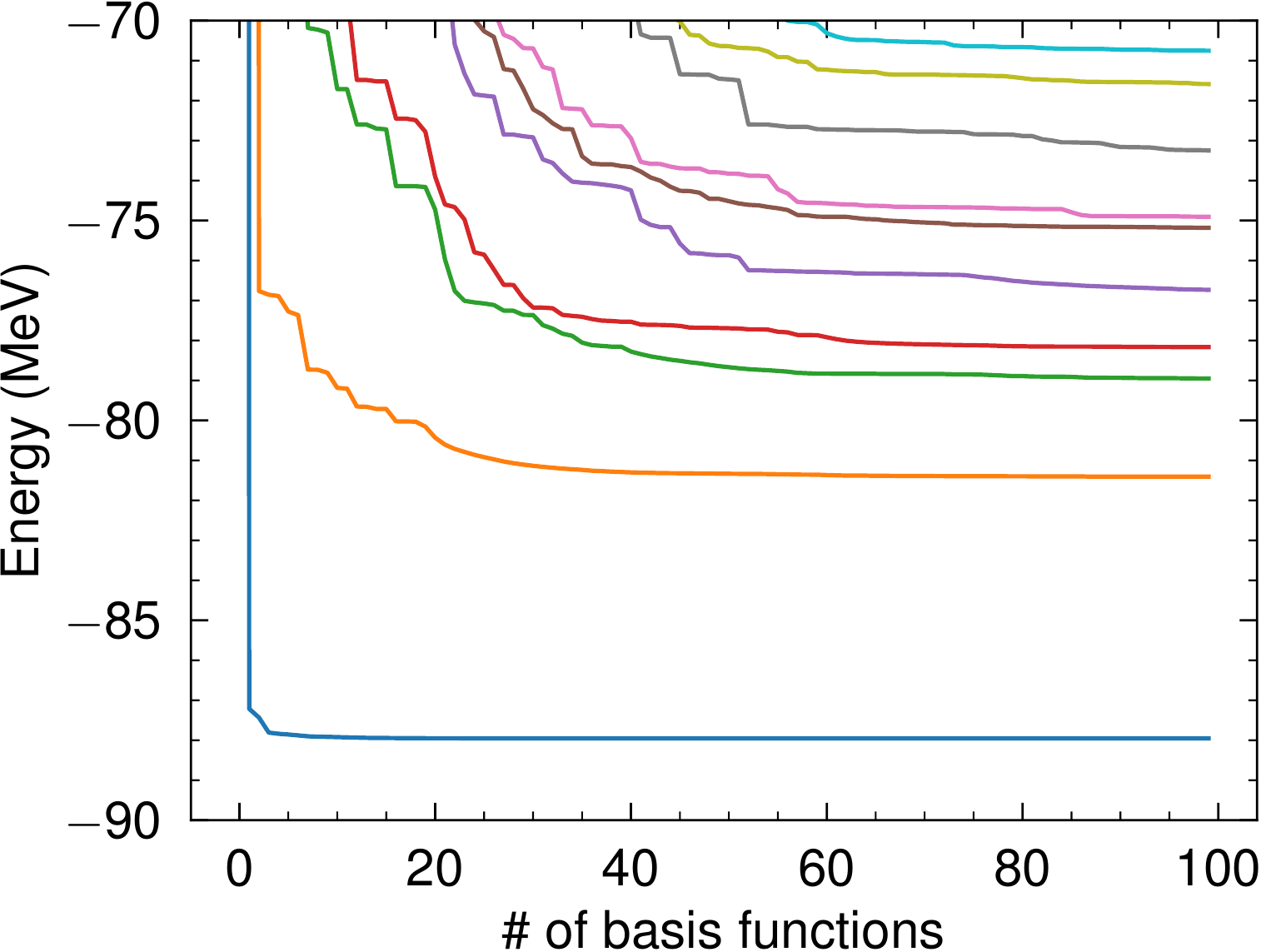}
\end{center}
\caption{Convergence of energies for the $0^+$ states of $^{12}$C as a function of the number of basis functions sampled with $T_L=2.5$ MeV.}
\label{conv}
\end{figure}

After sampling 500 basis functions, we calculate Hamiltonian and overlap matrix elements. Using the obtained matrix elements, we optimize energies from the ground to fourth excited states one by one. Each excited state is optimized for every 20 basis functions, that is, the total number of basis functions is 100. Figure \ref{conv} shows the convergence behavior of the ground and excited states with $T_L=2.5$ MeV thus obtained.
In the figure, we can see that the energies from the first to fifth $0^+$ states are well converged every 20 steps.
No unphysical small energy splits coming from the numerical accuracy of the angular-momentum projection
are found. Such split indeed occurs if we randomly and uniformly generate basis functions without the principal-axis transformation \cite{PhysRevC.83.061301,PhysRevC.86.031303}. 
The obtained energies of  the first five $0^+$ states are -$87.96$~MeV, $-81.37$~MeV, $-78.95$~MeV, $-78.11$~MeV, and $-76.80$~MeV. The latter four correspond to the excitation energies of
$E_{\rm ex} = 6.59$, 9.01, 9.85, and 11.16~MeV, measured from the ground $0^+$ state. In the present model, the three-$\alpha$ threshold energy is $-82.71$~MeV.

\begin{figure}[htbp]
\begin{center}
\includegraphics[keepaspectratio,width=0.9\linewidth]{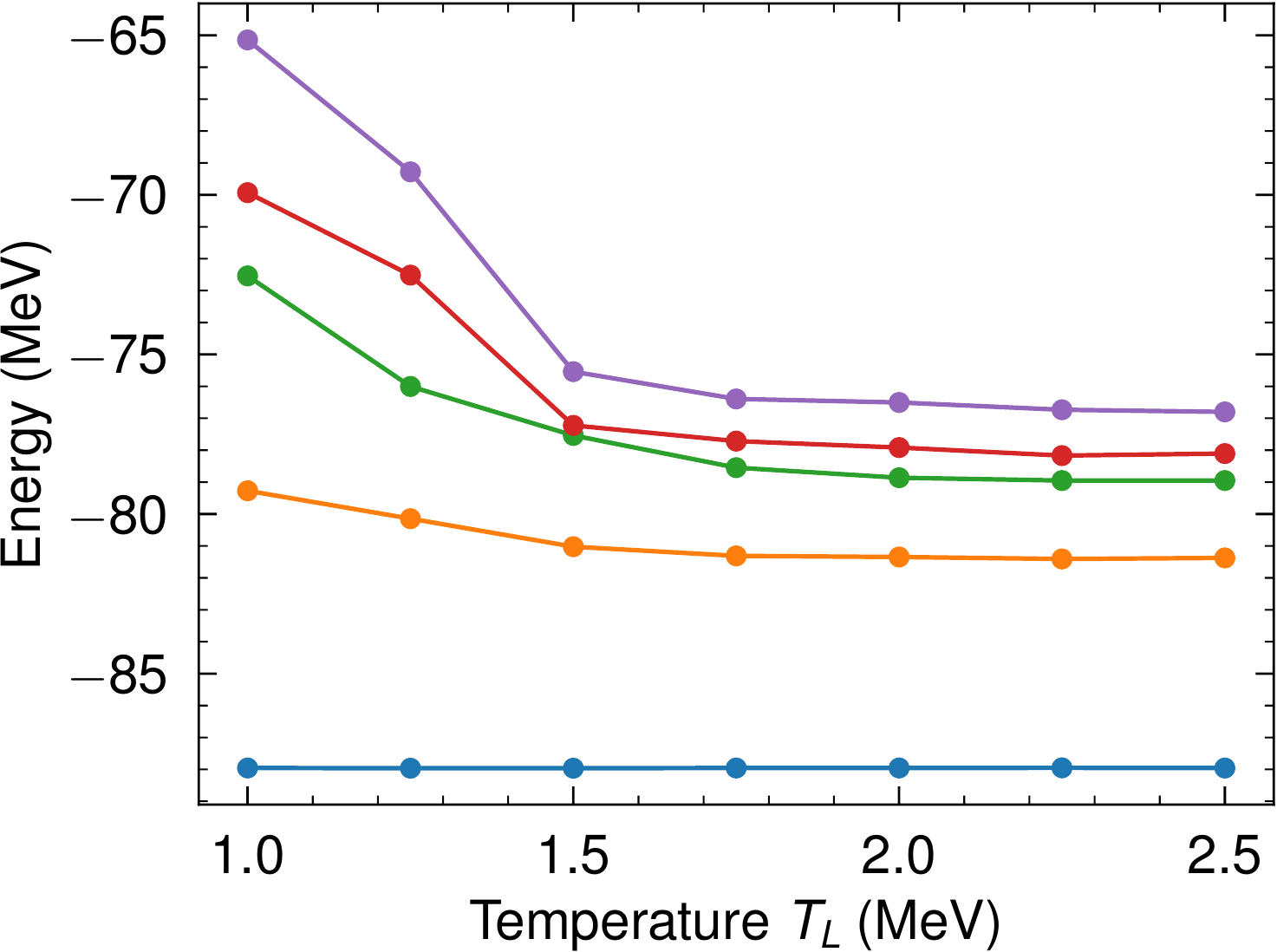}
\end{center}
\caption{Temperature dependence of the calculated energies for the $0^+$ excited states of $^{12}$C. At each temperature, the excited sates are calculated with the 100 optimized basis functions.}
\label{tmpedep}
\end{figure}

We also check the temperature dependence of the calculated results. Figure \ref{tmpedep} shows the obtained $0^+$ excited energies for $^{12}$C versus the temperature. We can see that the fourth and fifth $0^+$ states drastically decrease at $T_L=1.5$ MeV. This is because largely deformed basis functions with $\beta\gtrsim 2.0$ are sampled at temperatures higher than $T_L=1.5$ MeV (see Fig.~\ref{dis2}). These components play an important role in forming the fourth and fifth $0^+$ states.
Note that the excitation energy of $E_{\rm ex}=11.16$~MeV for the fifth $0^+$ state is similar to $E^*\sim 9.38$~MeV calculated from the microcanonical ensemble at $T_L=2.5$~MeV. It may be the best way to choose a temperature to coincide between the energy of a desirable excited state and the thermal excitation energy estimated from the canonical ensemble.

\begin{figure}[htbp]
\begin{center}
\includegraphics[keepaspectratio,width=0.7\linewidth]{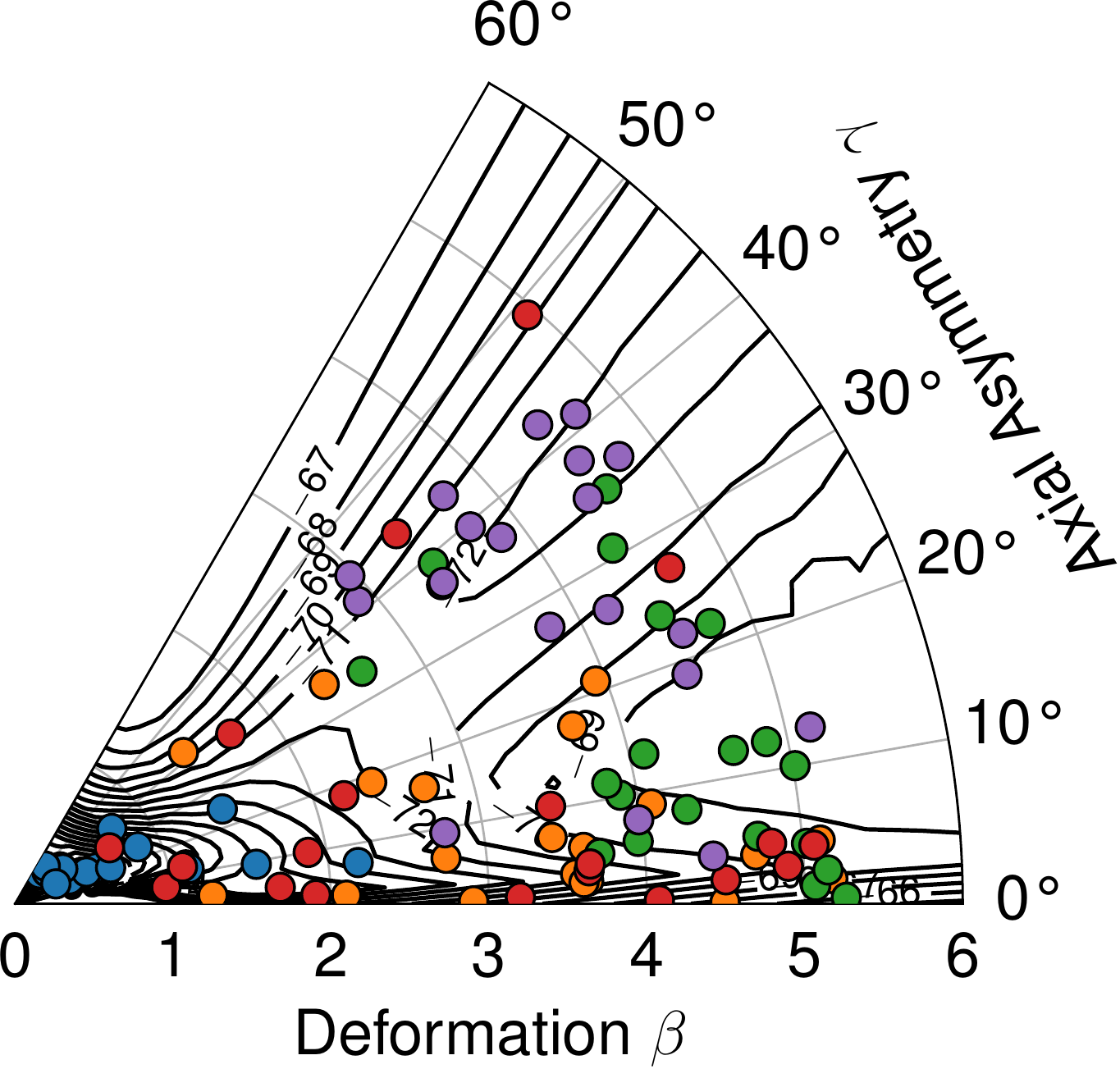}
\end{center}
\caption{Distribution of optimized 100 basis functions for the $0^+$ stets of $^{12}$C at $T_L=2.50$ MeV. The (blue, orange, green, red, and purple) solid circles indicate the basis functions thus obtained by the energy optimization for each excited state. (Each color of the symbols is the same as the corresponding excited state in Fig.~\ref{conv}.)}
\label{dist}
\end{figure}

We finally check the $\beta$-$\gamma$ distribution of the resulting 100 basis functions optimized by the energy of each excited state. Figure \ref{dist} shows the distribution of basis functions thus obtained at $T_L=$ 2.5 MeV.
We can see that the obtained basis functions are widely distributed in the PES.
However, there are no basis functions around the saddle point at $\beta=3.0$ and $\gamma\sim30^\circ$. There may be an excited state with these regions as the main component at much higher excitation energies.

\subsection{Calculated results}

\begin{table}
\caption{\label{tab1} Calculated $0^+$ excited states in $^{12}$C. $E$ and $E_{\rm ex}$ indicate the obtained energies and excitation energies measured from the first $0^+$ state. $r_{\rm rms}$ is the root-mean-squared radius. $B(\rm IS0)$ is the iso-scalar monopole strength from the ground state.}
\begin{ruledtabular}
\begin{tabular}{ccccc}
State & $E$ (MeV) & $E_{\rm ex}$ (MeV) & $r_{\rm rms}$ (fm)& $B(\rm IS0)$ (fm$^4$)\\
\hline
$0^+_1$ & $-87.96$ &0 &2.53&-\\
$0^+_2$ & $-81.37$ &6.59&4.00&399.6\\
$0^+_3$ & $-78.95$ &9.01&4.80&120.6\\
$0^+_4$ & $-78.11$ &9.85&4.48&113.6\\
$0^+_5$ & $-76.80$ &11.2&5.43&11.94\\
\end{tabular}
\end{ruledtabular}
\end{table}

\begin{figure}[htbp]
    \begin{center}
    \includegraphics[keepaspectratio,width=5cm]{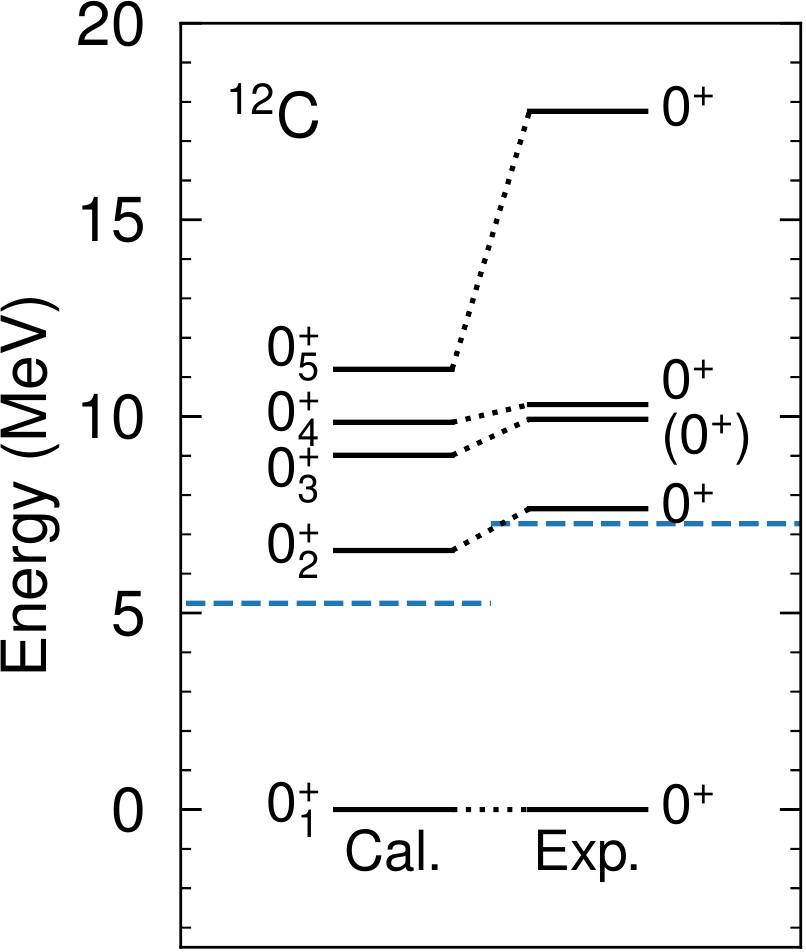}
    \end{center}
    \caption{Comparison between the calculated and experimental $0^+$ states. The experimental data are taken from the NuDat 2 database in National Nuclear Data Center \cite{NNDC}. The dashed lines indicate three $\alpha$'s decay threshold energy.}
    \label{exp}
\end{figure}

After 100 basis functions are optimized, we calculate $0^+$ states by diagonalizing the obtained matrix elements.
The energies, the root-mean-squared (RMS) radii, and the iso-scalar monopole transition strengths thus obtained are summarized in Table \ref{tab1}.
In Fig.~\ref{exp}, we also compare the calculated $0^+$ energies with the experimental data taken from NuDat2 of National Nuclear Data Center \cite{NNDC}.

We obtain the second $0^+$ state at an excitation energy of $E_{\rm ex}=6.59$~MeV just above the three $\alpha$'s decay threshold energy of 5.25~MeV. 
We also obtain a large iso-scalar monopole strength, $B(IS0)$, from this state to the ground state.  The obtained root-mean-squared (RMS) radius is 4.00 fm, which is significantly large value.
These results are consistent with other microscopic calculations \cite{KAMIMURA1981456,Schuck_Funaki_2016,Funaki_2015,Funaki_2016,Imai_Tada_Kimura_2019}. 
To compare the RMS radius with that of the resonating group method by Kamimura \cite{KAMIMURA1981456}, we calculate those with the same potential parameter set ($M=0.59$). In this parameter, the obtained energy at the ground state is -89.57 MeV. Then, the obtained RMS radius of the $0_2^+$ state is 3.93 fm, which is much larger than that of the RGM calculation by about 0.5 fm. 

The energies of the third and forth $0^+$ states are close to those of the experimental data, although the third $0^+$ state in the experimental data has not been identified.
Although the excitation energy of the fifth state is high ($E_{\rm ex}=11.2$ MeV), it is still lower than the lowest saddle point (Coulomb threshold). However, its energy is largely underestimated in comparison to the experimental data (see Fig.~\ref{exp}). There may still exist an unobserved state.

\begin{figure}[htbp]
    \begin{tabular}{c}
    \begin{minipage}{\hsize}
    \begin{center}
    \includegraphics[keepaspectratio,width=7.cm]{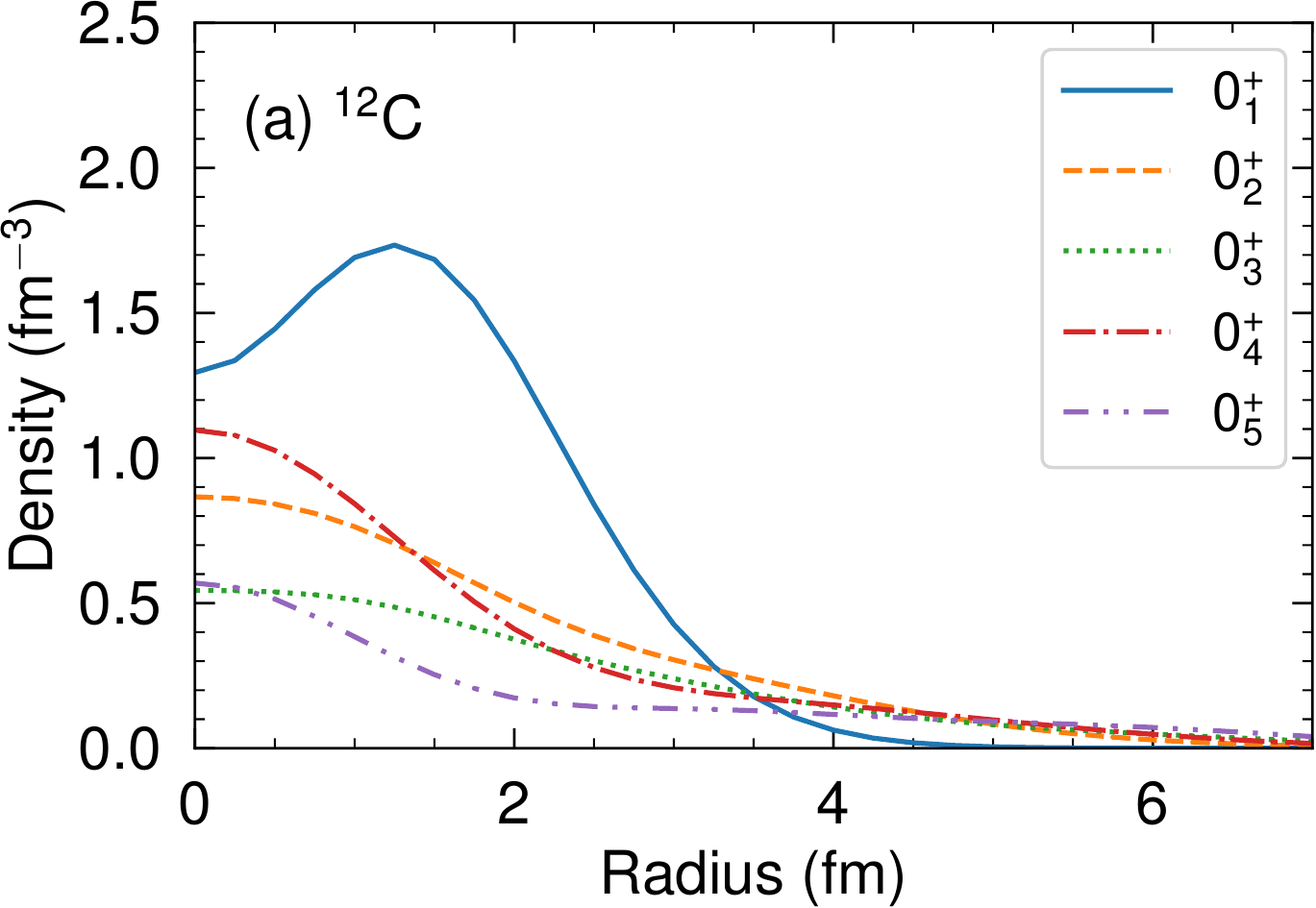}
    \end{center}
    \end{minipage} \\
    \vspace{0.1cm} \\
    \begin{minipage}{\hsize}
    \begin{center}
    \includegraphics[keepaspectratio,width=7.cm]{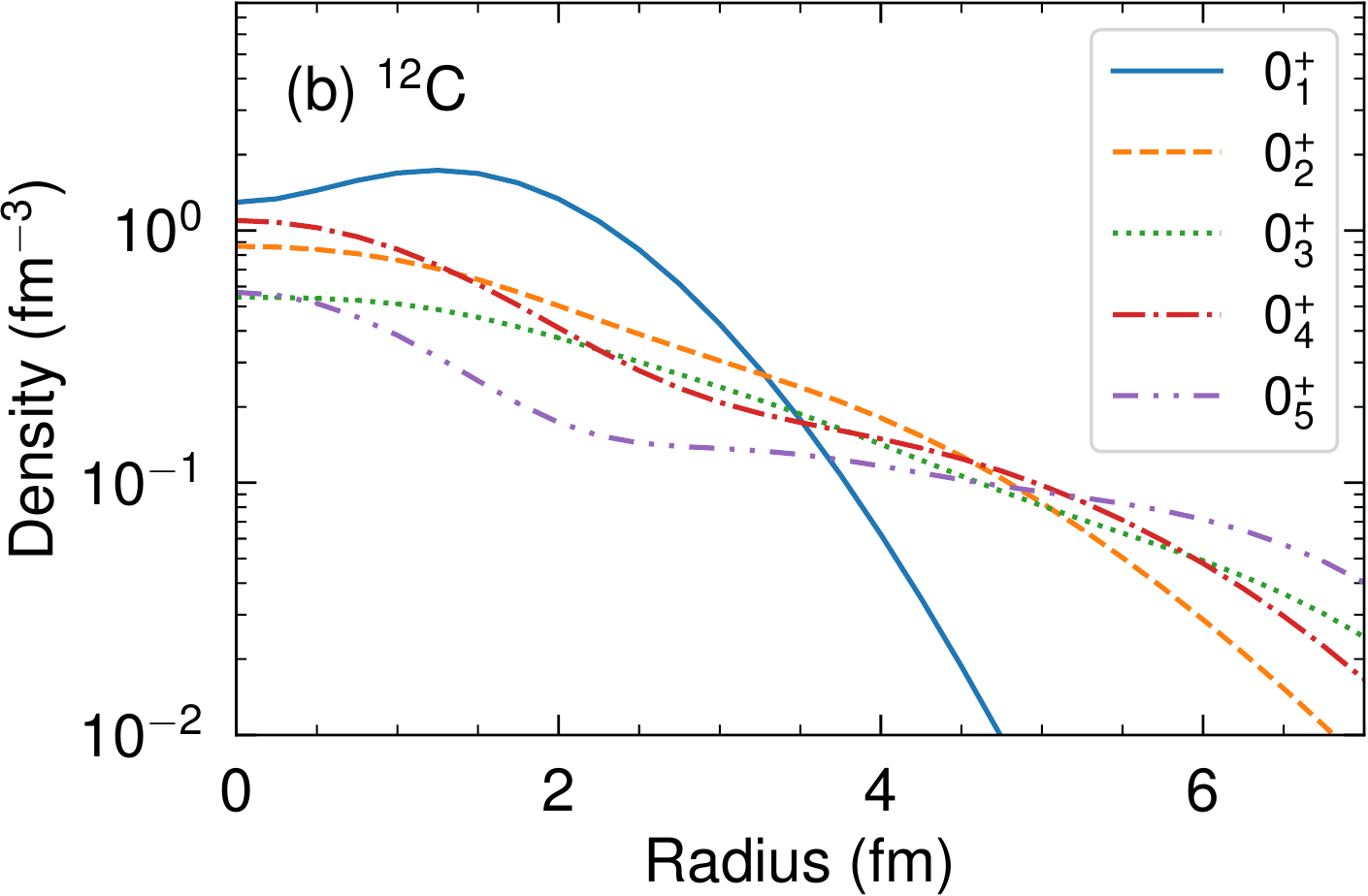}
    \end{center}
    \end{minipage} 
    \end{tabular}
    \caption{Density distribution of each state in the laboratory frame. The density distributions are represented by the normal and logarithm scales in the upper (a) and lower (b) panels, respectively. The solid, dashed, dotted, dot-dashed, dot-dot-dashed lines indicate the $0^+_1$, $0^+_2$, $0^+_3$, $0^+_4$, and $0^+_5$, respectively.}
    \label{den3}
\end{figure}

Figure~\ref{den3} shows the radial density distribution of the obtained $0^+$ excited states in the laboratory frame. 
The solid line is for the ground state. The density distribution shows a drop around the origin, which is a somewhat unnormal nuclear density distribution. However, this tendency is confirmed by the electron scattering experiments, and this may reflect the formation of the three $\alpha$ cluster structure with finite relative distances.  
The dashed line represents the density distribution of the second $0^+$ state, which shows a longer tail compared with the ground $0^+$ state.
Figure \ref{den3}(b) shows the densities in the logarithm scale, which allows us to confirm that the third, fourth, and fifth $0^+$ states have even longer tails.

\begin{figure}[htbp]
    \begin{tabular}{c}
    \begin{minipage}{\hsize}
    \begin{center}
    \includegraphics[keepaspectratio,width=7.cm]{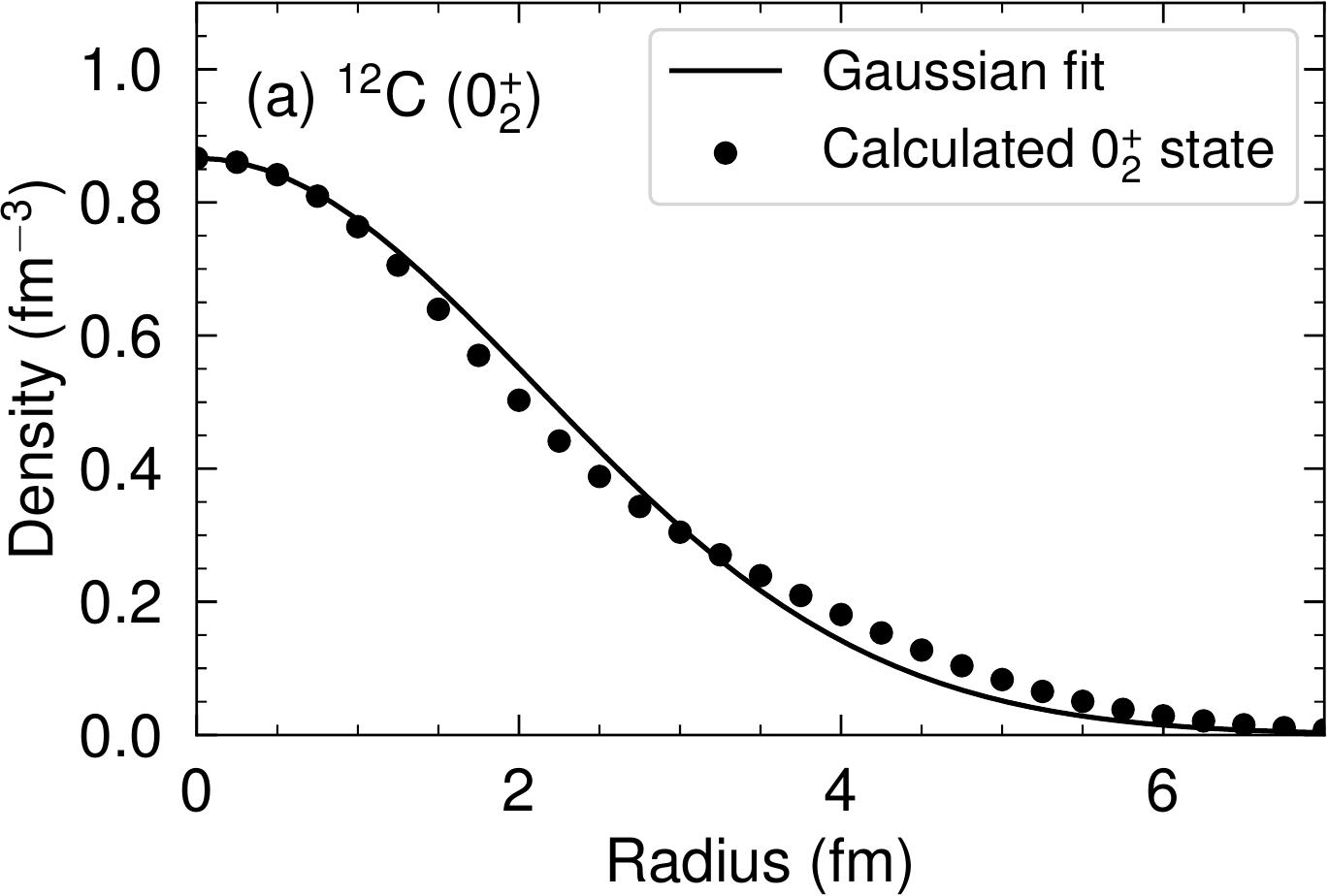}
    \end{center}
    \end{minipage} \\
    \vspace{0.1cm} \\
    \begin{minipage}{\hsize}
    \begin{center}
    \includegraphics[keepaspectratio,width=7.cm]{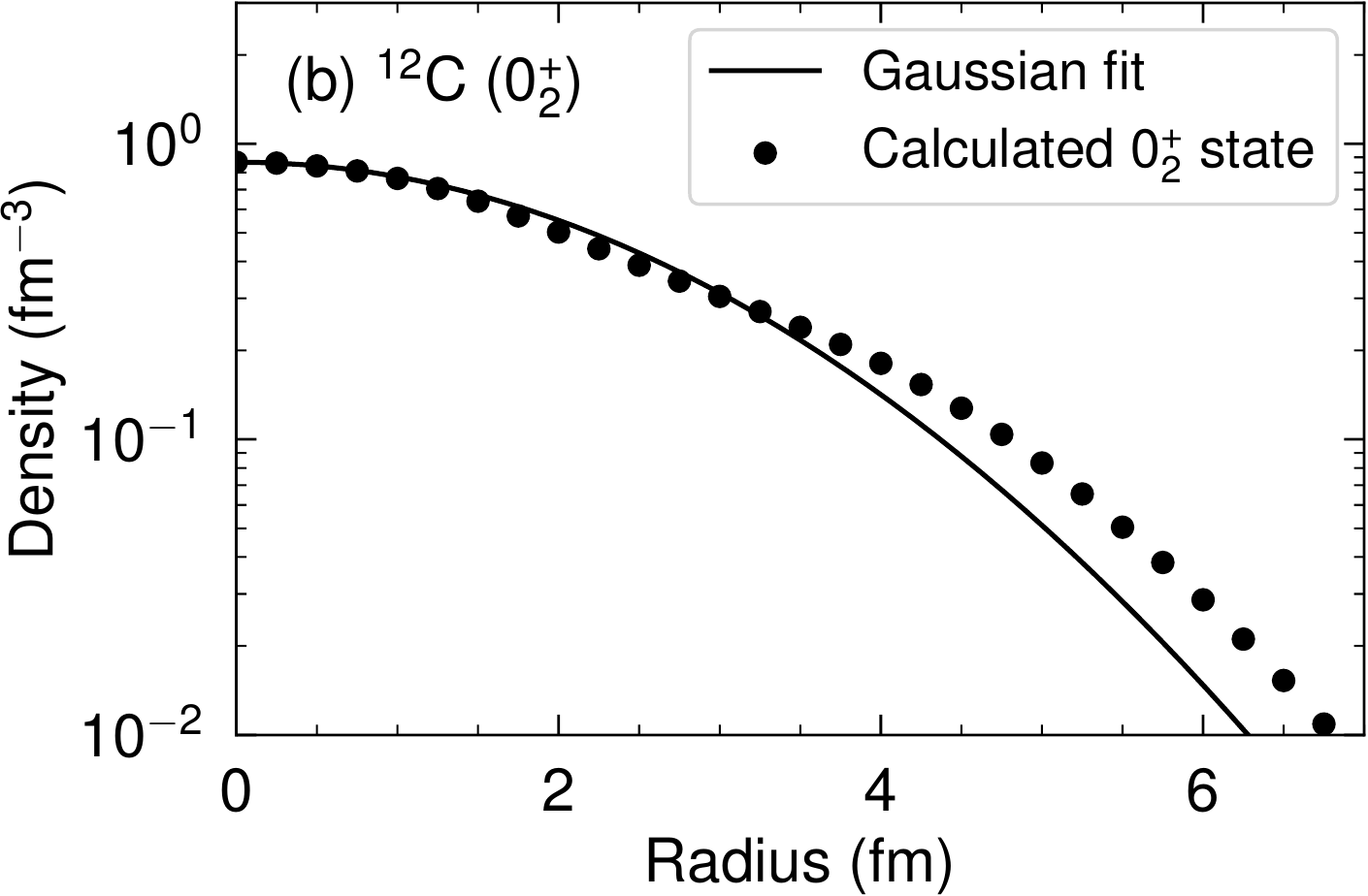}
    \end{center}
    \end{minipage} c
    \end{tabular}
    \caption{The Gaussian function fitted to the obtained density distribution of the gas-like state for $0^+_2$. The solid line indicates the obtained Gaussian distribution. The solid circle indicates the calculated density distribution of the $0^+_2$ state in the laboratory frame. The upper (a) and lower (b) panels are the same result, but the vertical scales are different. }
    \label{den4}
\end{figure}
    
In Fig.~\ref{den4}, we also compare the second $0^+$ states with a single Gaussian fitted by the calculated result. The ground-state density (Fig.~\ref{den4}(a)) is found to be rather well-fitted by a Gaussian distribution. 
However, the one of the second $0^+$ (Fig.~\ref{den4}(b)) is not, especially in the tail region. This is considered to be due to the Coulomb effect, which has an infinite range. The tail part of the wave function is slightly affected by the Coulomb interaction and deviates from the Gaussian form, unlike the ansatz of the spherical THSR wave function. At lease two gaussian components are necessary for obtaining a good agreement with the data.
The deformed THSR wave function would give a much better fitting to this long tail structure \cite{Funaki_Horiuchi_2009}.

\subsection{Analysis of each excited state}

\begin{figure*}[htbp]
\begin{tabular}{ccc}
\begin{minipage}{0.3\hsize}
\begin{center}
\includegraphics[keepaspectratio,width=\linewidth]{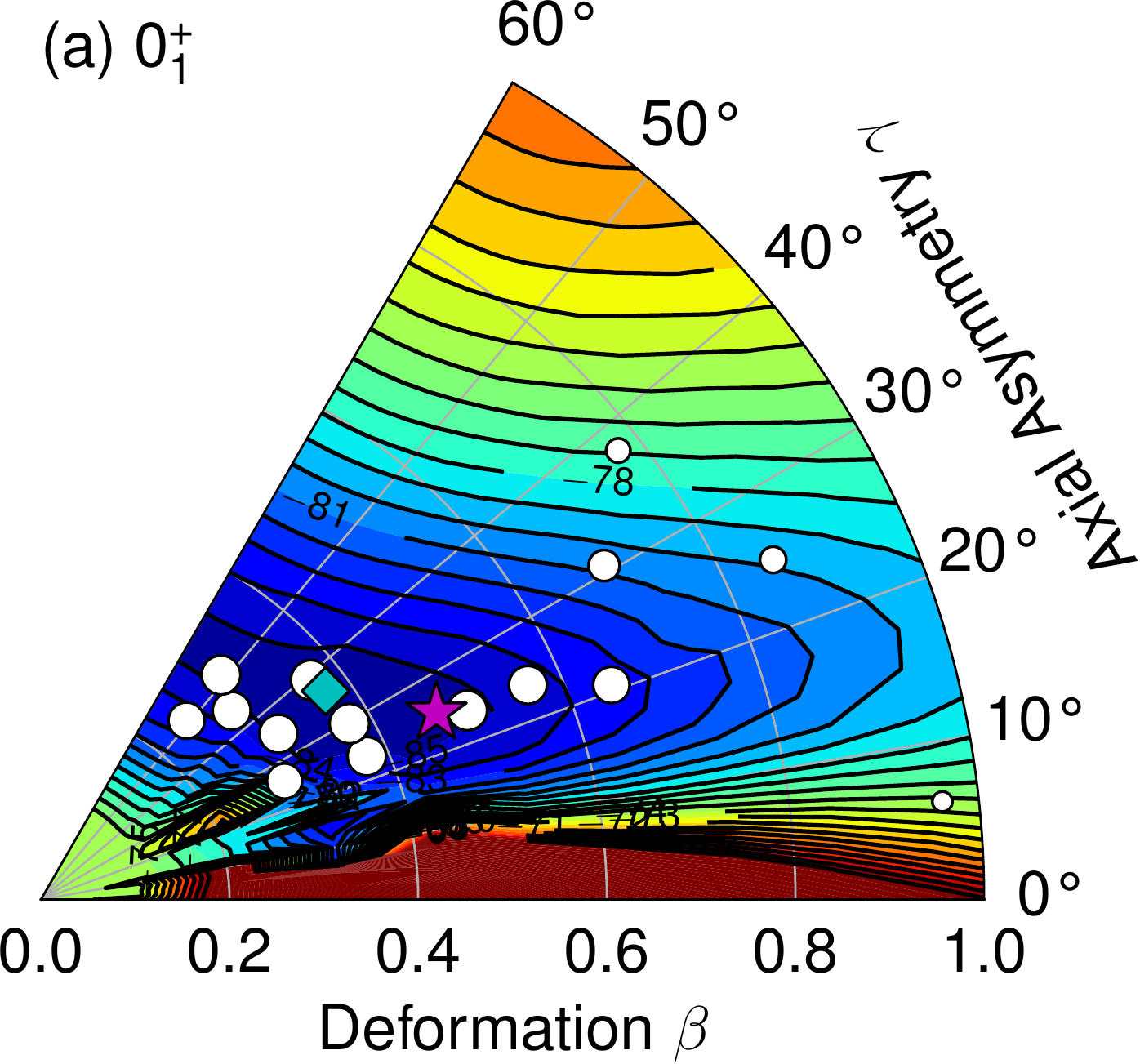}
\end{center}
\end{minipage} &
\begin{minipage}{0.3\hsize}
\begin{center}
\includegraphics[keepaspectratio,width=\linewidth]{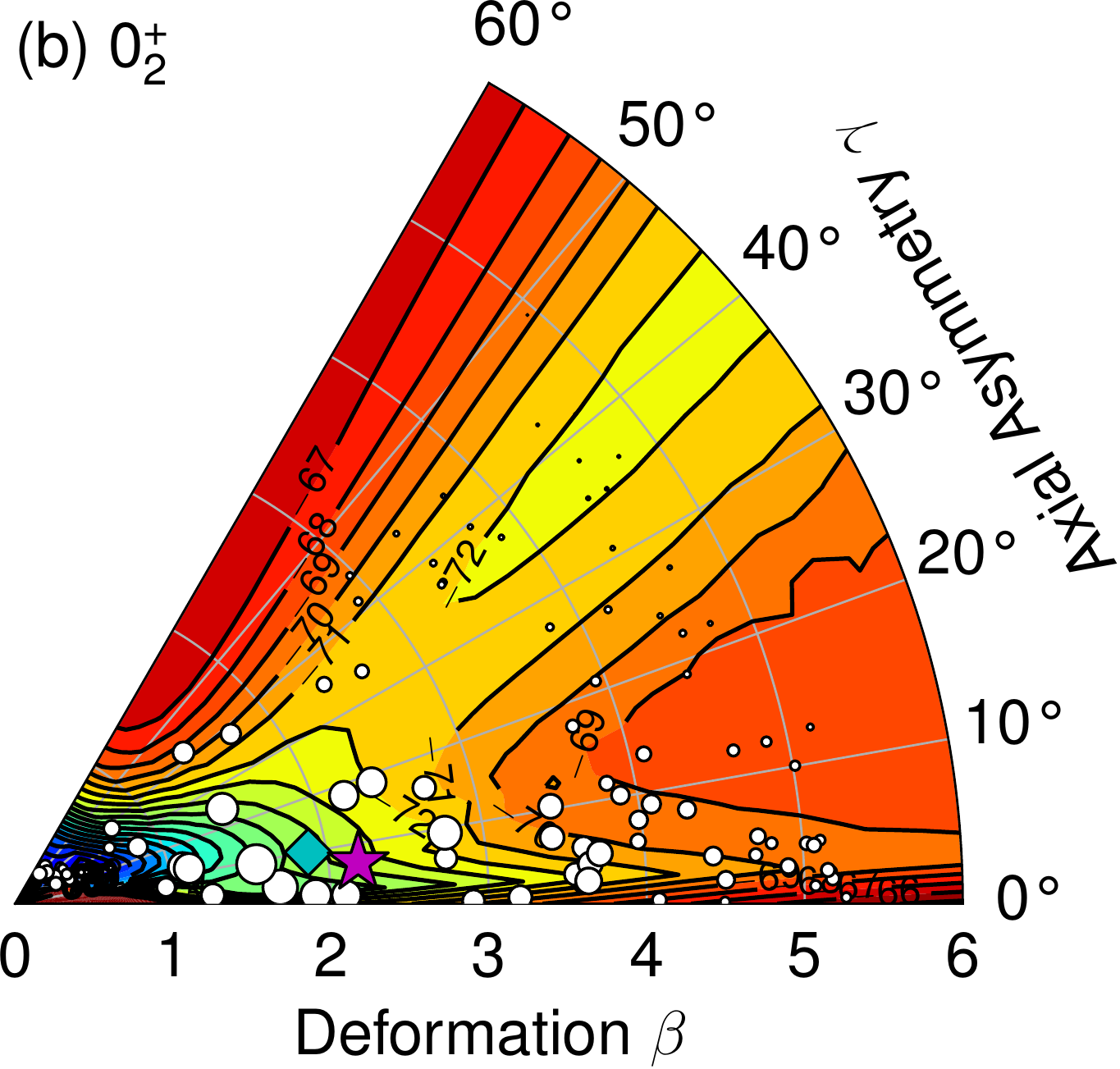}
\end{center}
\end{minipage} &
\begin{minipage}{0.3\hsize}
\begin{center}
\includegraphics[keepaspectratio,width=\linewidth]{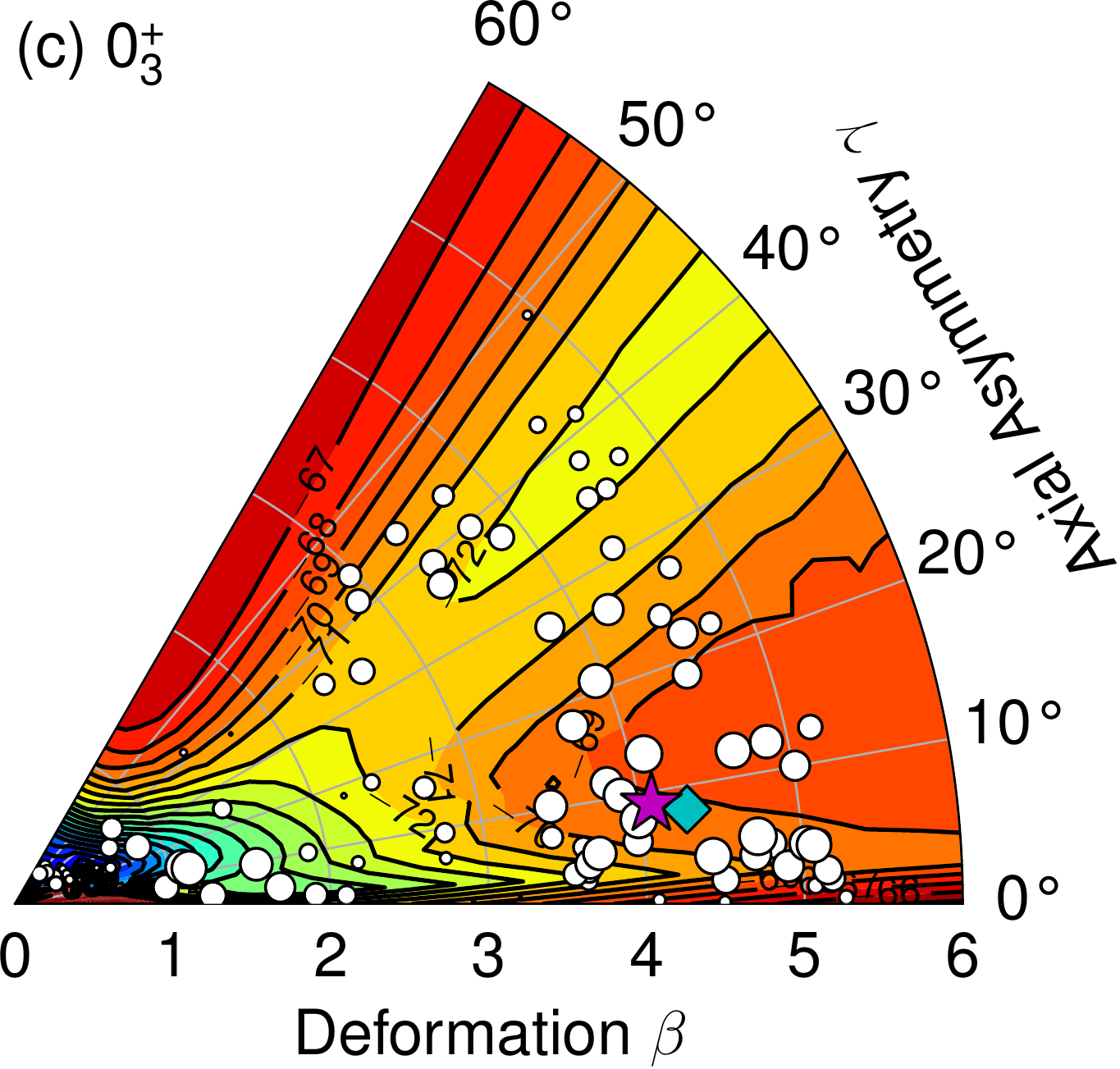}
\end{center}
\end{minipage}\\
\vspace{.1cm}\\
\begin{minipage}{0.3\hsize}
\begin{center}
\includegraphics[keepaspectratio,width=\linewidth]{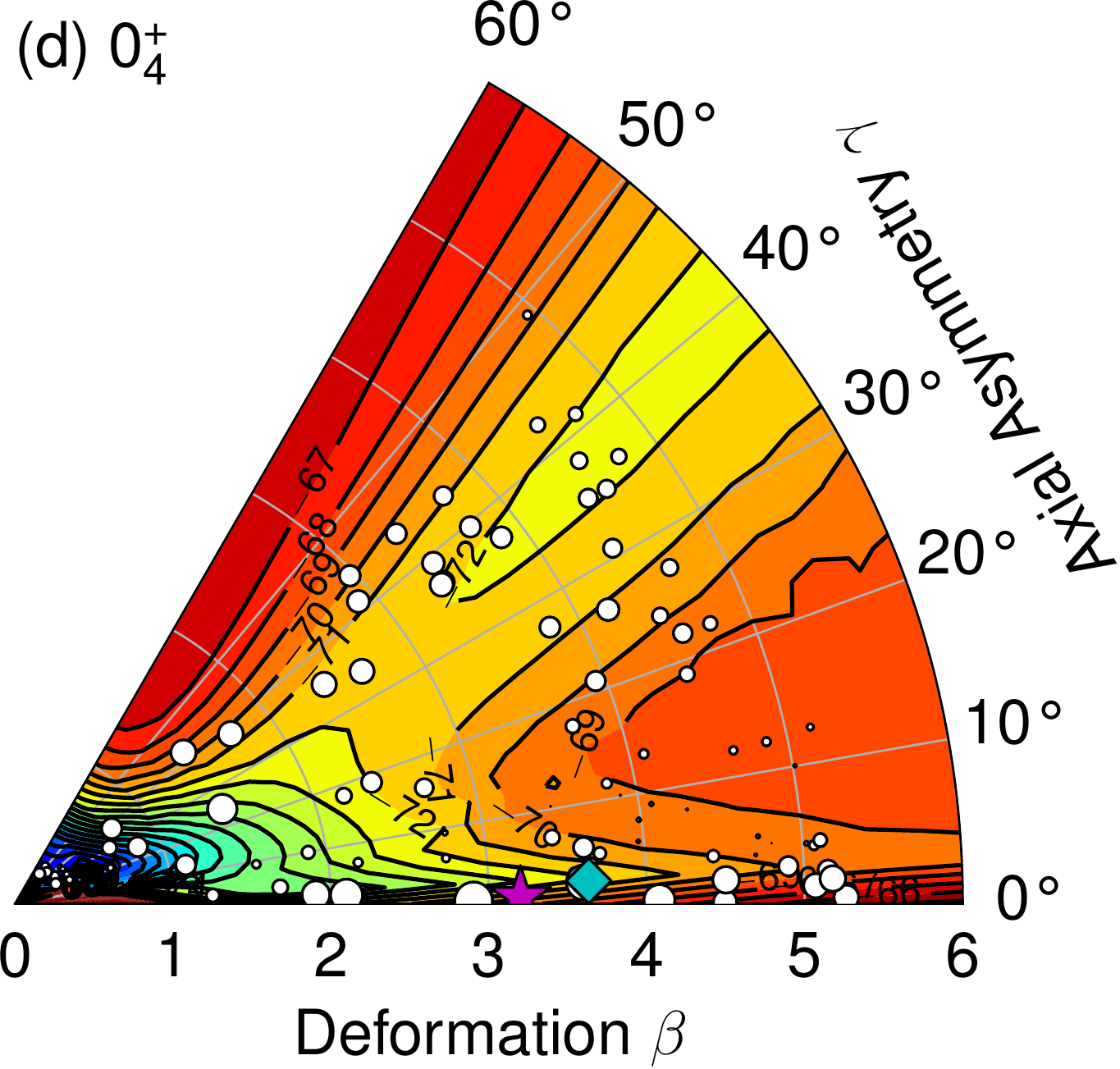}
\end{center}
\end{minipage} &
\begin{minipage}{0.3\hsize}
\begin{center}
\includegraphics[keepaspectratio,width=\linewidth]{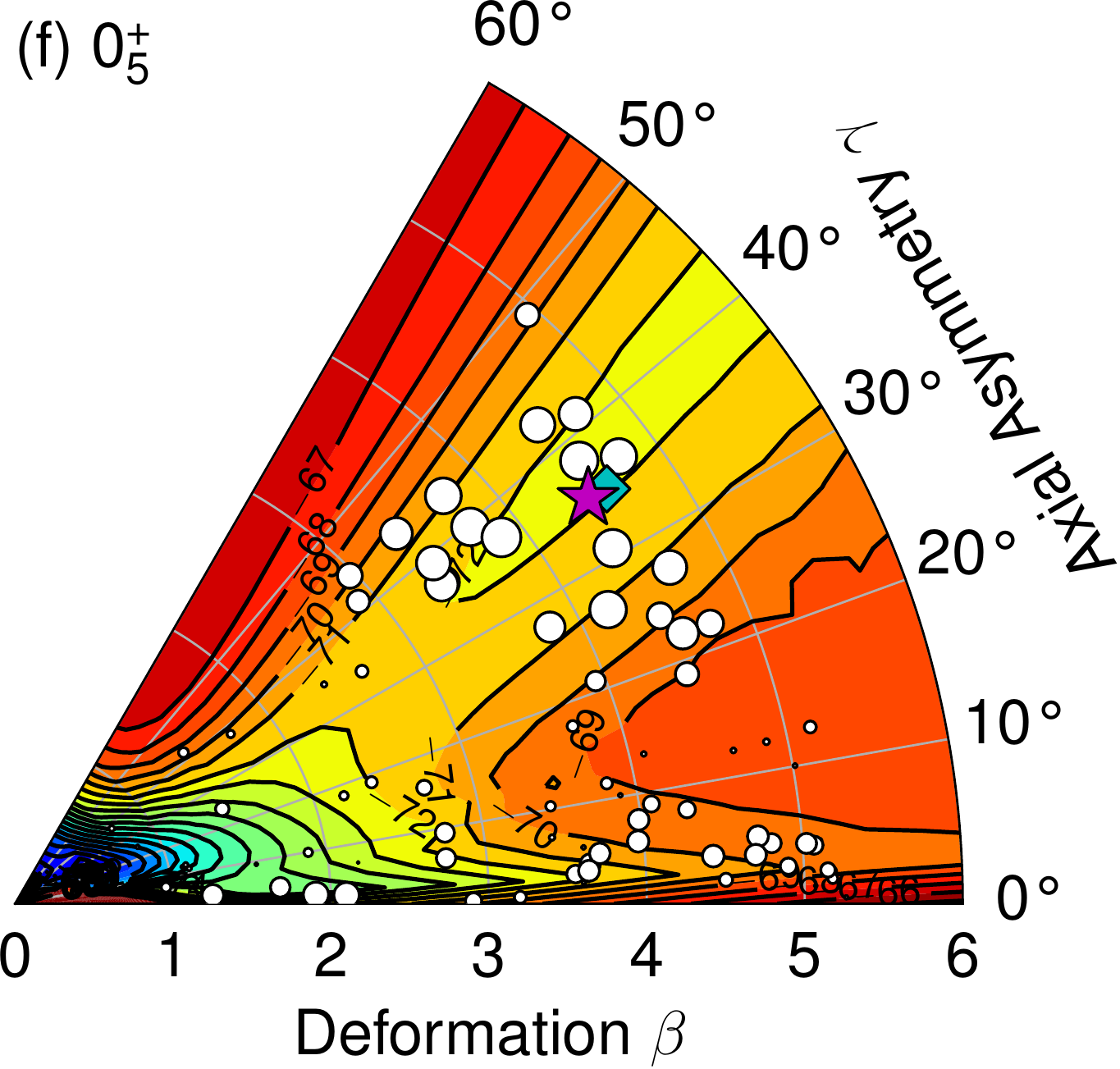}
\end{center}
\end{minipage} 
\end{tabular}
\caption{The squared overlap distribution of the basis functions for each excited state. The open circle indicate the basis function. The size of the open circles represents the squared overlap value for the total wave function of each state. Each circle is normalized by the maximum value. The star and diamond indicate the firstly and secondary largest components, respectively. }
\label{ovdis1}
\end{figure*}

\begin{figure*}[htbp]
\begin{tabular}{ccc}
\begin{minipage}{0.3\hsize}
\begin{center}
\includegraphics[keepaspectratio,width=\linewidth]{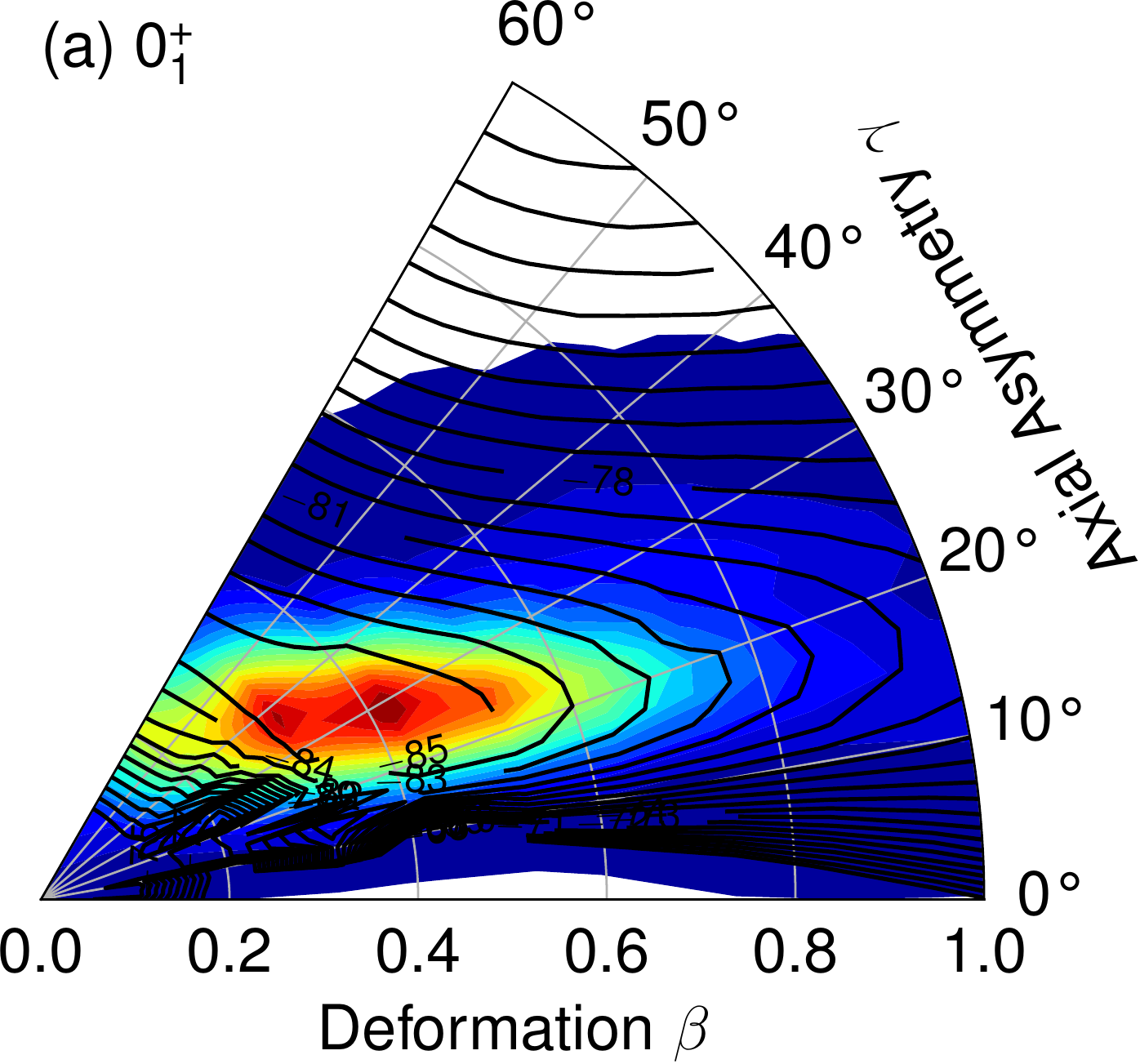}
\end{center}
\end{minipage} &
\begin{minipage}{0.3\hsize}
\begin{center}
\includegraphics[keepaspectratio,width=\linewidth]{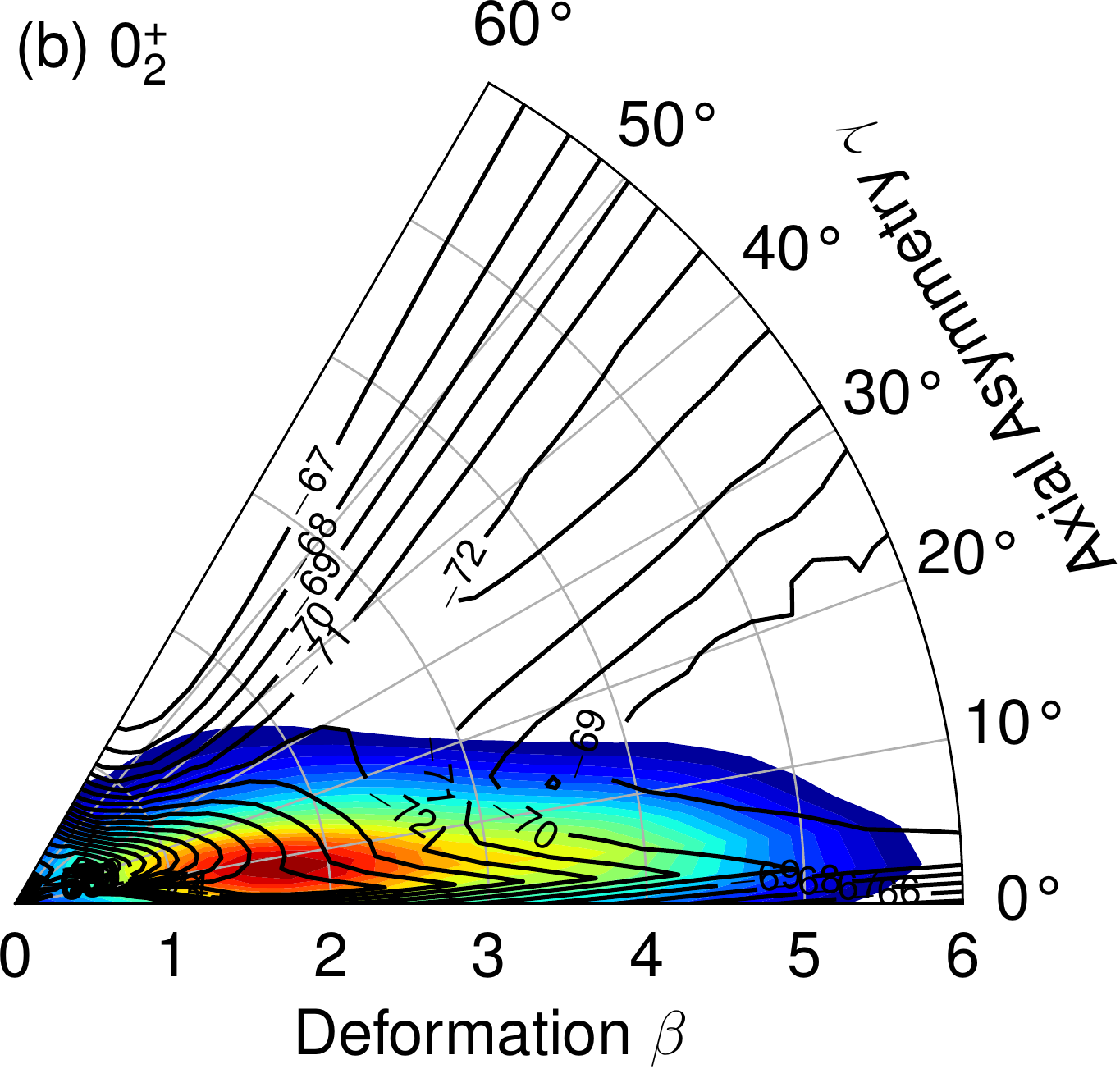}
\end{center}
\end{minipage} &
\begin{minipage}{0.3\hsize}
\begin{center}
\includegraphics[keepaspectratio,width=\linewidth]{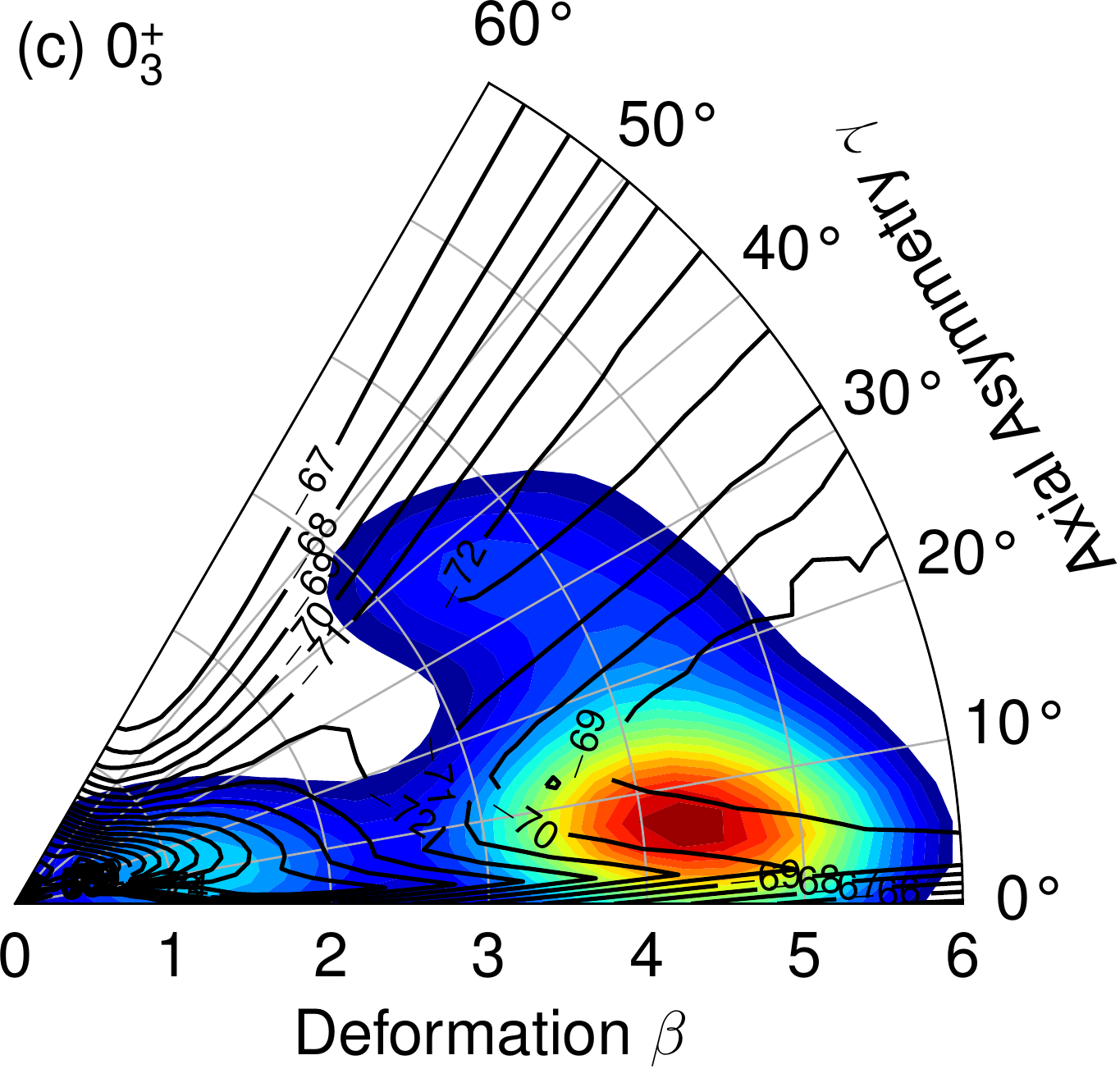}
\end{center}
\end{minipage}\\
\vspace{.1cm}\\
\begin{minipage}{0.3\hsize}
\begin{center}
\includegraphics[keepaspectratio,width=\linewidth]{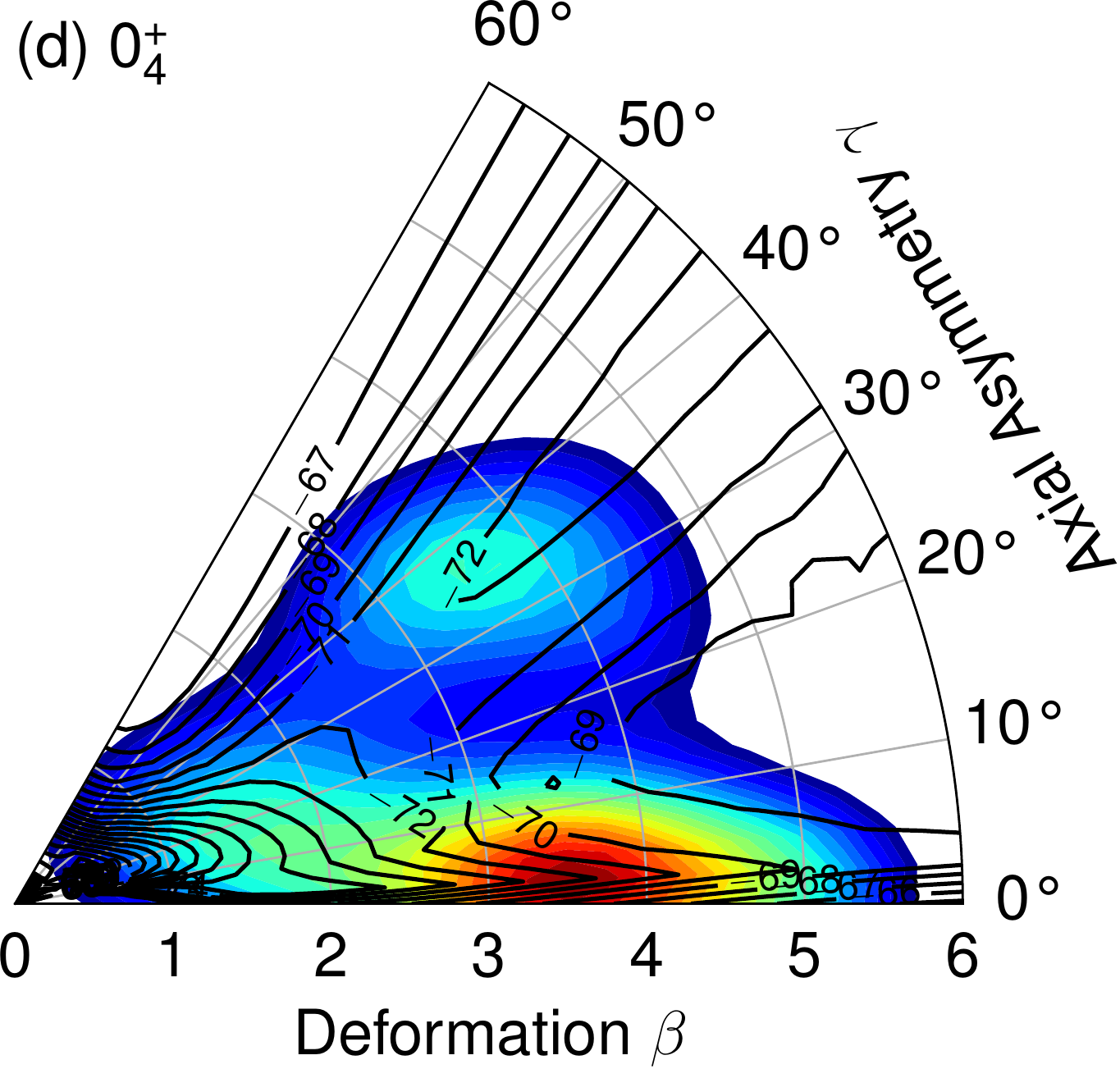}
\end{center}
\end{minipage} &
\begin{minipage}{0.3\hsize}
\begin{center}
\includegraphics[keepaspectratio,width=\linewidth]{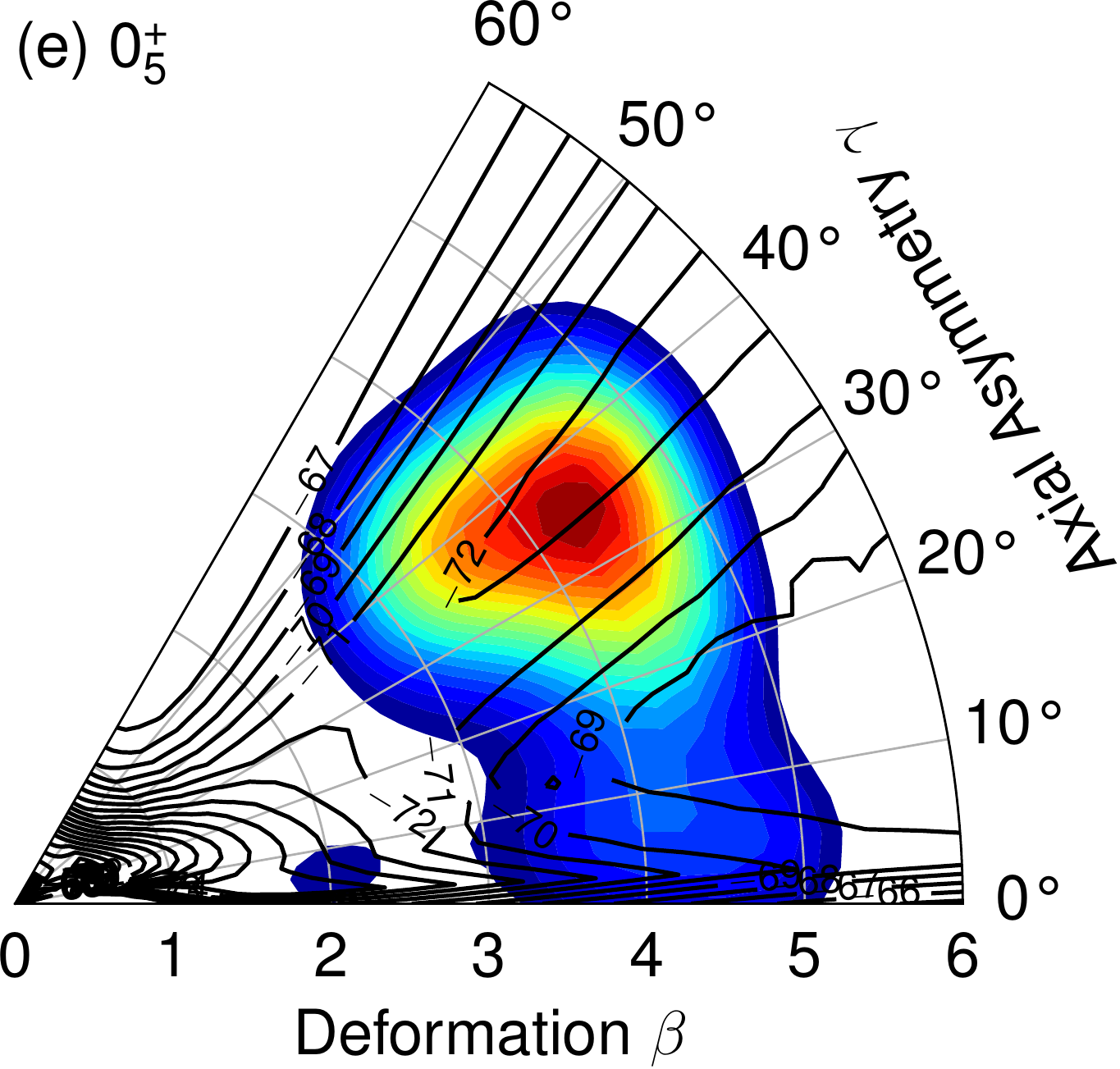}
\end{center}
\end{minipage} 
\end{tabular}
\caption{The squared overlap distribution of the basis functions for each excited state approximated by the weighed kernel density estimation. The color scale is normalized by the maximum overlap value in each excited state. }
\label{ovdis2}
\end{figure*}

\begin{figure*}[htbp]
\begin{tabular}{cccc}
\begin{minipage}{0.25\hsize}
\begin{center}
\includegraphics[keepaspectratio,width=\linewidth]{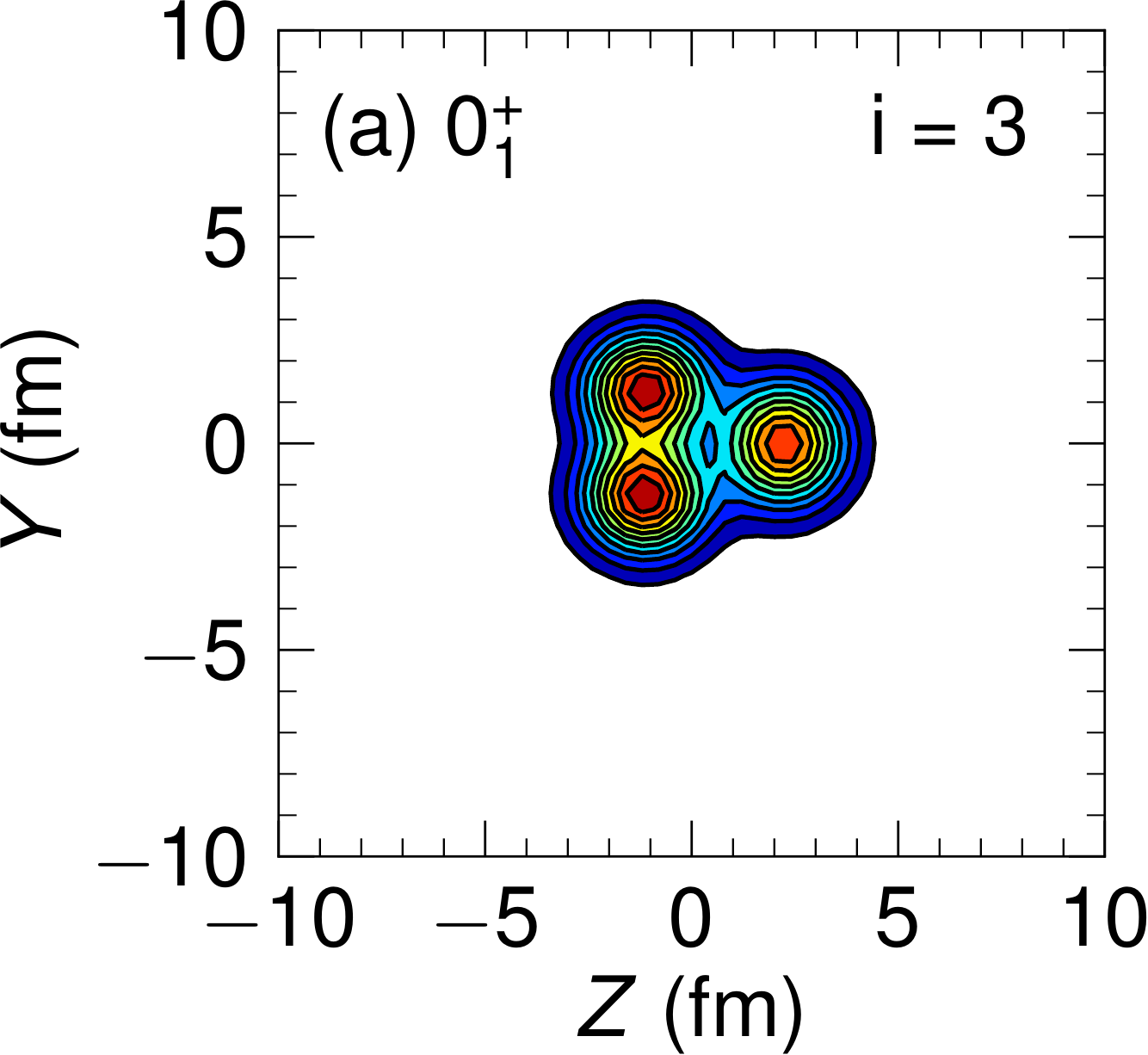}
\end{center}
\end{minipage} &
\begin{minipage}{0.25\hsize}
\begin{center}
\includegraphics[keepaspectratio,width=\linewidth]{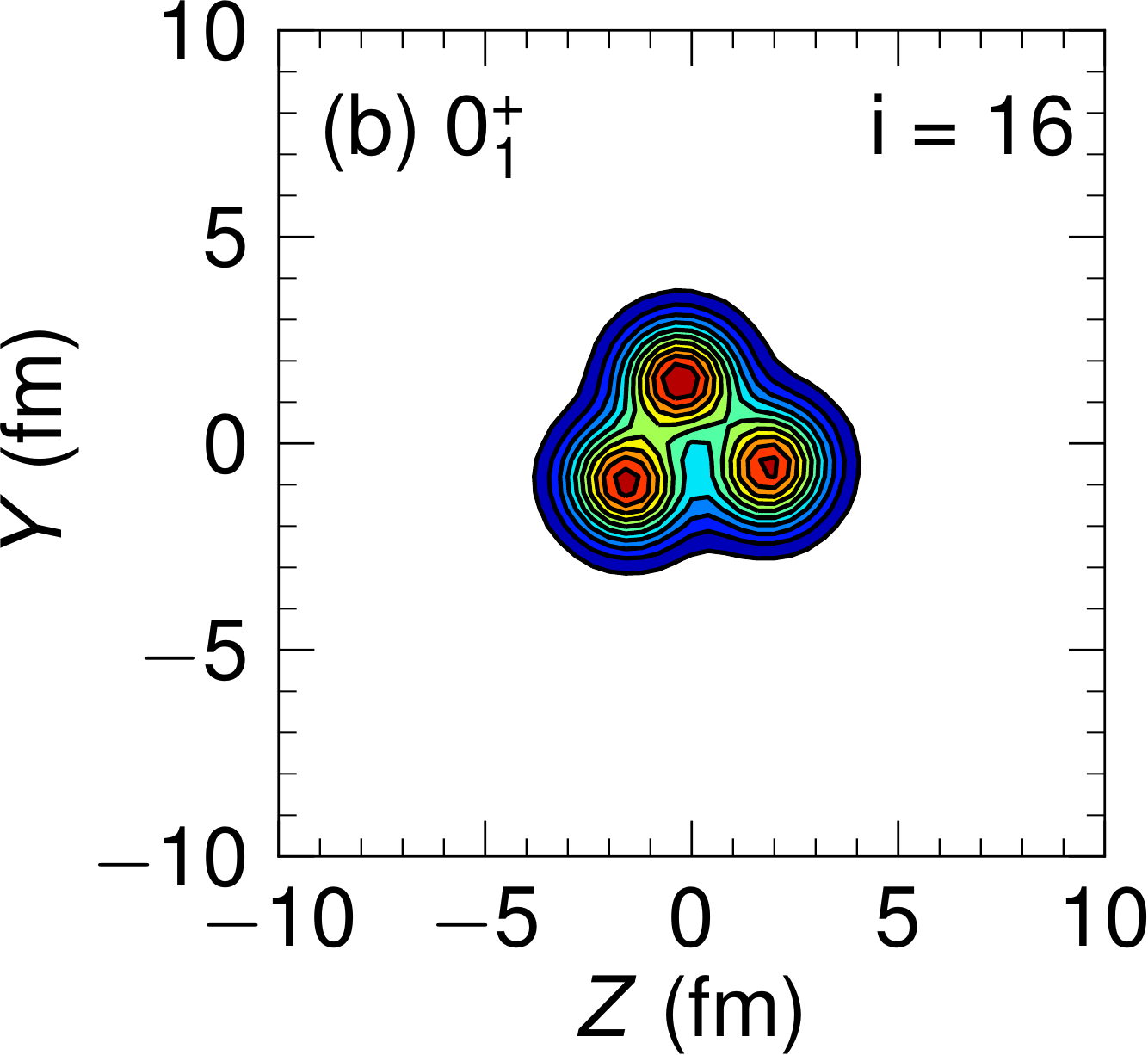}
\end{center}
\end{minipage} &
\begin{minipage}{0.25\hsize}
\begin{center}
\includegraphics[keepaspectratio,width=\linewidth]{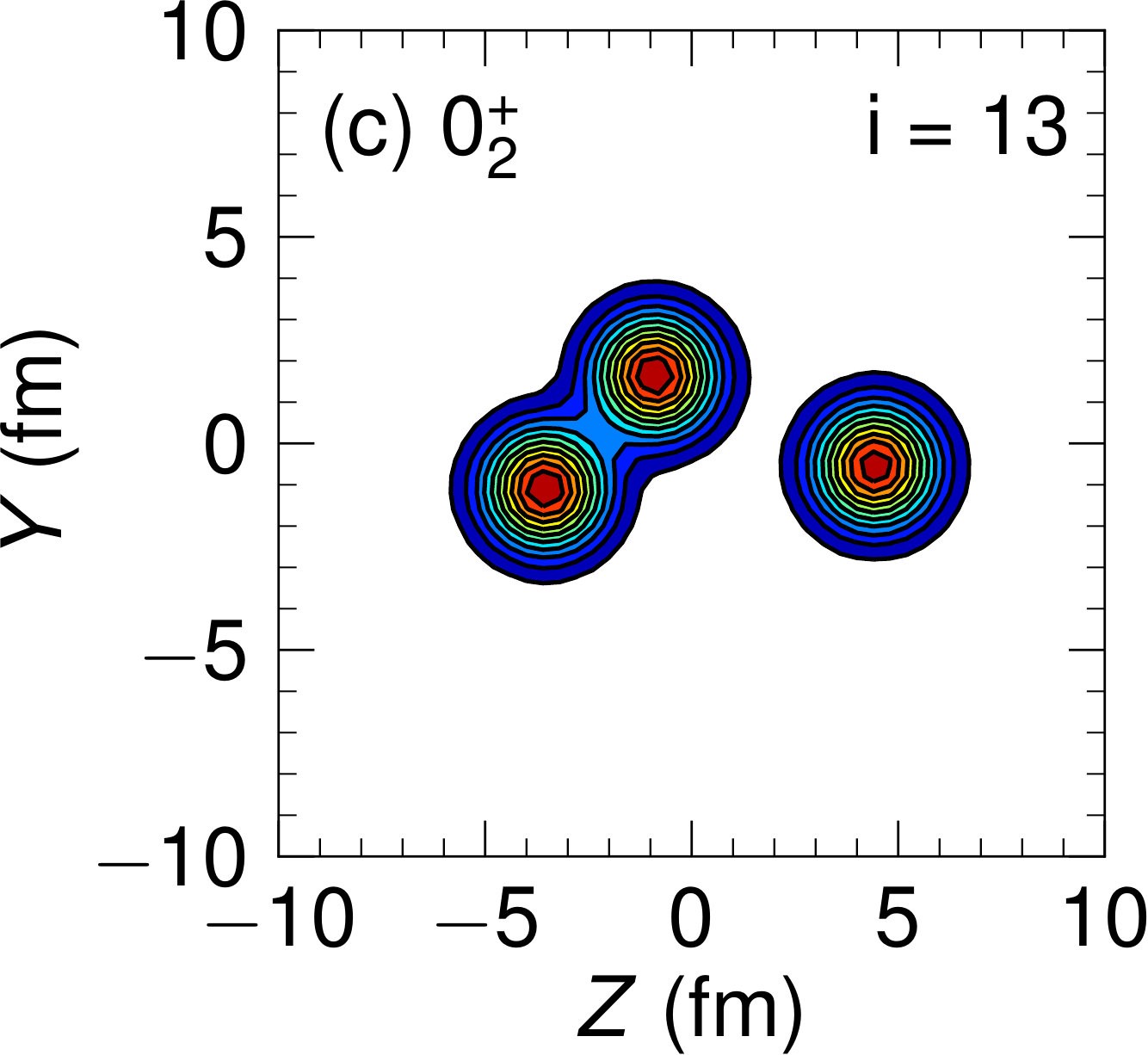}
\end{center}
\end{minipage}
\begin{minipage}{0.25\hsize}
\begin{center}
\includegraphics[keepaspectratio,width=\linewidth]{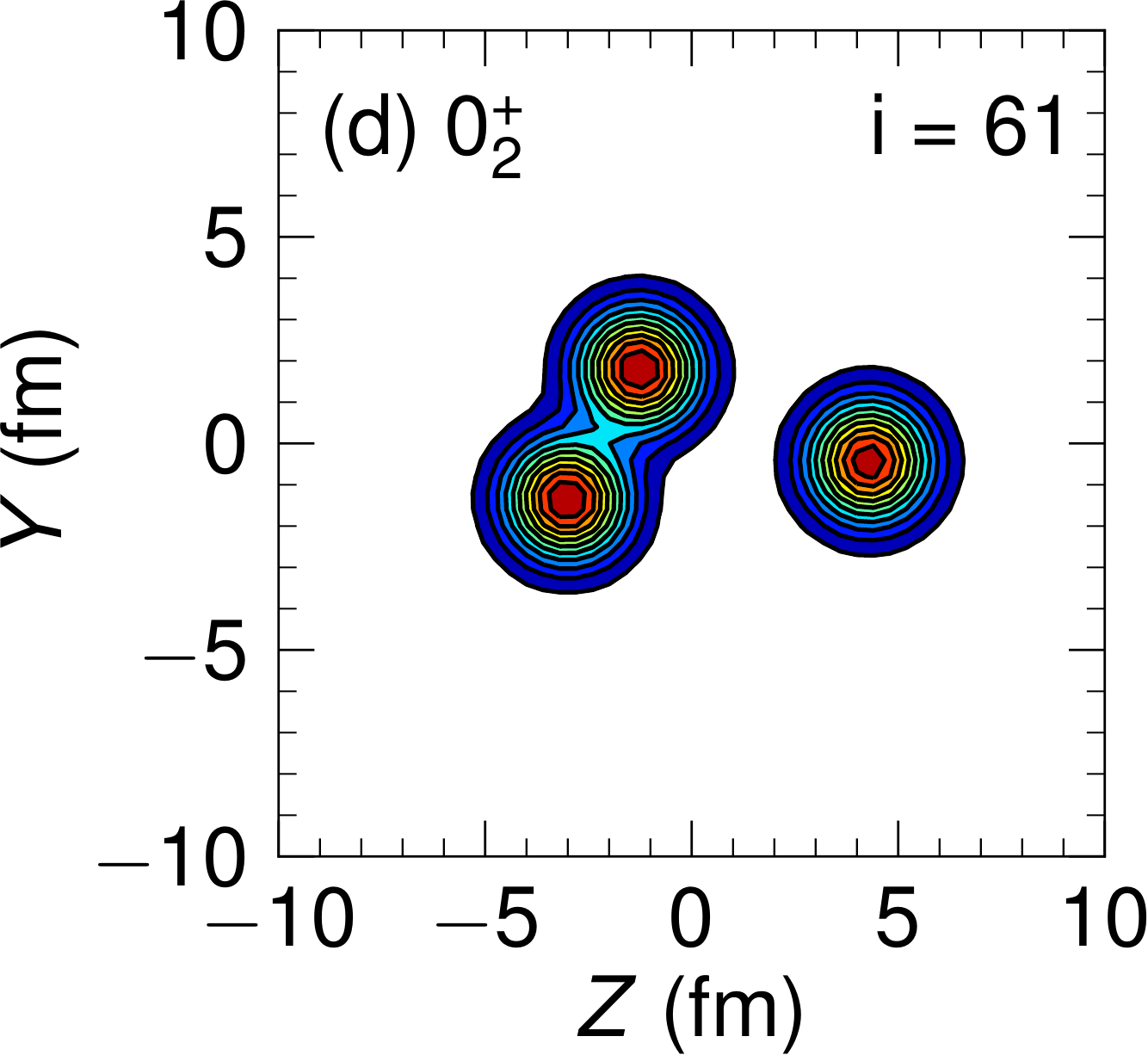}
\end{center}
\end{minipage} \\
\vspace{.1cm}\\
\begin{minipage}{0.25\hsize}
\begin{center}
\includegraphics[keepaspectratio,width=\linewidth]{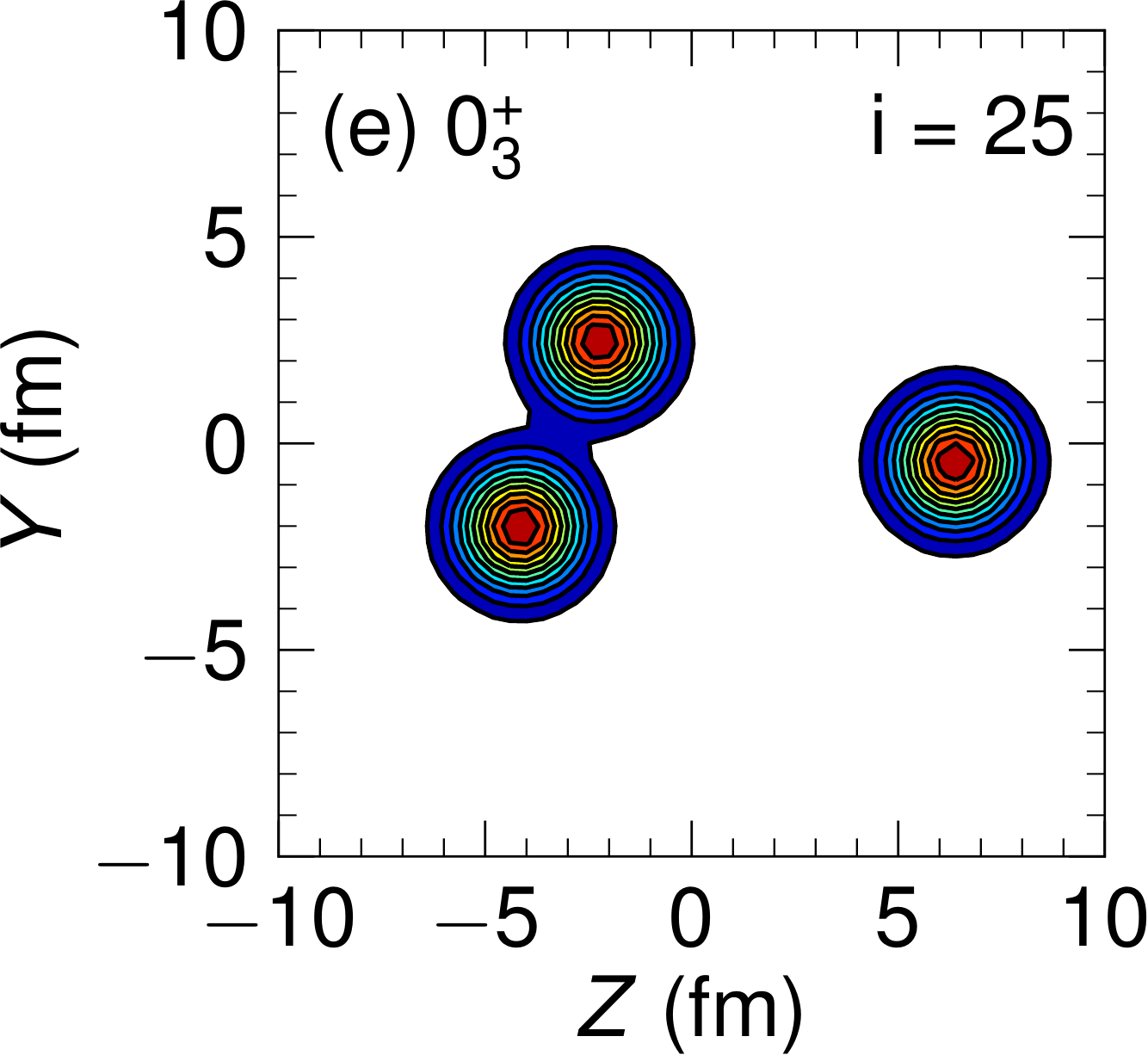}
\end{center}
\end{minipage} &
\begin{minipage}{0.25\hsize}
\begin{center}
\includegraphics[keepaspectratio,width=\linewidth]{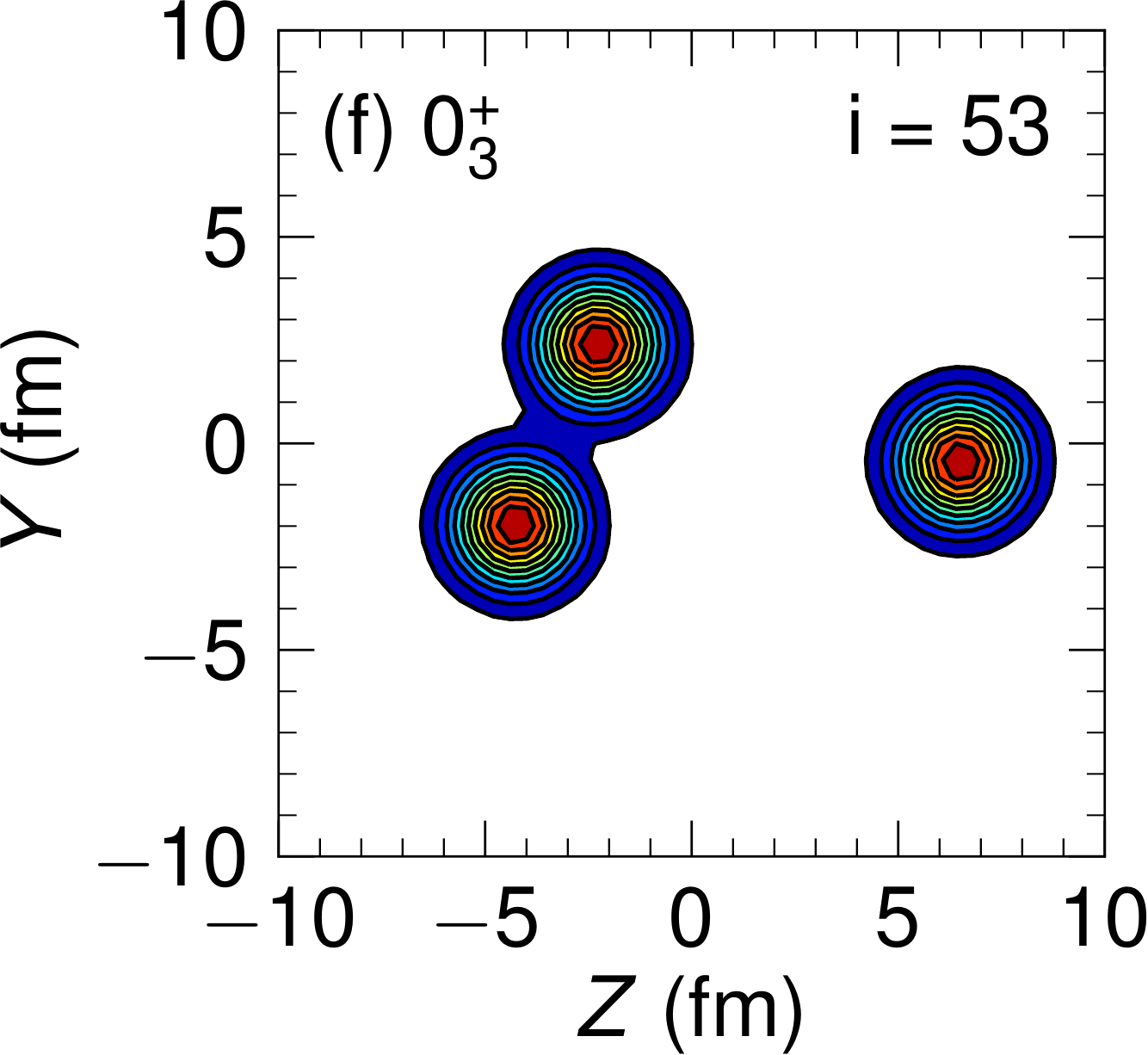}
\end{center}
\end{minipage} &
\begin{minipage}{0.25\hsize}
\begin{center}
\includegraphics[keepaspectratio,width=\linewidth]{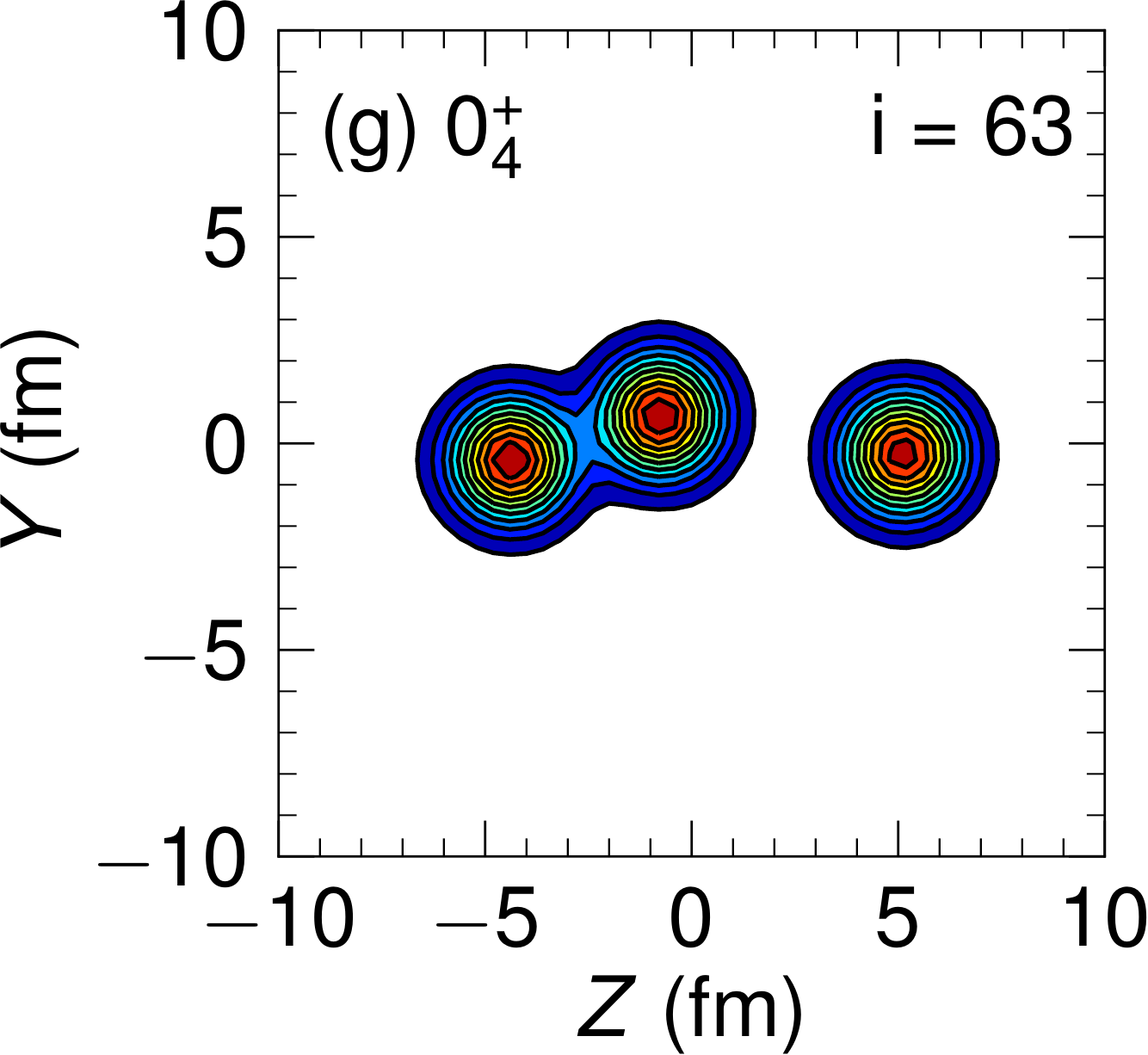}
\end{center}
\end{minipage}
\begin{minipage}{0.25\hsize}
\begin{center}
\includegraphics[keepaspectratio,width=\linewidth]{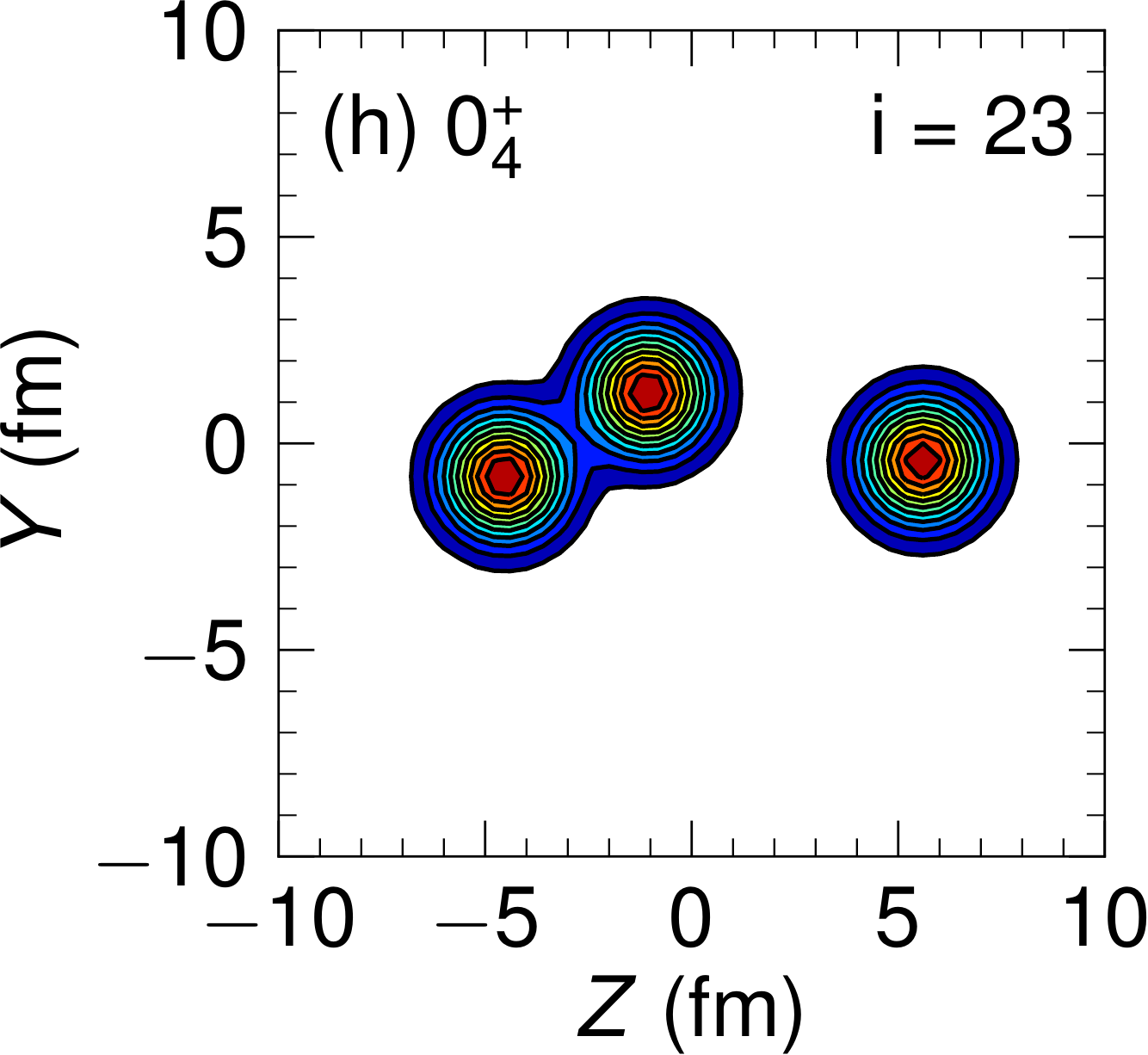}
\end{center}
\end{minipage} \\
\vspace{.1cm}\\
\begin{minipage}{0.25\hsize}
\begin{center}
\includegraphics[keepaspectratio,width=\linewidth]{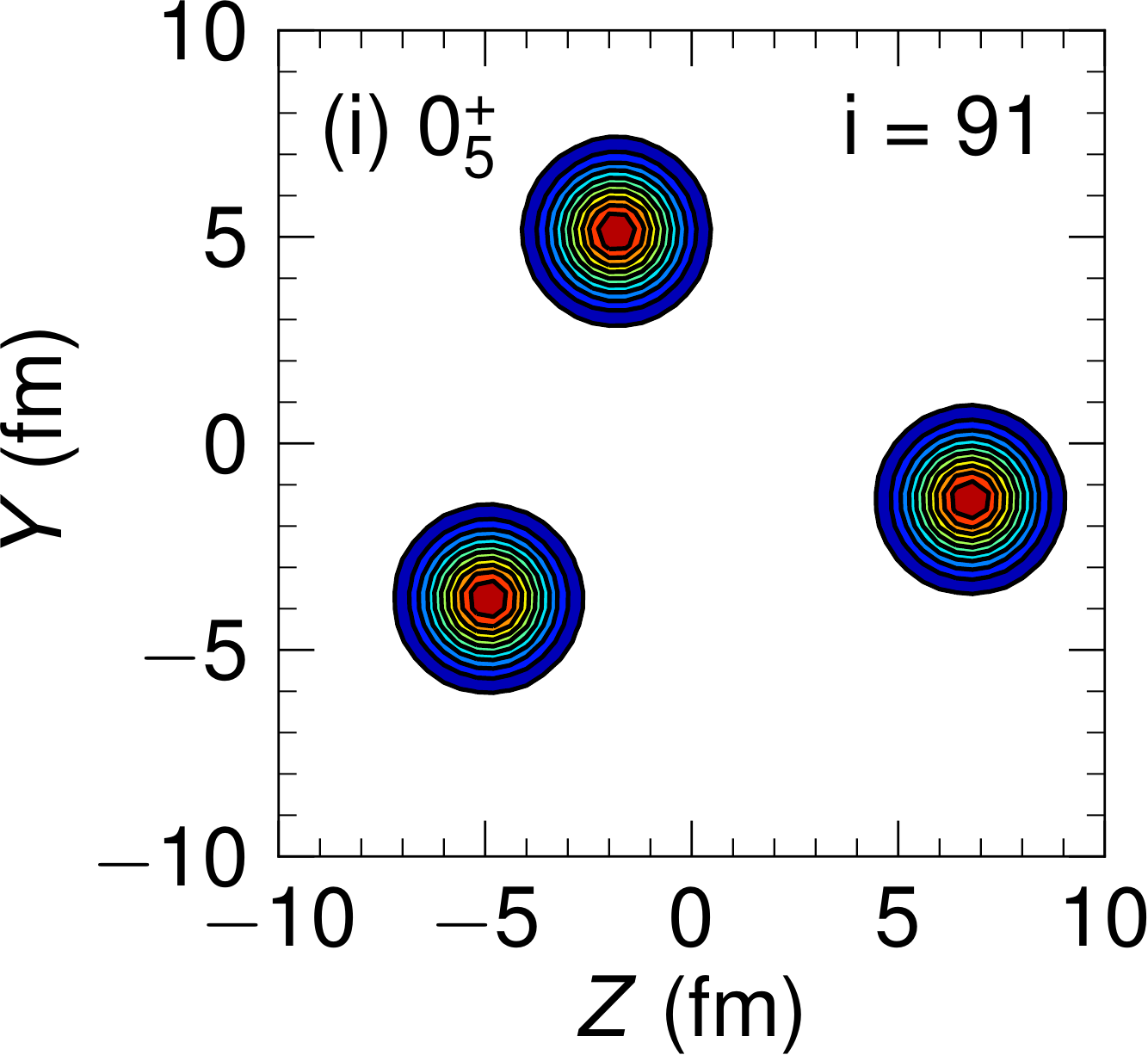}
\end{center}
\end{minipage} &
\begin{minipage}{0.25\hsize}
\begin{center}
\includegraphics[keepaspectratio,width=\linewidth]{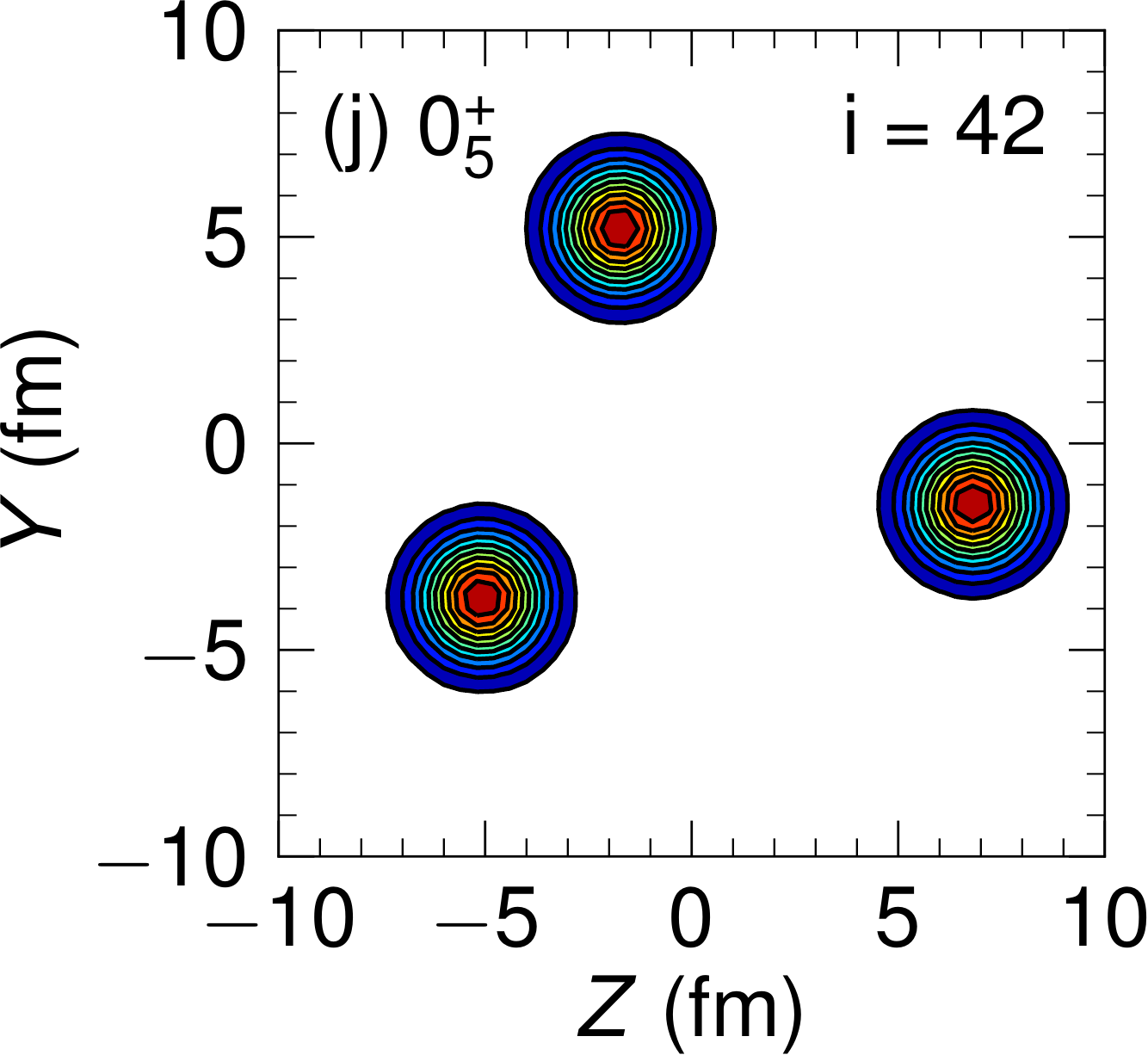}
\end{center}
\end{minipage} 
\end{tabular}
\caption{Density distribution of each excited state for the firstly [(a), (c), (e), (g), and (i)] and the secondary [(b), (d), (f), (h), and (j)] largest overlap values. }
\label{config}
\end{figure*}

\begin{table}
\caption{\label{tab2} Top three components with the largest squared overlap in each $0^+$ excited state of $^{12}$C. The number (\#) of bases indicate the order where they are chosen in the energy optimization. The symbol $w$ indicates the overlap weight. The symbols $\beta$ and $\gamma$ are the deformation parameter. The symbol $E$ is the expectation value of the energy for the basis function.}
\begin{ruledtabular}
\begin{tabular}{cccccc}
State & \# of bases& w & $\beta$ & $\gamma$ (deg)& $E$ (MeV)\\
\hline
-&0&-&0.377&29.7&$-86.11$\\
\hline
$0^+_1$ &3& 0.911 & 0.463 & 25.3 & $-85.64$ \\
&16&0.906& 0.374&36.2&$-85.82$\\
&14&0.905& 0.369&39.1&$-85.71$\\
\hline
$0^+_2$ &13& 0.449 & 2.18 & 6.84 & $-73.53$\\
&61&0.434&1.88&9.93&$-74.14$\\
&4&0.429&1.55&9.28&$-75.72$\\
\hline
$0^+_3$ &25& 0.278 & 4.07 & 8.91 & $-70.16$\\
&53&0.277&4.30&7.98&$-70.02$\\
&45&0.235&4.72&5.19&$-71.50$\\
\hline
$0^+_4$ &63& 0.326 & 3.20 & 0.732 & $-72.72$\\
&23&0.294&3.63&2.27&$-71.92$\\
&26&0.288&2.90&0.347&$-71.64$\\
\hline
$0^+_5$ &91& 0.396 & 4.44 & 35.3 & $-72.08$\\
&42&0.380&4.58&35.0&$-72.02$\\
&99&0.372&3.86&37.0&$-71.24$
\end{tabular}
\end{ruledtabular}
\label{maxov}
\end{table}

We here analyzed components of each excited state by plotting the squared overlap distribution of basis functions on the $\beta$-$\gamma$ plane. That is, 
the size of symbols in the scatter plot is proportional to the squared overlap between a component of basis functions and the total wave function for the corresponding excited state. This method has been introduced in Ref.~\cite{PhysRevC.89.031301} and revealed the shape-coexistence of shape isomers.
We also estimate the distribution function of the obtained overlap values using the weighted kernel density estimation.
Hereafter, we will discuss the results calculated with the basis functions sampled by $T_L=$ 2.5 MeV. The overlap weight is defined as $w_i^{(n)}=\left|\langle\Phi_i|\bar{\Psi}_n\rangle\right|^2$, where $n$ is the $n$th $0^+$ state and $i$ is the $i$-th component of basis functions.

The basis function with the maximum overlap weight in each excited state is tabulated in Table \ref{maxov}. 
Figure \ref{ovdis1} shows the scatter plots with the symbol size proportional to the overlap weight on the $\beta$-$\gamma$ plane in each excited state. 
In the figure, The component with the maximum overlap weight is indicated by the solid star.
Figure \ref{ovdis2} shows the overlap-weight distribution approximated with the weighed kernel density distribution.
The corresponding density distribution with the maximum overlap weight in each excited state is shown in Fig.~\ref{config}.

{\it The ground state.} We can see that large overlap-weight components are distributed around the bottom of the potential pocket [see Figs.~\ref{ovdis1}(a) and \ref{ovdis2}(a)]. This indicates that the samplings work very well. We emphasize that there are no assumptions in the samplings and did not use any results obtained by the constraint calculations.
The density configuration of the basis function with the maximum overlap weight is located at $\beta=0.46$ and $\gamma=25.3$ with $E=-85.64$ [see Fig.~\ref{config}(a)], which is slightly expanded from the lowest energy minimum [see Fig.~\ref{den1}(a)] in the $\beta$-$\gamma$ constraint calculation. This configuration shows the Be + $\alpha$ configuration. The basis functions with the second and third largest overlaps weights have the equilateral triangular configuration [see Fig.~\ref{config} (b)]. They would be seeds to enhance the iso-scalar monopole transition strength to the second $0^+$ state. 

{\it The second $0^+$ state.}
The main components of the second $0^+$ state is distributed around the local minimum at $\beta=1.75$ and $\gamma=5.0^\circ$ [see Figs.~\ref{ovdis1} (b) and \ref{ovdis2} (b)]. This state has been considered a gas-like state. Indeed, the iso-scalar monopole strength from the ground to this state is very large.
The density distributions of the main components are almost  isosceles triangular shapes [see Figs.~\ref{config} (c) and (d)]. In addition, stretched linear-chain-like configurations about $\beta\sim3.0$ are contained in state [see Fig.~\ref{ovdis2} (b)].

{\it The third $0^+$ state.}
In this state, the main components are distributed in the potential valley around $\beta=4.5$ and $\gamma=5^\circ$ [see Figs.~\ref{ovdis1} (c) and \ref{ovdis2} (c)]. This would indicate the bending vibrational mode of the linear-chain state along the direction perpendicular to the potential valley, as suggested by preceding 
works~\cite{PhysRevLett.81.5291,PhysRevLett.98.032501}. 
In this mode, the $^8$Be component vibrates for the last $\alpha$ by changing the angle [see Figs.~\ref{config} (e) and (f)]. 
This state is qualitatively consistent with the third $0^+$ state obtained using the THSR calculation by Funaki {\it et al.}~\cite{Funaki_Horiuchi_2009,Funaki_2015,Funaki_2016}. 
They have proposed that this state has the Be+$\alpha$ configuration. Our calculated result also supports this interpretation. 
This state is also very similar to the energy of the third $0^+$ state obtained with the real-time evolution method by Imai {\it et al.}~\cite{Imai_Tada_Kimura_2019}. 
Our obtained density distributions are also similar to their results. In contrast, it seems that this state is missing in the AMD calculation by Kanada-En'yo \cite{Kanada-Enyo_2007} due to the lack of the model space.
It is also interesting that the calculated energy of our third $0^+$ state may correspond to the fourth $0^+$ state obtained with the complex scaling method \cite{PhysRevC.71.021301,Ohtsubo-Kamimura}, which is large decay width below the traditional third $0^+$ state.

{\it The fourth $0^+$ state.}
This state would be the stretched vibrational mode of the linear-chain configuration. The amplitude of the vibration is between $\beta=1.5$ and 5.0 [see Figs.~\ref{ovdis1} (d) and \ref{ovdis2} (d)]. In this state, many linear-chain configurations with $\gamma\sim0^\circ$ are contained [see Fig.~\ref{config} (g) and (h)].
This state is the vibrational mode of the linear-chain configuration~\cite{PhysRevLett.104.212503}.
In general, three particles on a one-dimensional coordinate have two normal modes, symmetric and asymmetric vibrations. The present case could correspond to the former since the $\beta$ value distributes widely.
However, the search for the asymmetric vibration in higher energy would be intriguing, called billiard or Newton's cradle mode. 
The obtained energy and density distribution of the linear-chain configuration are very similar to other microscopic calculations such as Refs.~\cite{Funaki_2016,Imai_Tada_Kimura_2019,Suh10,Kanada-Enyo_2007}. 
Our fourth $0^+$ state corresponds to the third $0^+$ state obtained with the full microscopic AMD calculation by Kanada-En'yo \cite{Kanada-Enyo_2007}.

{\it The fifth $0^+$ state.}
The main components of this state are distributed around the saddle point at $\beta=2.5$ and $\gamma=30^\circ$ [see Figs.~\ref{ovdis1} (e) and \ref{ovdis2} (e)]. The many configurations in this states are large equilateral triangular shapes [see Fig.~\ref{config} (i) and (j)]. That is, it would be the breathing mode of the large equilateral triangle. Note that the energy of this state is $-76.80$~MeV, which is much lower than that of the saddle point by about 5~MeV. 
This state is directly connected to the triple $\alpha$ decay channel. 
That is, this state would play an important role in the synthesis of $^{12}$C. 
We emphasize that this state cannot be seen in other microscopic calculations. We, for the first time, find this state by expanding the model space. 

\subsection{Discussion}
\begin{figure}[htbp]
\begin{center}
\includegraphics[keepaspectratio,width=\linewidth]{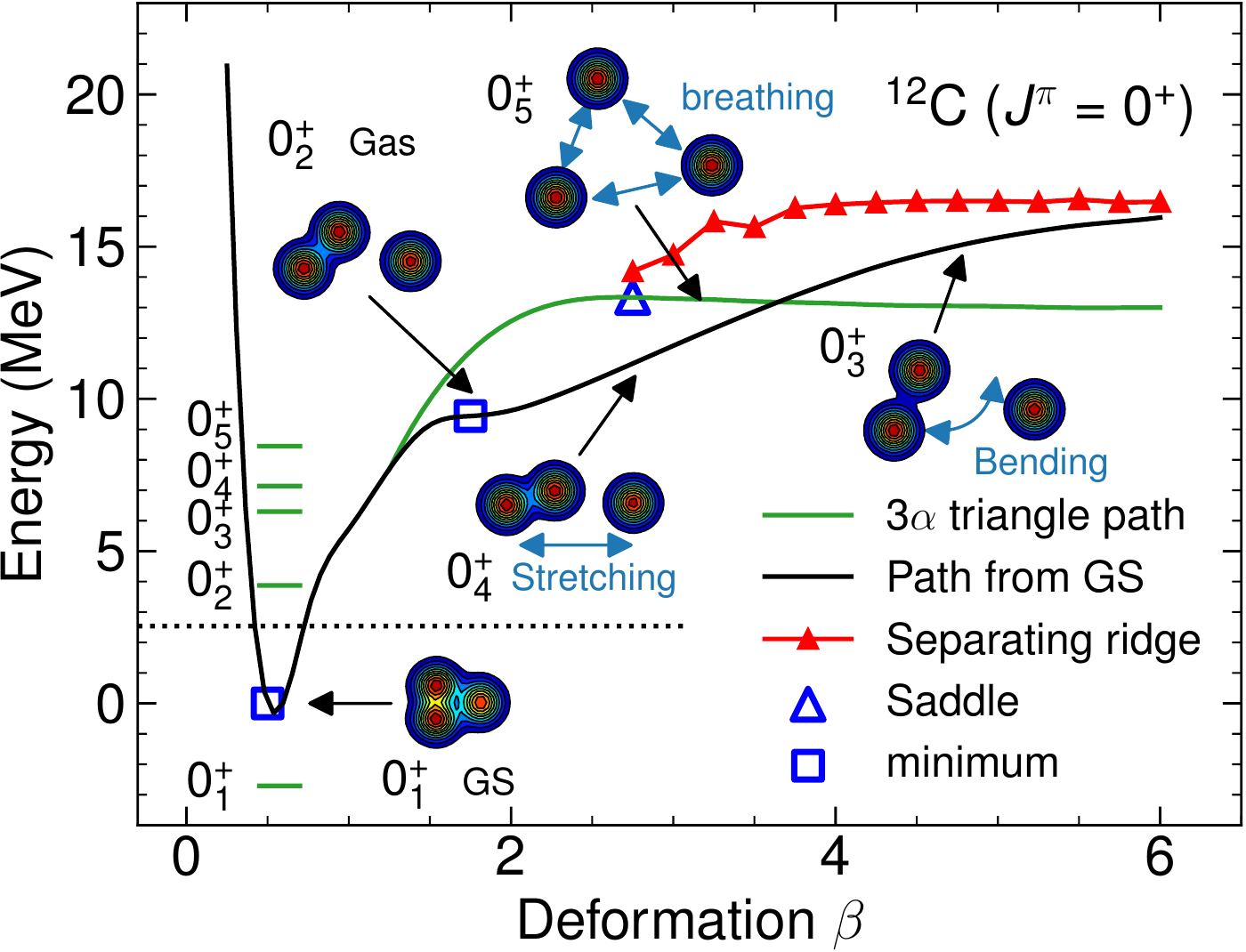}
\end{center}
\caption{Potential energy surface for $J^{\pi}=0^+$ of $^{12}$C. The origin of the potential energy is the minimum energy obtained by the $\beta$-$\gamma$ constraint calculations.  The solid black line indicates the minimum energy path from the ground state leading to the linear-chain decay channel. The solid green line indicates a path leading to the equilateral triangular decay channel. The solid line with triangles indicates the separating ridge. The dashed line indicates the three $\alpha$'s decay threshold energy. The open square and triangle represent the energy minimum and saddle point, respectively. The shape configuration of the main components in each state is depicted. The obtained energies of the $0^+$ states are also depicted.}
\label{1d}
\end{figure}

We summarize all calculated results in Fig.~\ref{1d}. An essential feature of the PES is that the minimum energy path is directly connected to the linear-chain configuration (see the solid line in Fig.~\ref{1d}), although the saddle point with the lowest energy is located at the three $\alpha$ triangular configuration (see the open triangle in Fig.~\ref{1d}). The gas-like state (second $0^+$ state) consists of the components distributed around the local minimum in the energy-minimum path connecting to the linear-chain decay threshold. 
This shallow potential pocket would emerges due to the large shell gap at the 1:3 deformation in Nillson's diagram using the deformed harmonic oscillator potential \cite{Bohr75}. 
The bending and stretched (symmetric) 
vibrational modes of the linear-chain configurations
also emerge with the components distributed in this energy-minimum path (third and fourth $0^+$ states). 
Although the saddle point (threshold energy) for the path to the linear chain is much higher than that of the lowest energy located at the three $\alpha$ triangle path, these two potential-energy valleys 
(triangle and linear)
are well separated by the ridge structure (see the solid line with triangles in Fig.~\ref{1d}). This ridge structure is a key to the emergence of not only the gas-like state but also the linear-chain vibrational states. An interesting point is that the main components of the fifth $0^+$ state are distributed just around the lowest-energy saddle point in the three $\alpha$'s triangular path. There is a very shallow PES in this channel.  
These shallow and ridge potential structures after a saddle point can be also seen in lighter mass nuclei such as $^{28}$Si and $^{32}$S \cite{PhysRevC.83.054319}.

In comparison to other microscopic calculations, we succeed in describing all the states including ones often missing in other microscopic calculations in a unified way for the first time.
Our calculated results, especially for $0_3^+$ and $0_4^+$ states, are consistent with that of Funaki {\it et al.}~\cite{Funaki_Horiuchi_2009, Funaki_2015,Funaki_2016}. In our calculation, an $\alpha$ cluster at the edged of the linier-chain-like configuration in the $0_4^+$ state bends, which is consistent with that of the $0_3^+$ state obtained by Kanada-En'yo \cite{Kanada-Enyo_2007}. However, in her calculation, our $0_3^+$ state is missing.
Our calculated results of the $0_3^+$ and $0_4^+$ states are also consistent with that of Imai {\it et al.} \cite{Imai_Tada_Kimura_2019}, although they did not show the $0_5^+$ state in their results.
It seems that their real-time evolution path may be trapped in the potential valley leading to the linear-chain decay channel. To obtain the $0_5^+$ state, it is necessary to explore the potential valley leading to three $\alpha$ decay channel by overcoming the separating ridge. The classical real-time evolution path, including the time-dependent Hartree-Fock path, without the quantum tunneling often suffers the trapping in a well-developed potential valley. 

Although the second $0^+$ state has been understood as the gas-like state described by $(0s)^3$ for three $\alpha$ clusters, our results suggests that it belongs to a member of the linear-chain states, rather than the equilateral triangular states. Indeed, this state consists of many isosceles triangular configurations emerging at a shallow local minimum in the potential valley directly leading to the linear-chain decay channel (see the solid line in Fig.~\ref{1d}). 
This state is not connected to the potential valley leading to the equilateral triangular decay channel (see the green solid line in Fig.~\ref{1d}). In this respect, the fifth $0^+$ state containing many components in the equilateral triangular path may play an important role in the synthesis of $^{12}$C with the three-body reaction.

Recent experimental data have shown that the direct three $\alpha$ decay width of the second $0^+$ state is negligible \cite{Del17,Bishop19}.
Itoh {\it et al.} further found that the $0_3^+$ state only decays into $^8$Be($0^+$)+$\alpha$ channel and the $0_4^+$ state prefers
$^8$Be($2^+$)+$\alpha$ decay \cite{Itoh_Akimune_2013}, which can support the above argument, since a linear-chain structure is considered to contain non-negligible partial wave components other than an S-wave. 
These data would support our calculated results.

\section{Summary}

We have proposed a novel efficient RXMC method to sample important Slater determinants from a given Hamiltonian. Using the RXMC method, we can sample Slater determinants following the Boltzmann distribution on the PES constructed by the Hamiltonian. We have also shown how the obtained Slater determinants are sorted out by optimizing the energy of each excited stated. 
To eliminate unphysical splits of excited states due to numerical errors, we transform all configurations so as to align the principal axis.
To analyze the obtained excited states, we have also projected the distribution of the sampled basis functions embedded onto the $\beta$-$\gamma$ PES. We have applied this method to the ground and excited states of the $^{12}$C nucleus.

We employed the Bloch-Brink $\alpha$ cluster model with the Volkov No.~2 interaction,
where the Majorana exchange parameter was chosen to reproduce the $\alpha$-$\alpha$ 
scattering phase shifts.
to describe three $\alpha$ cluster states in $^{12}$C. We performed the RXMC samplings with diffident temperatures. Then, we found that the basis functions sampled with a temperature of $T_L=2.5$ MeV describe well the excited states so that the energies from the ground to fifth $0^+$ states are well converged.
We also investigated the structure of the PES calculated with the $\beta$-$\gamma$ constraint method. We found two prominent valleys: one is the path with the lowest saddle point connecting to the equilateral triangular decay channel. Another is the path connecting to the linear-chain configuration. The ridge structure well separates these two valleys. 

To analyze the properties of the obtained $0^+$ states, we depicted the weight distribution of the basis functions overlapping with the obtained states. By embedding the calculated weight distributions onto the $\beta$-$\gamma$ PES, we investigated the main components of the basis functions in each excited state.
We found that the ground state consists of small equilateral triangular configurations. The second $0^+$ state, often called the gas-like state, is built on the shallow local minimum at the beginning of a potential valley directly connected to the linear-chain decay channel. The third $0^+$ state emerges in the linear-chain potential valley and has the bending vibrational model in this valley structure. The fourth $0^+$ state has the stretched vibrational mode in the linear-chain potential valley. The fifth $0^+$ state has many large equilateral triangular components, which is directly connected to the ground state.

As shown in this paper, the RXMC method can efficiently sample important Slater determinants from a huge model space and eliminate the continuum solutions. In many previous calculations, their results strongly depend on individual research because its sampling was often performed by hand. The RXMC method completely removes this defect. The great advantage of the RXMC method can be applied to not only the cluster model but also other general models, including Antisymmetrized quasi cluster model (AQCM),
where the effect of the spin-orbit interaction is incorporated by replacing $\alpha$ clusters to
quasi clusters~\cite{Itagaki11}.
We hope that this method is helpful to describe much-complicated fission dynamics in heavy mass nuclei.
 
The numerical calculations were carried out on
Yukawa-21 at YITP in Kyoto University.

\bibliography{cluster.bib}

%

\end{document}